\documentclass[fleqn,usenatbib]{mnras}

\usepackage{newtxtext,newtxmath}

\usepackage[T1]{fontenc}

\DeclareRobustCommand{\VAN}[3]{#2}
\let\VANthebibliography\thebibliography
\def\thebibliography{\DeclareRobustCommand{\VAN}[3]{##3}\VANthebibliography}

\usepackage{graphicx}	
\usepackage{amsmath}	
\usepackage{subcaption}
\usepackage{adjustbox}

\usepackage{algorithm}
\usepackage{algpseudocode}
\usepackage{amsmath}

\usepackage{pdflscape}

\title[COBALT-BLUE]{COBALT-BLUE: a fast and accurate model for gravitational self-lensing binary systems}

\author[D.J. Dixon et al.]{
D. J. Dixon,$^{1}$\thanks{E-mail: D.J.Dixon2@newcastle.ac.uk}
Adam Ingram,$^{1,2}$
Matthew J. Middleton,$^{3}$
Adam McMaster,$^{3}$
Grzegorz Wiktorowicz,$^{4}$
\newauthor \;
Hugh Dickinson,$^{5}$
Allison Crossland$^{6}$
\& Eric C. Bellm$^{7}$
\\
$^{1}$School of Mathematics, Statistics, and Physics, Newcastle University, Newcastle upon Tyne NE1 7RU, UK\\
$^{2}$Centre for Extragalactic Astronomy, Department of Physics, Durham University, South Road, Durham DH1 3LE, UK\\
$^{3}$Department of Physics and Astronomy, University of Southampton, Highfield, Southampton SO17 1BJ, UK\\
$^{4}$Nicolaus Copernicus Astronomical Center, Polish Academy of Sciences, Bartycka 18, PL-00-716 Warsaw, Poland \\
$^{5}$ School of Physical Sciences, The Open University, Milton Keynes, MK7 6AA, UK \\
$^{6}$Michigan State University, 567 Wilson Rd, East Lansing MI 48824, USA \\
$^{7}$DIRAC Institute, Department of Astronomy, University of Washington, 3910 15th Avenue NE, Seattle, WA 98195, USA
}

\date{Accepted XXX. Received YYY; in original form ZZZ}

\pubyear{\the\year{}}

\begin{document}
\label{firstpage}
\pagerange{\pageref{firstpage}--\pageref{lastpage}}
\maketitle

\begin{abstract}
We present \texttt{COBALT-BLUE}, a fast public model for the optical light curve of a luminous star in a non-accreting binary with a compact object. The model includes gravitational self-lensing, Doppler beaming, ellipsoidal variations, eccentric orbital geometry, and wavelength-dependent limb darkening. In sufficiently edge-on systems, it predicts sharp periodic self-lensing flares together with phase-locked orbital modulation outside the flare.
We improve upon earlier models by including non-linear limb darkening laws based on stellar atmosphere simulations to calculate the self-lensing flare profile. This enables light curves to be self-consistently calculated in multiple wavebands from one set of physical parameters, thus offering the prospect of constraining binary parameters such as compact object mass from photometry alone if multiple self-lensing flares are observed. 
We also include the option to utilise simpler limb darkening prescriptions to save computational expense. As a proof of principle, we fit our model to ZTF data of 14 candidate self-lensing sources. The model describes the flare morphology well for most candidates, but we find that the best fitting parameters are not physically plausible. We conclude that the observed flares are most likely stellar flares and not self-lensing events. In future, we plan to use the model to systematically search optical survey data for signatures of binary motion.
\end{abstract}

\begin{keywords}
binaries: general -- stars: black holes -- stars: neutron -- gravitational lensing: micro
\end{keywords}



\section{Introduction}

Stellar evolution models predict there to be $\sim 10^8$ stellar black holes (BHs) in our Galaxy \citep{Wiktorowicz2019,Olejak2020}. However, most of these are undetectable since BHs do not emit radiation (or, more accurately, they are thought to emit only a tiny luminosity of Hawking radiation; \citealt{Hawking1975}). The Galactic BHs that we know of are nearly all in X-ray binary systems. These are close binaries between a BH and a luminous companion star in which material from the companion falls onto the BH to generate a bright X-ray signal during transient episodes known as outbursts. Around 70 such systems have been discovered in this way \citep{blackcat}. There are orders of magnitude more BHs in the Galaxy predicted to be in binaries too wide for mass transfer to occur, and even more still that are isolated \citep{Wiktorowicz2019,Wiktorowicz2021}.

So how can Galactic BHs be detected without accretion from a binary partner? For isolated BHs, there are only two potential ways. First, the expected very low level of accretion onto the BH from the interstellar medium may produce a large enough radio signal to be detected in future by sensitive radio telescopes \citep{Fender2013,Gaggero2017}. Second, flares in the brightness of background stars are caused by gravitational lensing when a foreground BH passes in front of it as both objects move through the Galaxy. Thousands of such chance-alignment micro-lensing events have been detected by optical surveys \citep{OGLE,Bond2001,KMTNet}, but no confirmed BHs have been discovered from photometry alone due to the degeneracy between the lens mass and distance \citep{Bennett2002, Mao2002}. The degeneracy can be broken by including astrometry; i.e. directly observing a deflection in the apparent path of the source due to the lens. Recently, the first BH was discovered via this method, with a mass of $\approx 7~M_\odot$ \citep{Sahu2025}.

There are several methods available to detect BHs in non-interacting binary systems. One is simply measuring radial velocity curves for many stars, which has yielded several detections of dark binary partners \citep{Thompson2019,Giesers2019,Jayasinghe2021,An2025}. Although this method can be effective, it is very expensive, requiring time-resolved optical spectroscopy for many stars. Another avenue is discovery via astrometry; i.e. directly observing the binary motion of a luminous star in the presence of a dark binary partner. \textit{Gaia} is predicted to be able to detect $\sim$hundreds of BHs in this way \citep{Yamaguchi2018}. At the time of writing, three have been detected, known as \textit{Gaia} BH1 \citep{GaiaBH1}, BH2 \citep{GaiaBH2} and BH3 \citep{GaiaBH3}. The first two have masses of $\approx 9~M_\odot$, whereas BH3 has a very large mass of $\approx 33~M_\odot$. In addition, a $\sim 3-4 M_\odot$ object has been reported in the so-called `mass gap' between BHs and neutron stars \citep{Wang2024}. In future, it will be possible to detect close binary systems from the gravitational wave signal they generate, primarily at twice the orbital frequency \citep[e.g.][]{McMillan2026}.

Here we focus on gravitational self-lensing events \citep{Maeder1973}, whereby an $\sim$ edge-on observer sees light from the luminous star gravitationally lensed by the BH (or neutron star) once per orbital period. A major advantage of self-lensing events over chance-alignment events is that they repeat each orbital cycle and so signal-to-noise can be stacked up over many events, and candidates can be followed up with, for example, radial velocity measurements. A further advantage is that accompanying effects to the microlensing events help to break parameter degeneracies. For instance, a Doppler modulation in the observed flux from the luminous star results from its orbital motion. 

Current and upcoming large photometric optical surveys such as \textit{the Transiting Exoplanet Survey Satellite} (TESS; \citealt{TESS}), \textit{the Zwicky Transient Facility} (ZTF; \citealt{ZTF}), and \textit{the Legacy Survey of Space and Time} (LSST; \citealt{LSST}) are ideal for searching for such self-lensing systems. \cite{Wiktorowicz2021} predicted that $\sim 90$, $\sim 1800$ and $\sim 8000$ self-lensing systems containing either a BH or a neutron star will be visible in TESS, ZTF and LSST respectively, and \cite{Wiktorowicz2025} showed that the observed distributions can be used to constrain supernova physics (also see \citealt{Sajadian2025} for more detailed TESS predictions). However, although 5 self-lensing binaries containing a white dwarf have been discovered in \textit{Kepler} data \citep{Kruse2014,Kawahara2018,Masuda_etal2019,Sorabella2024}, no BHs or neutron stars have yet been discovered with this method \citep{Yamaguchi2024}. Recently, \cite{Crossland2023} identified 19 candidate self-lensing flares in a high cadence subset of ZTF survey data. Only one flare was detected per object, meaning that alternative explanations such as stellar flares are viable, but the candidate events are worthy of further investigation.

Self-lensing flares were first explored theoretically $\sim 20-30$ years ago \citep{Gould1995,Qin1997,Beskin2002,Marsh2001,Rahvar2011}. Since then, models have been increasing in sophistication. \cite{Sahu2003} were the first to include the effect of stellar limb-darkening -- stars appearing brighter in their centres compared to their edges -- which impacts the shape of the predicted lensing flares considerably. \cite{Agol2002} and \cite{Han2016} included occultation in their calculations, which is important for the case of a white dwarf lens. \cite{MasudaHotokezaka2019} also considered the Doppler modulation resulting from orbital motion of the luminous star, and ellipsoidal variations caused by tidal deformation of the luminous star, but only applied a very simple lensing prescription. \cite{Sorabella2022} improved on the earlier models by additionally considering eccentric orbits and stellar limb darkening, albeit only with a simple linear prescription. \cite{Sajadian2024} also considered a linear limb darkening prescription, but no Doppler modulation or ellipsoidal variations.

Here we present a new, fast, accurate self-lensing model called \texttt{COBALT-BLUE} (Compact Object Binaries Accessed via Lensing Techniques -- Boosted Lensing Under Eccentricity). We include the Doppler modulation, ellipsoidal variations and eccentric orbits, and we employ the most advanced treatment of limb darkening to date, utilising the results of stellar atmosphere calculations. We then fit our new model to the candidate ZTF flares presented in \cite{Crossland2023}. We describe our model in Section \ref{sec:model}, present the fits to ZTF data in Section \ref{sec:fits}, before discussing our results in Section \ref{sec:discussion} and drawing conclusions in Section \ref{sec:conclusions}.

\section{The Model}
\label{sec:model}

Here we describe the \texttt{COBALT-BLUE} model. The input model parameters are listed in Table \ref{tab:parameters}.

\begin{table}
\begin{tabular}{|l|c|c|c|}
\hline
\textbf{Parameter} & \textbf{Symbol} & \textbf{Unit} & \textbf{Default} \\ \hline \hline
 \multicolumn{4}{|c|}{Compact object parameters} \\ \hline \hline
BH Mass & \(M_{\rm co}\) & $M_\odot$ & 25\\ \hline \hline
\multicolumn{4}{|c|}{Luminous Star (LS) parameters} \\ \hline \hline
LS Mass & $M_{\rm \star}$& $M_\odot$ & 1.0\\  \hline
LS Radius* & \(R_{\rm \star}\)& $R_\odot$ & 1.0\\\hline
LS Effective Temperature*& \(T_{\text{eff}}\) & K & 6017\\ \hline
LS Surface Gravity$^\dagger$& \(\log g\) & dex (cm.s$^{-2}$) & 4.44\\ \hline
Metallicity & $\log_{10}[M/H]$ &  - & 0.0\\ \hline \hline
\multicolumn{4}{|c|}{Orbital parameters} \\ \hline \hline
Orbital period                 & $P$ & days & 3.0\\ \hline
Inclination                 & $i$ & rad & $\pi/2$\\ \hline
Eccentricity                 & $\varepsilon$ & - & 0.0 \\ \hline
Phase of Inferior Conjunction& $\phi_c$ & rad & 0.0 \\ \hline
Time of Inferior Conjunction& $t_c$ & d& 0.0 \\ \hline
\end{tabular}
\caption{Parameters and their default values for the binary lensing system. $\dagger$ denotes a quantity that is not a model parameter, but is instead derived from model parameters. Starred parameters can optionally be derived from the mass assuming main sequence relations. There is an additional normalisation parameter to convert magnification to observed flux.}
\label{tab:parameters}
\end{table}

\begin{figure}
\centering 
\includegraphics[width=\columnwidth,trim=0.0cm 0.0cm 0.0cm 0.0cm,clip=true]{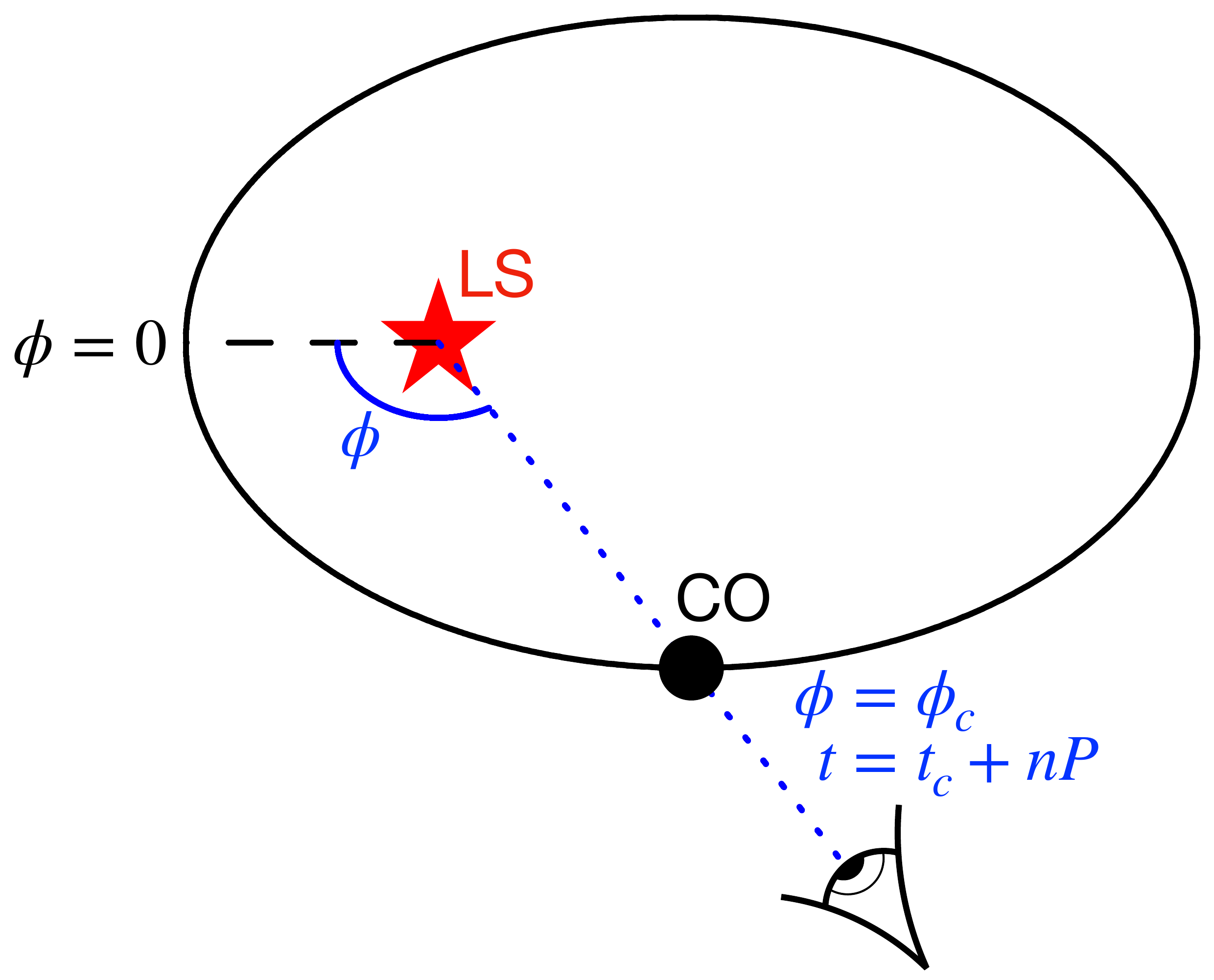}
\caption{Schematic of the assumed geometry with the luminous star (LS) at the origin and the compact object (CO) on an elliptical orbit around it with periastron at orbital phase $\phi=0$. The compact object crosses the projection of the observer's line of sight onto the binary plane at orbital phase $\phi=\phi_c$.}
\label{fig:schem}
\end{figure}

\subsection{Geometry}
\label{sec:geom}

Fig \ref{fig:schem} demonstrates our assumed coordinate system, with the luminous star at the origin and the compact object following an elliptic orbit in the $x$-$y$ plane. Under these definitions, the luminous star crosses the x-axis of the binary plane at orbital phases $\phi = 0$ (periapsis) and $\phi=\pi$ (apoapsis). The observer's position is defined by the two angles $i$ and $\phi_c$. The inclination angle $i$ is the angle between the observer's line of sight and the orbital axis, whereas $\phi_c$ is the phase of inferior conjunction (i.e. closest approach of the compact object to the observer).

We define the orbit by the following input parameters: orbital period $P$, eccentricity $\varepsilon$, compact object mass $M_{\rm co}$, and the mass of the luminous star $M_{\rm \star}$. The orbital separation is
\begin{equation}
    r_{\rm bin}(\phi) = \frac{ a ( 1 - \varepsilon^2) }{ 1 + \varepsilon \cos\phi },
\end{equation}
where $a$ is the semi-major axis, which is related to the orbital period and the masses of the two binary components by Kepler's law
\begin{equation}
    a^3 = P^2 \frac{ G M_{\rm co} (1+q) }{ (2\pi)^2 },
\end{equation}
and $q = M_{\rm \star} / M_{\rm co}$ is the binary mass ratio. In our coordinate system, the projected separation on the observer's sky between the luminous star and the compact object is
\begin{equation}
B(\phi) = r_{\rm bin}(\phi) \sqrt{ 1 - \sin^2 i ~\cos^2(\phi-\phi_c) }.
\label{eqn:proj}
\end{equation}
The orbital angular frequency is
\begin{equation}
    \frac{d\phi}{dt} = \frac{[1+\varepsilon \cos\phi]^2}{[1-\varepsilon^2]^{3/2}} \frac{2\pi}{P}.
    \label{eqn:dphibydt}
\end{equation}

We specify that inferior conjunction occurs at time $t=t_c$ (or, more accurately, one particular instance of inferior conjunction). To implement this, we convert the time of each flux measurement $t$ to instantaneous orbital phase $\phi(t)$ by solving Equation (\ref{eqn:dphibydt}) with boundary condition $\phi(t_c) = \phi_c$. For circular orbits, the trivial analytical solution is $\phi = (2\pi/P)(t-t_c)+\phi_c$. For eccentric orbits, we instead solve Equation (\ref{eqn:dphibydt}) numerically using the fourth order Runge-Kutta algorithm (the \texttt{numpy} implementation).

\subsection{Gravitational Self-Lensing}
\label{sec:SL}

For high inclination angles, a gravitational self-lensing flare will occur when the compact object transits its stellar companion at orbital phase $\phi = \phi_c$. The lensing magnification depends on the impact parameter $b=B/R_E$, where $B$ is the projected separation and $R_E$ is the Einstein radius \citep{Agol2002,Agol2003,D'Orazio2018}. In general, the Einstein radius is given by
\begin{equation}
    R_E = \sqrt{ 4 R_g D_\ell (D_s-D_\ell)/D_s },
\end{equation}
Where $R_g = G M_{\rm co} / c^2$ is the gravitational radius of the lens (compact object) and $D_\ell$ and $D_s$ are the distances from the observer to the lens and source respectively. For the case of self-lensing, we can set $D_\ell \approx D_s$ to a very good approximation, simplifying the expression for the Einstein radius to $R_E = \sqrt{ 4 R_g (D_s-D_\ell) }$. In our geometry, we can write the orbital phase-dependent Einstein radius as
 \begin{equation}
    R_E(\phi) =  \sqrt{4~R_g~r_{\rm bin}(\phi) \sin i~\cos(\phi-\phi_c) }.
\end{equation}
Note that the Einstein radius is only real for $-\pi/2 < \phi - \phi_c < \pi/2$, since gravitational lensing only occurs when the lens is closer to the observer than the source is. The impact parameter is therefore given by
\begin{equation}
    b(\phi) = \sqrt{ \frac{ r_{\rm bin}(\phi) [ 1 - \sin^2 i \cos^2(\phi-\phi_c) ] } { 4 R_g \sin i \cos( \phi - \phi_c ) } },
\end{equation}
and this is again only real for $-\pi/2 < \phi - \phi_c < \pi/2$.

The lensing magnification (ratio of observed to intrinsic flux) of a spherical, isotropically emitting source with radius $R_{\star}$ is given by \citep{Witt1994,Agol2002,Wiktorowicz2021}
\begin{equation}
\mathcal{M}_{\rm iso} = \frac{1}{\pi} \bigg[ c_F F(k) + c_E E(k) + c_{\Pi} \Pi(n,k) \bigg],
\label{eqn:MSL}    
\end{equation}
where \( F, E \) and \(\Pi\) are complete elliptic integrals of the first, second and third kind respectively,
\begin{eqnarray}
    c_F &=& - \frac{b-r}{r^2} \frac{ 4 + (b^2-r^2) / 2 }{ \sqrt{ 4 + (b-r)^2 } } \nonumber \\
    c_E &=& \frac{b+r}{2r^2} \sqrt{ 4 + (b-r)^2 } \nonumber \\
    c_\Pi &=& \frac{ 2(b-r)^2 }{ r^2(b+r) } \frac{ 1 + r^2 } { \sqrt{ 4 + (b-r)^2 } } \nonumber \\
    n &=& \frac{ 4 b r }{ (b+r)^2 } \nonumber \\
    k &=& \sqrt{ \frac{ 4 n } { 4 + (b-r)^2 } },
\end{eqnarray}
and $r=R_{\star}/R_E$. 

\begin{figure*}
    \centering
    \includegraphics[width=1\linewidth]{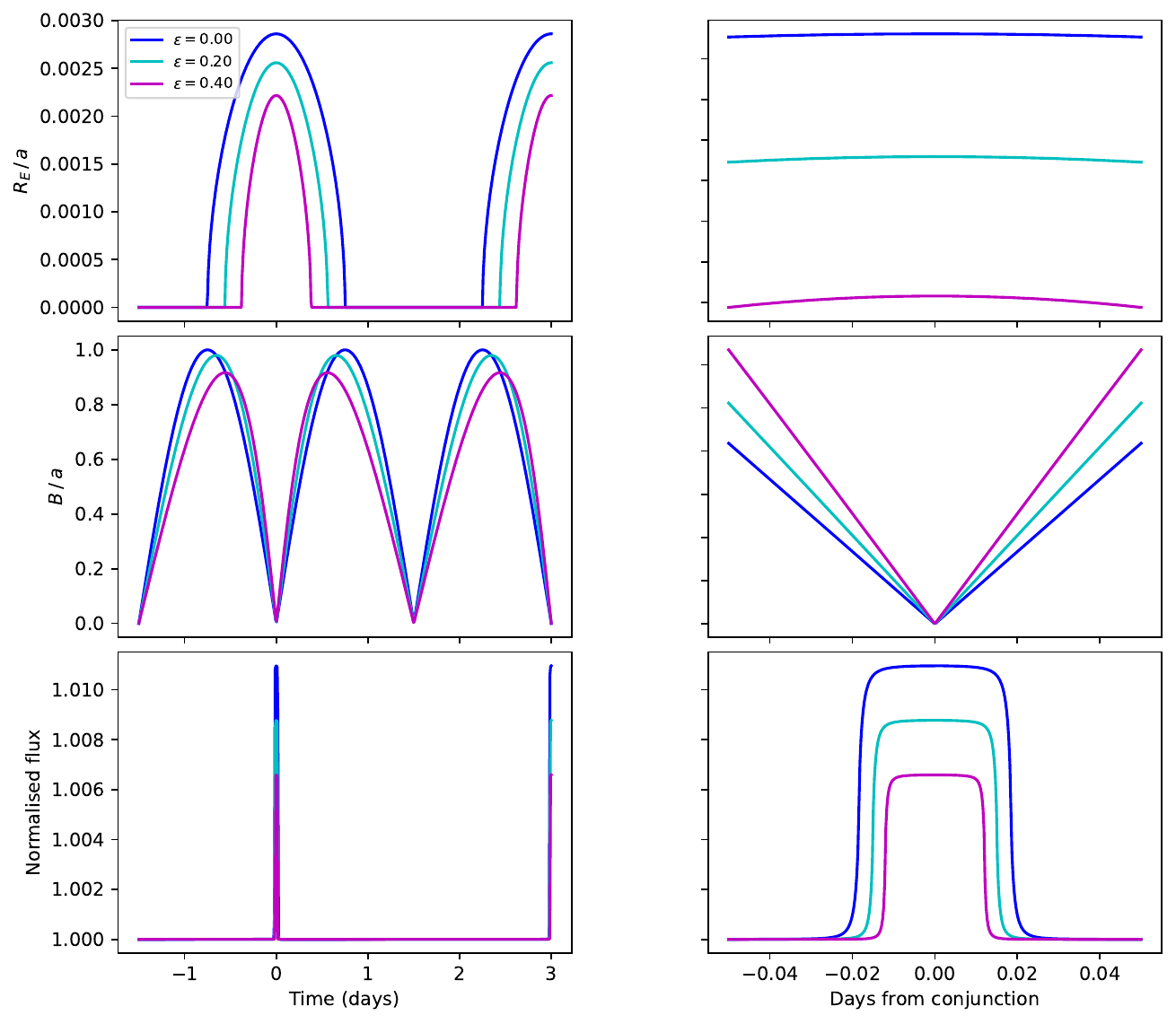}
    \caption{Einstein radius (top), projected separation (middle), and isotropic self-lensing magnification (bottom) as functions of time for $\varepsilon=0.0$, $0.2$, and $0.4$, assuming the default model parameters. The left-hand panels show several orbital cycles and the right-hand panels show a zoom around inferior conjunction. The Einstein radius and projected separation are given in units of the semi-major axis, $a$. For the case shown here with $\phi_{c}=0$, increasing eccentricity shortens and suppresses the self-lensing flare.}
    \label{fig:b_re_mag}
\end{figure*}

Fig \ref{fig:b_re_mag} shows $R_E$ (top), $B$ (middle) and $\mathcal{M}_{\rm iso}$ (bottom) versus time for three different eccentricities (as labelled). All other parameters take the default values listed in Table \ref{tab:parameters}. We see that the lensing flare has a `top hat' shape, and significant magnification occurs for impact parameters of $b \lesssim 1$. The assumed eccentricity influences the duration of the flare due to the orbital dependence of angular velocity for eccentric orbits (Equation \ref{eqn:dphibydt}). In this case, the eccentric orbits both have $\phi_c=0$, and thus the flare is shorter for higher eccentricities because the compact object crosses our sight line to the luminous star faster during inferior conjunction. Conversely, an eccentric orbit with $\phi_c=180^\circ$ would produce longer flares.

\subsection{Limb darkening}
\label{sec:LD}

Realistic stellar surfaces are not uniformly bright: the specific intensity increases from the limb to the centre because of radiative–transfer effects in the atmosphere (\emph{limb darkening}). In a self–lensing event, the lens magnifies different patches of the anisotropic stellar surface as it crosses our sight line to the luminous star. The assumed form of limb darkening therefore impacts the profile of the self-lensing flare. Following \citet{Witt1994, MandelAgol2002}, we calculate the total limb-darkened magnification as
\begin{equation}
\label{eq:Msl_cont}
\mathcal{M}_{\rm SL}
= \frac{\displaystyle \int_0^{R_{\rm \star}} I(R)\,\frac{d\!\left[R^2\,\mathcal{M}_{\rm iso}(R)\right]}{dR}\,dR}
       {\displaystyle 2 \int_0^{R_{\rm \star}} I(R)\,R\,dR } \, ,
\end{equation}
for $-\pi/2<\phi-\phi_c<\pi/2$ and $\mathcal{M}_{\rm SL}=1$ otherwise. To evaluate the above, we use concentric–ring quadrature,
\begin{equation}
\label{eq:Msl_disc}
\mathcal{M}_{\rm SL}
= \frac{\displaystyle \sum_{j=1}^J I(R_j)\,\Big[ R_j^2\,\mathcal{M}_{\rm iso}(R_j) - R_{j-1}^2\,\mathcal{M}_{\rm iso}(R_{j-1}) \Big]}
       {\displaystyle 2 \sum_{j=1}^J I(R_j)\,R_j\,\Delta R_j } \, .
\end{equation}
with $J=100$ limbs. The luminous star radius $R_{\rm \star}$ can either be set as a model parameter or be calculated from $M_{\rm \star}$ assuming the empirical main sequence mass-radius relation of \citet{DemircanKahraman1991}. The function used to describe $I(R)$ is known as the limb darkening law, and it is typically parameterised in terms of
\begin{equation}
\label{eq:mu_def_app}
\mu \equiv \cos\theta = \sqrt{1 - \left(\frac{R}{R_{\rm \star}}\right)^2}\, ,
\end{equation}
which is the cosine of the angle between the line of sight and the normal to the stellar surface, such that $\mu=1$ corresponds to the centre of the star ($R=0$) and $\mu=0$ to the limb ($R=R_{\rm \star}$).

We include several limb darkening laws in our code as options for the user \citep{eddington1926,Hestroffer1997,Claret2000,MandelAgol2002}
\begin{align}
&{\rm \it Isotropic:}~~ &\frac{I(\mu)}{I(\mu=1)} &=& &1 \nonumber \\
&{\rm \it Eddington:}~~ &\frac{I(\mu)}{I(\mu=1)} &=& &\frac{3}{5} \left[ \mu + \frac{2}{3} \right], \nonumber \\
&{\rm \it Linear:}~~ &\frac{I(\mu)}{I(\mu=1)} &=& &1 - u_1 (1-\mu), \nonumber \\
&{\rm \it Quadratic:}~~ &\frac{I(\mu)}{I(\mu=1)} &=& &1 - u_1(1-\mu) - u_2(1-\mu)^2, \nonumber \\
&{\rm \it Power-2:}~~&\frac{I(\mu)}{I(\mu=1)} &=& &1 - g( 1 - \mu_{\star}^{h} ), \nonumber \\
&{\rm \it Four-parameter:}~~&\frac{I(\mu)}{I(\mu=1)} &=& &1 - \sum_{k=1}^4 a_k ( 1 - \mu_{\star}^{k/2} ), \label{eqn:limblaw}
\end{align}
where $\mu_{\star}=\sqrt{ 1 - ( R / R_{\rm limb} )^2 }$, $R_{\rm limb} = R_{\rm \star} \sqrt{ 1 - \mu_{\rm cri} }$, and $u_k$, $g$, $h$, $a_k$ and $\mu_{\rm cri}$ are the limb darkening coefficients. The coefficient $\mu_{\rm cri}$ captures the drop off in intensity in stellar atmosphere models just inside of the limb, such that the intensity drops to zero at $R_{\rm limb} \leq R_{\rm \star}$. It is calculated as the $\mu$ value for which $|dI/dR|$ reaches a maximum \citep{Claret2025}.

The two simplest limb darkening laws (isotropic and Eddington) have no free parameters. For the other laws, we calculate the limb darkening coefficients for the user-defined photometry filter from the effective temperature $T_{\rm eff}$, metallicity $\log_{10}[M/H]$ and surface gravity $g=G M_{\rm \star} / R_{\rm \star}^2$ of the luminous star. $\log_{10}[M/H]$ is a model parameter, whereas $T_{\rm eff}$ can either be set as a parameter or derived from the Stefan-Boltzmann law and the empirical main sequence luminosity-mass relation of \citet{DemircanKahraman1991}.

To calculate the limb darkening coefficients, we use the Limb Darkening Toolkit \citep[LDTK:][]{ParviainenLDTK2015}, which fits the considered limb darkening law to the intensity profile generated by \textsc{PHOENIX} stellar atmosphere calculations \citep{Husser2013,Claret2022, Claret2023}, returning both best-fitting coefficients and uncertainty estimates. LDTK offers several methods to conduct this fit. One is Monte Carlo sampling, for which we use 250 samples, to enable the uncertainties on the limb-darkening coefficients fit to be propagated through to the overall posterior probability distribution inferred from fitting \texttt{COBALT-BLUE} to data. Alternatively, we offer the option to only use the best fitting values of the limb darkening coefficients, which reduces runtime but, for the case of exoplanet transit modelling, has been shown to lead to under-estimated uncertainties when the data quality is very high \citep{Espinoza_Jordán_2015}. We recommend the use of Monte Carlo sampling for a detailed analysis of high signal to noise data, whereas the latter option is sufficient for survey screening.

\begin{figure*}
    \centering
    \includegraphics[width=1\linewidth]{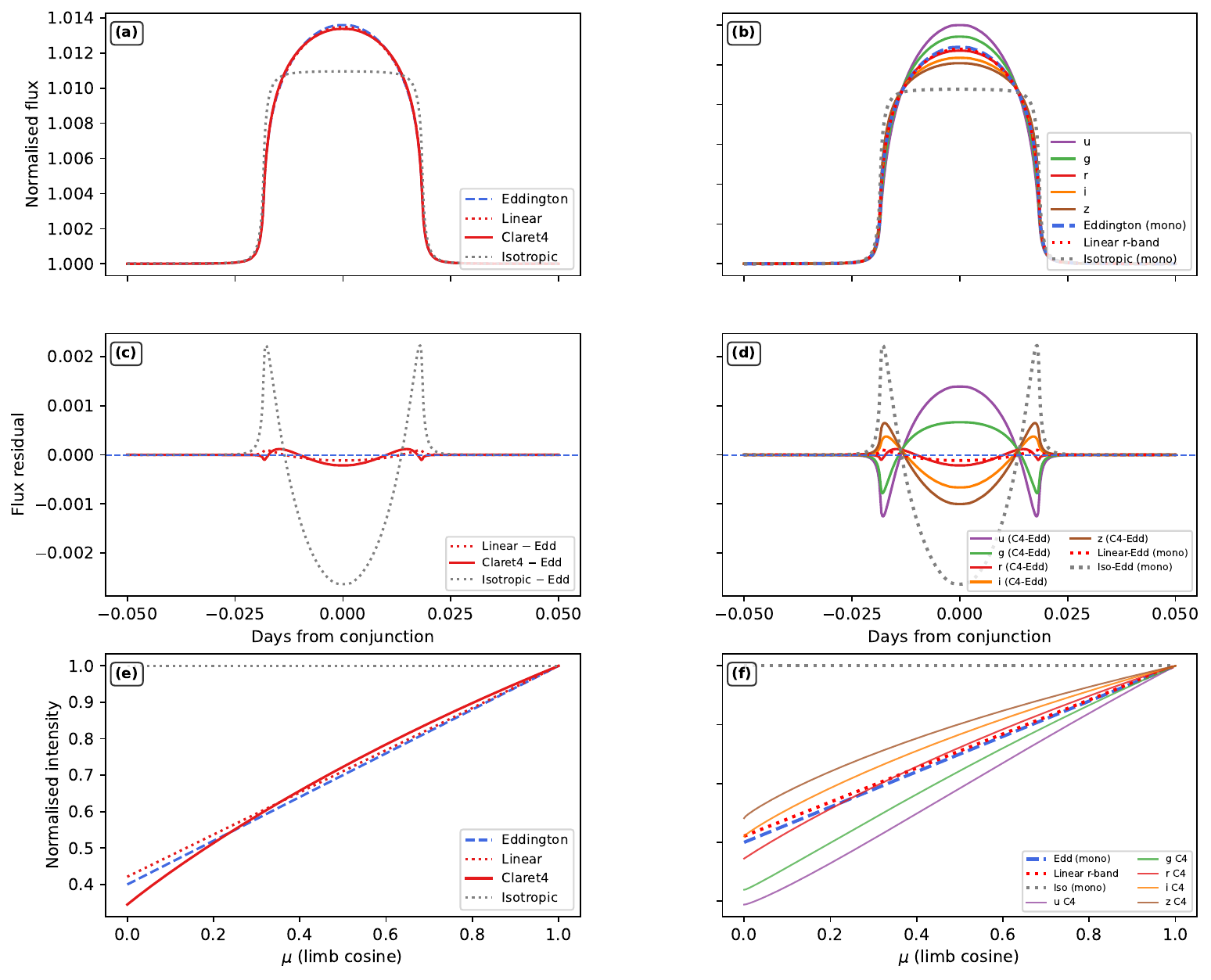}
    \caption{Self-lensing flare profiles and limb-darkening intensity laws for the default system parameters. Panels (a), (c), and (e) compare monochromatic $r$-band predictions for the isotropic, Eddington, linear, and four-parameter (Claret-4) limb-darkening laws. Panel (a) shows the normalised flare profile, panel (c) the residuals relative to the Eddington law, and panel (e) the corresponding normalised intensity profiles as functions of $\mu=\cos\theta$. Panels (b), (d), and (f) show the wavelength dependence obtained with the four-parameter law for the $u$, $g$, $r$, $i$, and $z$ bands, together with monochromatic reference models. More realistic limb-darkening prescriptions yield a more rounded, centrally peaked flare, and the deviations from simplified models are strongest in the bluer bands.}
    \label{fig:LDmag}
\end{figure*}

Fig \ref{fig:LDmag} demonstrates how the assumed radial dependence of intensity (panels e and f) influences the predicted self-lensing magnification profile (panels a and b), which we also illustrate by plotting the magnification as a ratio to the Eddington case (panels c and d). The left hand panels (a, b and c) demonstrate the influence of our choice of limb darkening law. Whereas the isotropic law yields a flat `top hat' profile, more realistic limb darkening laws produce a more curved and centrally peaked `bowler hat' profile. The four-parameter law is typically considered to describe the results of stellar atmosphere calculations most accurately \citep{Claret2000}. We see that the Eddington and linear models subtly differ from the four-parameter law, which will introduce small biases (we do not plot the other non-linear limb-darkening laws, which return very similar results to the four-parameter law). Since the four-parameter law is no more computationally intensive to fit than the other non-linear laws, we recommend its use for a detailed analysis. However, for survey screening, the Eddington law provides a good balance between physical realism and computational speed, avoiding the need for further modelling assumptions while introducing only minor bias that can be ironed out later after self-lensing candidates are identified. 

The right hand panels of Fig \ref{fig:LDmag} (b, d and f) highlight the wavelength dependence of limb darkening (employing the four-parameter law). Bluer bands (u, g) exhibit stronger limb darkening, resulting in higher flux peaks and larger residuals relative to simplified models, while redder bands (i, z) show weaker limb darkening and smaller deviations.

\subsection{Doppler Modulation}
\label{sec:DM}

Our formulation accounts for the wavelength-dependent relativistic Doppler modulation in the flux observed from the luminous star due to its motion around the barycentre. For an orbital velocity \textbf{v} with component ${\rm v}_{\rm los}$ in the observer's line of sight, the observed wavelength $\lambda_{\rm obs}$ is related to the emitted wavelength $\lambda_{\rm emit}$ as $\lambda_{\text{obs}} = \lambda_{\text{emit}}/\mathcal{D}$, where
\begin{equation}
  \mathcal{D} = \frac{ \sqrt{ 1 - ({\rm v}/c)^2 } }{ 1 - {\rm v}_{\rm los} / c },
  \label{eqn:doppler}
\end{equation}
is the \textit{Doppler factor}. In our coordinate system, the magnitude of the velocity is
\begin{equation}
    {\rm v} = \frac{1}{1+q} \frac{2\pi a}{P\sqrt{1-\varepsilon^2} } \sqrt{ 1 + \varepsilon^2 + 2 \varepsilon \cos\phi },
\end{equation}
and the line-of-sight velocity of the luminous star towards the observer is
\begin{equation}
    {\rm v}_{\rm los} = \frac{\sin i}{1+q} \frac{2\pi a}{P\sqrt{1-\varepsilon^2}} \bigg[ \sin(\phi-\phi_c) - \varepsilon \sin\phi_c \bigg].
    \label{eqn:vlos}
\end{equation}

The observed specific intensity relates to the emitted specific intensity as \citep[e.g.][]{Misner1973}
\begin{equation}
    I^{\rm obs}_{\lambda_{\rm obs}} = \mathcal{D}^5 I^{\rm emit}_{\lambda_{\rm emit}}
\end{equation}
Following \cite{Sorabella2022}, we represent the specific intensity in the reasonably narrow photometry band as a power law $I_\lambda \propto \lambda^{-\beta}$. In this case, the effective magnification from Doppler boosting becomes
\begin{equation}
    \mathcal{M}_{\rm DB} = \frac{I^{\rm obs}}{I^{\rm emit}} = \mathcal{D}^{5-\beta}.
\end{equation}
Note that this is exactly equivalent to setting the magnification to $\mathcal{D}^{3-m}$, where $I_\nu \propto \nu^m$. Assuming that the stellar spectrum is a blackbody with temperature $T_{\rm eff}$, we approximate the power law index as \citep{Mihalas1978}
\begin{equation}
\beta = 5 - \frac{hc}{\lambda k_B T_{\rm eff}}\frac{e^{hc/\lambda k_B T_{\rm eff}}}{e^{hc/\lambda k_B T_{\rm eff}} - 1},
\label{eq:spectral_index_eddington}
\end{equation}
where $\lambda$ is the centroid wavelength of the photometry band.

\subsection{Ellipsoidal Variations}
\label{sec:EV}

Tidal distortion of the luminous star by the compact object produces periodic photometric variability as the projected stellar shape changes with orbital phase. At leading order in eccentricity, the ellipsoidal signal is dominated by a modulation at approximately half the orbital period, while eccentricity introduces an additional phase dependence through the varying orbital separation. We therefore write
\begin{equation}
\mathcal{M}_{\rm EV}
=
1 - A_{\rm EV}\cos\!\left[2(\phi-\phi_c)\right].
\end{equation}

At leading order in eccentricity, the amplitude may be approximated as \citep[e.g.][]{MasudaHotokezaka2019}
\begin{equation}
A_{\rm EV}(\phi)
=
\alpha\,\sin^2 i\,
\frac{M_{\rm co}}{M_{\rm \star}}
\left(\frac{R_{\rm \star}}{a}\right)^3
\left(\frac{1+\varepsilon \cos\phi}{1-\varepsilon^2}\right)^3.
\end{equation}

The appearance of the inverse mass ratio relative to the Doppler case reflects the underlying physics: Doppler modulation depends on the motion of the luminous star about the barycentre, whereas ellipsoidal variations are driven by the tidal field imposed by the companion. The constant $\alpha$ encapsulates gravity-darkening and limb-darkening effects \citep{Morris1993}. For simplicity we set $\alpha = 1$, following \citet{MasudaHotokezaka2019} and \citet{Sorabella2022}. This treatment is approximate but sufficient for the present work. In the eccentric case, the ellipsoidal amplitude acquires an additional phase dependence through the varying orbital separation, leading to a modulation that is strongest near periapsis and weakest near apoapsis. For our proof-of-principle ZTF fits, we adopt the circular-orbit form with constant amplitude but the general form provides a natural extension for future applications of \texttt{COBALT-BLUE}.


\begin{figure*}
    \centering
    \includegraphics[width=1\linewidth]{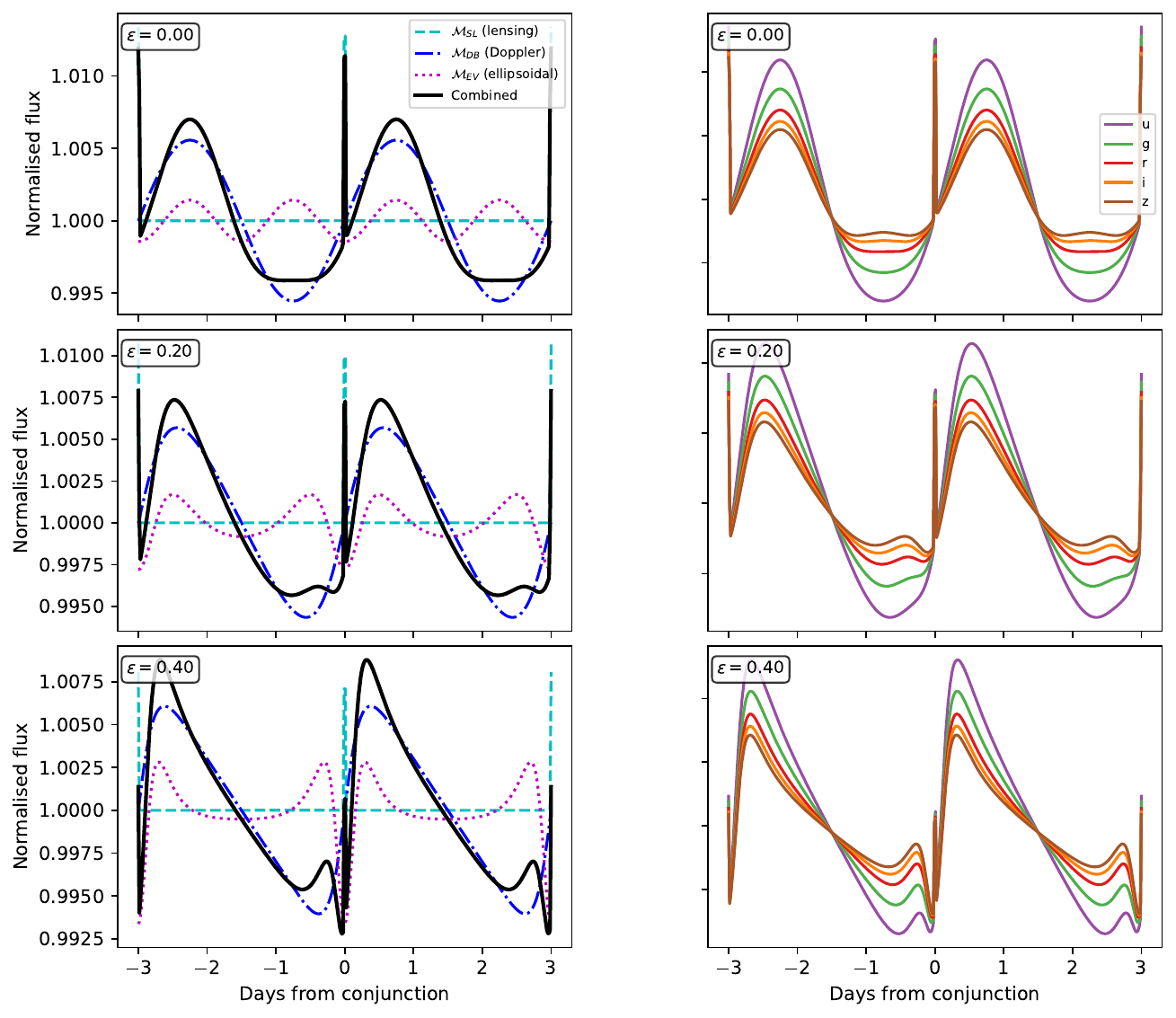}
     \caption{Decomposition of the \texttt{COBALT-BLUE} light-curve model and its wavelength dependence for the default system parameters, shown for eccentricities $\varepsilon=0.00$, $0.20$, and $0.40$ (top to bottom). The left-hand panels show the separate self-lensing, Doppler-boosting, and ellipsoidal-variation components in the $r$ band, together with the combined flux. The right-hand panels show the corresponding total flux in the $u$, $g$, $r$, $i$, and $z$ bands. Increasing eccentricity alters the shape of the orbital modulation and the lensing flare, while the chromatic dependence is strongest in the bluer bands.}
    \label{fig:alleffects}
\end{figure*}

\subsection{All Effects Combined}
\label{sec:all}

In the \texttt{COBALT-BLUE} framework, the observed flux from the luminous star is modelled as the product of the three aforementioned photometric effects and an intrinsic baseline flux. Hence, we express the total observed flux as the product
\begin{equation}
F(t) =
F_0 \times
\mathcal{M}_{\rm DB}(t)\times
\mathcal{M}_{\rm EV}(t)\times
\mathcal{M}_{\rm SL}(t)
\end{equation}
where $\mathcal{M}_{\rm DB}$, $\mathcal{M}_{\rm EV}$, and $\mathcal{M}_{\rm SL}$ denote the Doppler modulation, ellipsoidal variation, and self-lensing magnification terms, respectively. The constant $F_0$ is a flux normalisation parameter representing the baseline stellar flux in the absence of modulation.


Figure~\ref{fig:alleffects} illustrates the relative contributions of the three effects for representative system parameters. For our default model parameters, Doppler modulation is stronger than ellipsoidal variations, while the self-lensing signal appears as a sharp enhancement near inferior conjunction. Orbital eccentricity modifies both the phase and amplitude of the Doppler and ellipsoidal modulations, and the timing and strength of the lensing flare. The right-hand panel of Figure~\ref{fig:alleffects} demonstrates the wavelength dependence of the full model, with Doppler modulation, ellipsoidal variation and self-lensing exhibiting strong chromatic behaviour.

As a physical feasibility check, we compute the Roche-lobe radius $R_{\rm RL}$ of the luminous star and require that the Roche-lobe filling factor $f_{\rm RL} = R_{\rm \star} / R_{\rm RL}$
satisfies $f_{\rm RL} \leq 1$ for a physically viable configuration. We calculate $R_{\rm RL}$ using the \citet{Eggleton1983} approximation,
\begin{equation}
R_{\rm RL} =
a \,
\frac{0.49 q^{2/3}}
{0.6 q^{2/3} + \ln(1 + q^{1/3})}.
\end{equation}
The above expression formally assumes a circular orbit. For eccentric systems, the Roche-lobe radius varies with orbital phase \citep{Leahy1983}, and physical systems must satisfy $R_{\rm \star}$ smaller than the minimum Roche-lobe radius attained over the orbit. Our calculation therefore allows us to rule out parameter sets with $f_{\rm RL} > 1$, although some eccentric configurations with $f_{\rm RL} < 1$ may still be unphysical if Roche-lobe overflow occurs near periastron.

\section{Example fits to ZTF candidates}
\label{sec:fits}

Here we apply the \texttt{COBALT-BLUE} forward model to candidate self-lensing flares identified in high-cadence $r$-band ZTF data by \citet{Crossland2023}.

\subsection{Candidate selection}

\citet{Crossland2023} identified 19 candidate self-lensing sources from a 14-day subset of continuous-cadence $r$-band ZTF data obtained during August 2018 \citep{Kupfer2021}. Each candidate exhibits a single isolated flare, leaving stellar flares and other non-periodic phenomena as plausible alternative explanations. \citet{Crossland2023} estimated stellar masses and radii from photometry using main-sequence scaling relations, and they inferred compact-object masses from simplified self-lensing formulae. Here, we instead adopt luminous mass, radii and effective temperature estimates from the TESS Input Catalog (TIC; \citealt{Stassun2019}) and \textit{Gaia} Data Release 3 (DR3; \citealt{GaiaDR3}) where available. When an effective temperature estimate is not available, we estimate $T_{\rm eff}$ from the stellar luminosity and radius using the Stefan--Boltzmann relation; where a luminosity estimate is not available, we estimate the luminosity from the stellar mass using an empirical main-sequence mass--luminosity relation before calculating $T_{\rm eff}$. Compact-object masses reported by \citet{Crossland2023} are not adopted, since we infer $M_{\rm co}$ directly by fitting the \texttt{COBALT-BLUE} forward model. We also note that the analysis presented by \citet{Crossland2023} contains inconsistencies: in particular, the flare durations listed in their Table~1 are incorrect, and the compact-object masses reported in that table are not solutions of the equations presented, whether using the tabulated durations or the true flare widths.

We use the same reduced high-cadence $r$-band ZTF light curves described by \citet{Crossland2023}. We convert magnitudes to flux after subtracting the median magnitude, such that the baseline flux is approximately unity outside the flare. We discard candidates with peak normalised flux $\geq 2.0$, which is unrealistically large for self-lensing, and we omit targets for which we do not have TIC or \textit{Gaia} DR3 estimates for the stellar mass or radius. It may be possible to treat $M_\star$ and $R_\star$ as free parameters if multiple wavebands are being considered, each containing multiple self-lensing flares, but for the current dataset these parameters will be highly degenerate with other model parameters. After these cuts, 14 systems remain and constitute the fitted sample.

\subsection{Fitting procedure}

We employ a simplified set of assumptions for these proof of principle fits, assuming Eddington limb darkening and fixing $\varepsilon=0$, $\phi_c=0$ and $i=90^\circ$. We set the photometry band to the r-band, which influences the Doppler modulation, but not the lensing flare due to our use of the Eddington limb darkening law. We fix $M_{\rm \star}$ and $R_{\rm \star}$ to the values adopted from catalogues. The free parameters are $P$, $M_{\rm co}$, $t_c$ and the flux normalisation $F_0$. We fit to the flux time series using Gaussian likelihood with flux uncertainties $\sigma$ propagated from the magnitude uncertainties, and report reduced $\chi^2$ ($\chi^2/{\rm d.o.f.}$) for goodness-of-fit. For the fit, we only consider a maximum time window of $\pm0.5$ days around the flare peak. In practice, the time window fitted for is often smaller than this 1 day window due to data gaps.

For each candidate, we initialise the fit by first manually adjusting the free parameters to provide a reasonable visual description of the data. We find that this initial manual anchoring is necessary because the parameter space is highly complex and thus fully automated optimisation from arbitrary initial conditions is prone to converging to unphysical local minima. We then minimise $\chi^2$ using the Levenberg-Marquardt algorithm (the \texttt{curve\_fit} implementation).

To estimate parameter uncertainties and assess fit stability, we use a sub-sampling procedure. For each candidate, we generate 100 random sub-samples, each containing 90\% of the data points, drawn without replacement (i.e. we randomly choose $10\%$ of the data points to ignore for each sub-sample), and refit the model to each subsample. Each sub-sampled fit is initialised from the best-fitting solution obtained using the full fitted window. We summarise the resulting empirical parameter distributions by reporting the median as the central estimate, together with lower and upper uncertainties defined by the 16th and 84th percentiles. This procedure naturally yields asymmetric uncertainty intervals and also provides a direct measure of fit stability through the fraction of subsampled fits that converge successfully.

\begin{figure*}
\centering
\includegraphics[width=0.49\textwidth]{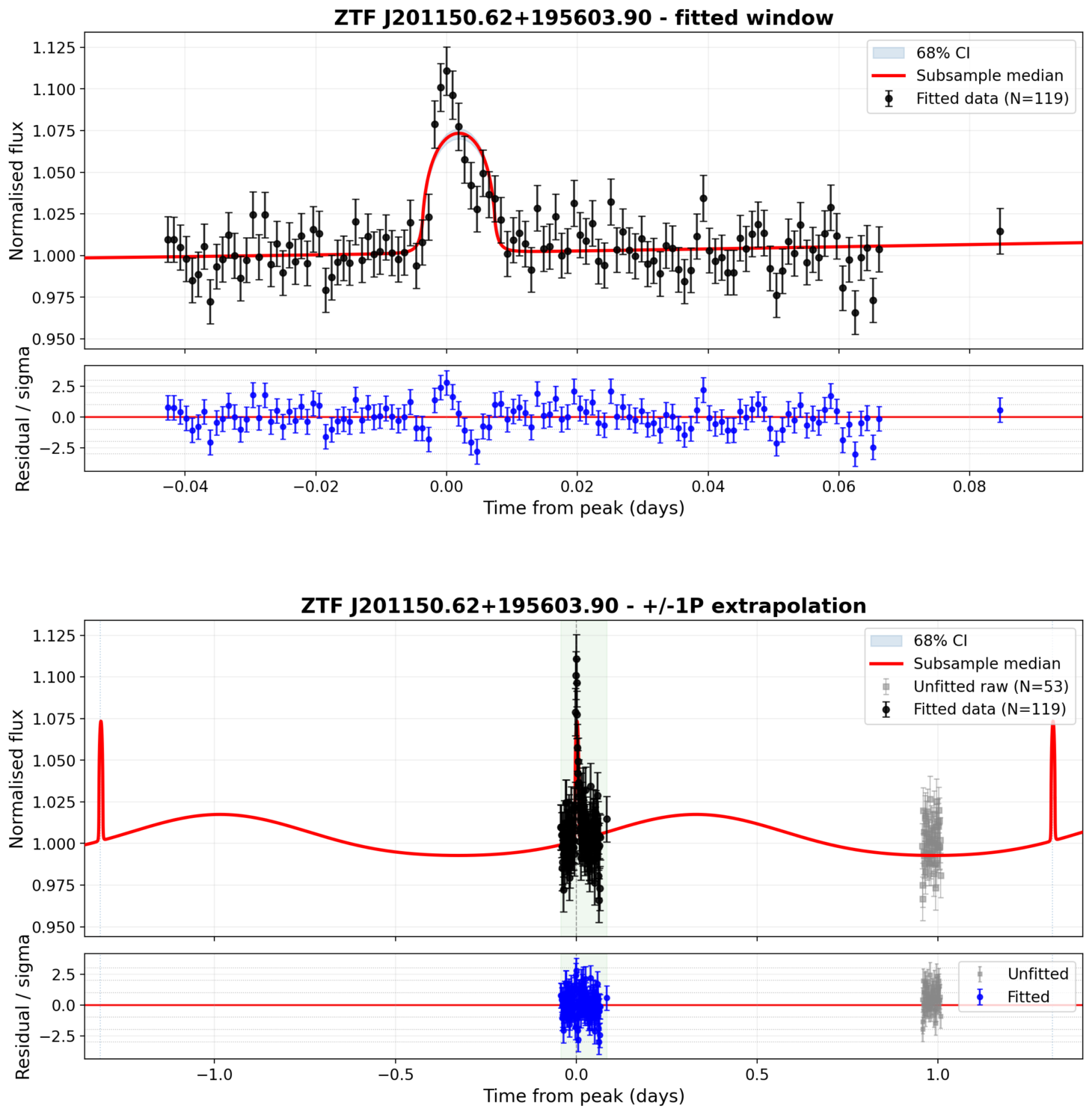}
\hfill
\includegraphics[width=0.47\textwidth]{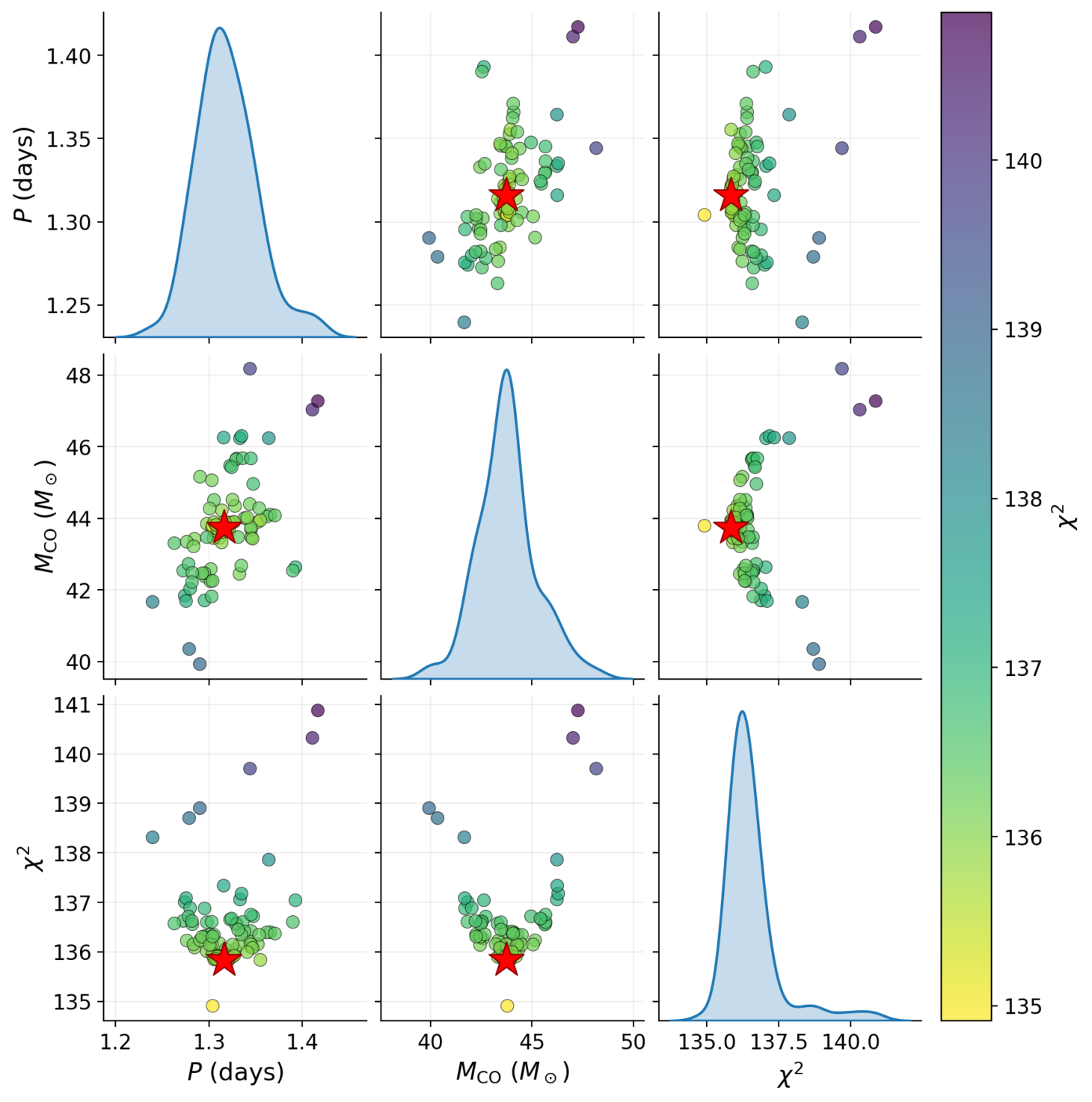}

\caption{Results of fitting the model to ZTF J201150.62+195603.90. Data (black/grey), model (red) and residuals (blue/grey) are shown on the left, both in the fitting window (top) and extrapolated to an extended window (bottom). The posterior distribution, estimated from the sub-sampling procedure, is shown on the right. The red star represents the median.}
\label{fig:J201150_full}
\end{figure*}

\begin{figure*}
\centering
\includegraphics[width=0.49\textwidth]{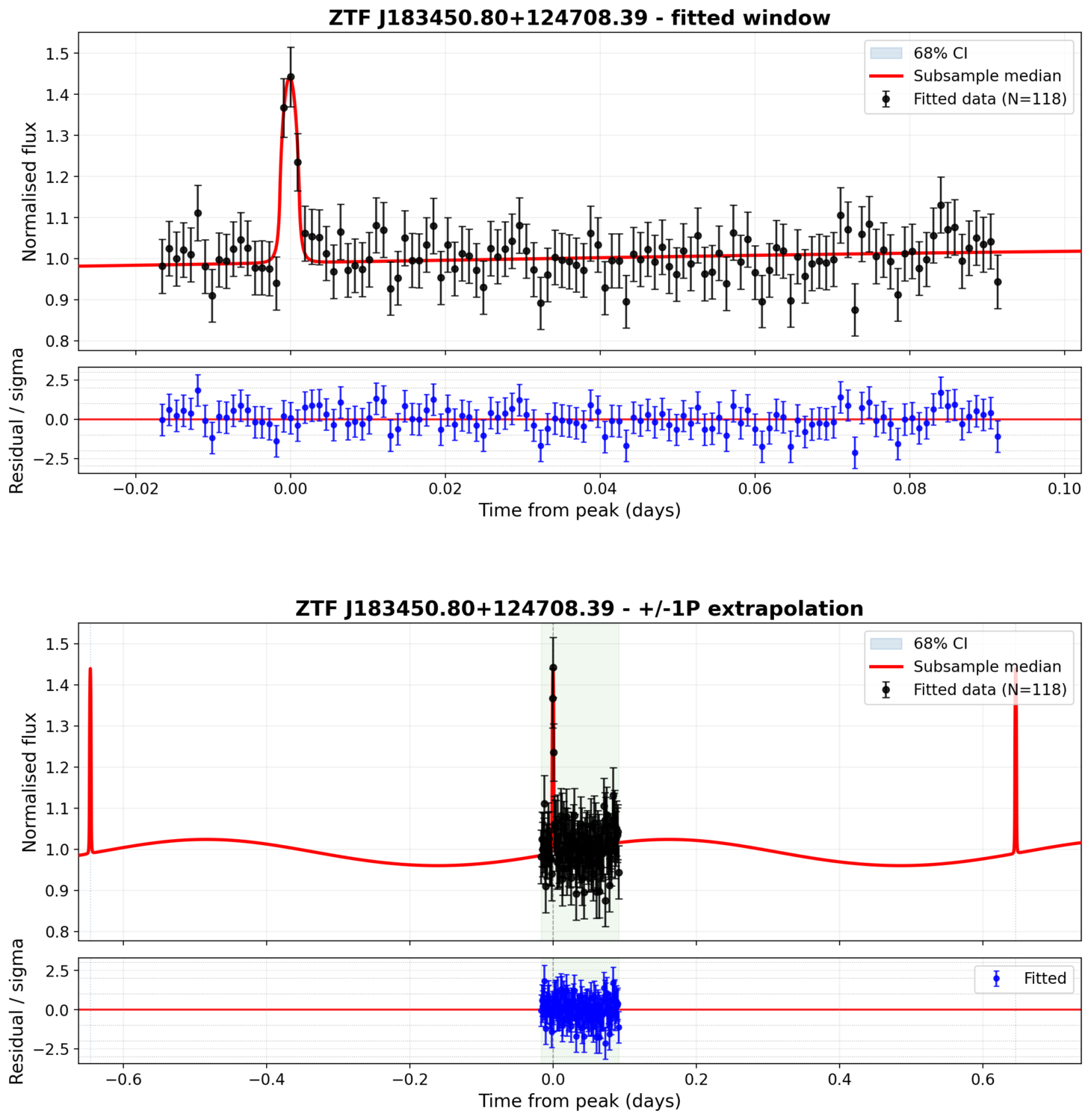}
\hfill
\includegraphics[width=0.47\textwidth]{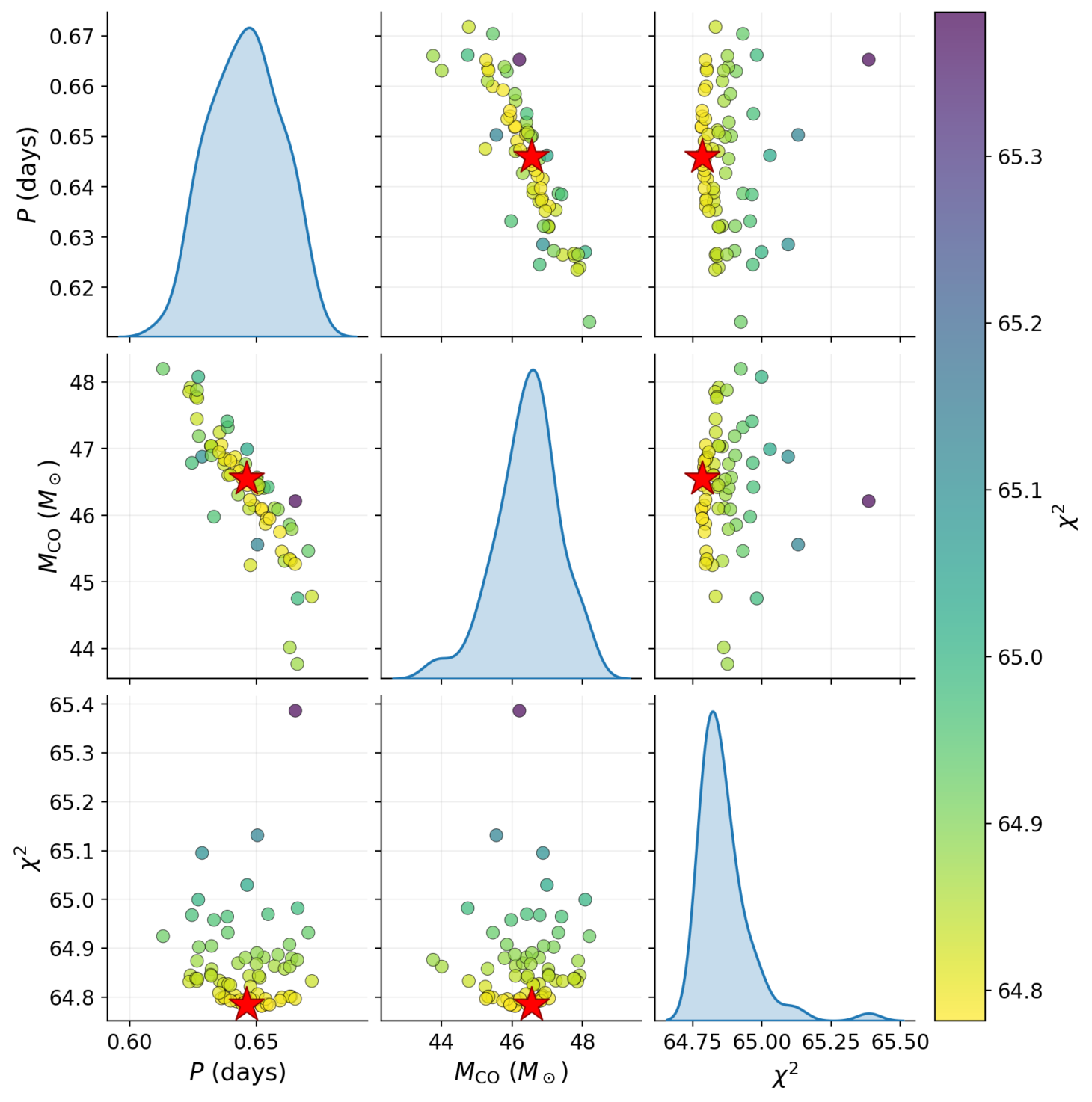}

\caption{Results of fitting the model to ZTF J183450.80+124708.39. The format is identical to that of Fig \ref{fig:J201150_full}.}
\label{fig:J183450_full}
\end{figure*}

\begin{table*}
\centering
\caption{Results for the 14 fitted candidates using 100 random subsampled fits, each containing 90\% of the fitted-window data points drawn without replacement. Reported values are medians of the subsampled fit distributions, with asymmetric uncertainties defined by the 16th and 84th percentiles. The quoted uncertainties reflect stability under subsampling rather than a complete physical posterior, since several parameters are fixed in this proof-of-principle analysis. 
The final column gives the goodness of fit explicitly as $\chi^2/\rm d.o.f.$, where $\rm d.o.f.$ is the number of degrees of freedom. The convergence column gives the fraction of the 100 subsampled fits that converged successfully.}
\label{tab:blueI_results}
\setlength{\tabcolsep}{4pt}
\renewcommand{\arraystretch}{1.08}

\begin{tabular}{lccccccccc}
\hline
Short name & $P$ [d] & $M_{\rm co}$ [$M_\odot$] & $M_\star$ [$M_\odot$] & $R_\star$ [$R_\odot$] & $a$ [$R_\odot$] & $f_{\rm RL}$ & $F_0$ & $\chi^2/ \rm{d.o.f.}$ & Conv. \\
\hline
J201150 & $1.32^{+0.03}_{-0.03}$ & $43.7^{+1.4}_{-1.3}$ & $0.4668$ & $0.4692$ & $17.9^{+0.4}_{-0.4}$ & $0.252^{+0.004}_{-0.003}$ & $1.0033^{+0.0004}_{-0.0005}$ & $136/107$ & $84\%$ \\
J183450 & $0.65^{+0.02}_{-0.01}$ & $46.6^{+0.8}_{-0.9}$ & $0.1128$ & $0.1391$ & $11.3^{+0.1}_{-0.1}$ & $0.190^{+0.003}_{-0.003}$ & $0.991^{+0.001}_{-0.002}$ & $65/106$ & $70\%$ \\
J182258 & $0.55^{+0.01}_{-0.005}$ & $75.7^{+0.9}_{-5.6}$ & $0.5843$ & $0.5953$ & $12.0^{+0.2}_{-0.2}$ & $0.528^{+0.003}_{-0.009}$ & $1.0101^{+0.00006}_{-0.0006}$ & $122/107$ & $100\%$ \\
J191320 & $6.82^{+0.02}_{-0.02}$ & $125.8^{+0.2}_{-0.2}$ & $0.5400$ & $0.5561$ & $75.9^{+0.1}_{-0.1}$ & $0.094^{+0.002}_{-0.002}$ & $0.99824^{+0.00009}_{-0.00009}$ & $124/105$ & $55\%$ \\
J190822 & $1.256^{+0.004}_{-0.006}$ & $132.9^{+0.8}_{-0.8}$ & $0.2400$ & $0.2800$ & $25.0^{+0.1}_{-0.1}$ & $0.1907^{+0.0006}_{-0.0005}$ & $1.001^{+0.002}_{-0.003}$ & $74/107$ & $64\%$ \\
J190041 & $8.3^{+0.2}_{-0.2}$ & $162.6^{+3.4}_{-4.4}$ & $0.6681$ & $0.7065$ & $94.1^{+1.2}_{-1.0}$ & $0.0976^{+0.002}_{-0.002}$ & $0.994^{+0.001}_{-0.001}$ & $329/107$ & $93\%$ \\
J200714 & $0.430^{+0.009}_{-0.009}$ & $178.1^{+2.4}_{-1.9}$ & $0.1900$ & $0.2200$ & $13.5^{+0.2}_{-0.2}$ & $0.330^{+0.004}_{-0.004}$ & $1.032^{+0.003}_{-0.002}$ & $114/103$ & $77\%$ \\
J194939 & $2.03^{+0.05}_{-0.04}$ & $283.0^{+5.8}_{-3.8}$ & $0.5500$ & $0.6000$ & $44.3^{+0.5}_{-0.5}$ & $0.225^{+0.003}_{-0.003}$ & $0.985^{+0.002}_{-0.002}$ & $74/105$ & $80\%$ \\
J195500 & $0.41^{+0.05}_{-0.04}$ & $317^{+13}_{-15}$ & $0.4203$ & $0.4258$ & $15.8^{+0.3}_{-0.5}$ & $0.51^{+0.02}_{-0.04}$ & $1.021^{+0.001}_{-0.001}$ & $76/105$ & $100\%$ \\
J185533 & $1.44^{+0.02}_{-0.03}$ & $349^{+30}_{-32}$ & $1.8100$ & $2.6880$ & $37.9^{+1.4}_{-1.5}$ & $0.86^{+0.01}_{-0.01}$ & $1.062^{+0.003}_{-0.001}$ & $276/88$ & $100\%$ \\
J182644 & $0.350^{+0.002}_{-0.003}$ & $355.3^{+1.6}_{-1.8}$ & $0.2446$ & $0.2717$ & $14.8^{+0.6}_{-0.5}$ & $0.43^{+0.02}_{-0.02}$ & $1.030^{+0.002}_{-0.002}$ & $124/105$ & $66\%$ \\
J191032 & $0.97^{+0.02}_{-0.03}$ & $391^{+19}_{-20}$ & $0.4900$ & $0.8975$ & $30.1^{+0.6}_{-0.5}$ & $0.57^{+0.01}_{-0.01}$ & $1.006^{+0.002}_{-0.002}$ & $87/106$ & $79\%$ \\
J185515 & $1.850^{+0.002}_{-0.002}$ & $729.4^{+0.3}_{-0.3}$ & $1.3330$ & $2.5968$ & $57.1^{+0.3}_{-0.3}$ & $0.77^{+0.05}_{-0.05}$ & $1.05229^{+0.000084}_{-0.000084}$ & $116/106$ & $100\%$ \\
J192607 & $0.54^{+0.02}_{-0.04}$ & $803^{+21}_{-110}$ & $1.0100$ & $0.6325$ & $25.9^{+0.3}_{-1.1}$ & $0.47^{+0.02}_{-0.02}$ & $1.016^{+0.002018}_{-0.002303}$ & $78/106$ & $100\%$ \\
\hline
\end{tabular}
\end{table*}

\subsection{Results}

Details for the 14 fitted candidates are summarised in Table~\ref{tab:blueI_results}, where we employ a short name for each candidate. We achieve statistically acceptable fits for all objects except J190041 and J185533, for which $\chi^2/{\rm d.o.f.} > 1.5$. We report median and $68\%$ confidence uncertainties for each model parameter using the sub-sampling procedure described in the previous sub-section. We also report the fraction of sub-sampled fits that converge successfully (i.e. those for which a meaningful covariance matrix can be populated), which provides a direct diagnostic of fit robustness and model identifiability. In addition to the fitted model parameters, we plot the semi major axis for physical intuition, and the Roche-Lobe filling factor of the luminous star as a first physically-motivated plausibility check. We see that this remains below unity for every candidate, as is physically expected for a detached binary. We find that the inferred compact object masses are all very high, with several in the $\sim 50-130~M_\odot$ pair-instability mass gap \citep{Belczynski2016}. We thus plot our fit results in increasing order of compact object mass, such that later entries correspond to less physically plausible parameters.


In Fig.~\ref{fig:J201150_full} we show the fit for J201150, which has the lowest best-fitting compact object mass. The top-left panel shows the data and best-fitting model over the fitted time range, while on the right we plot the posterior distributions derived from the sub-sampling procedure. We plot the parameter and $\chi^2$ values for each converged sub-sample fit, as well as the initial fit to the full dataset. Note that each sub-sample fit only considers $90\%$ of the data points, but the $\chi^2$ plotted here is calculated by comparing the model to all of the data, therefore the initial fit always returns the lowest $\chi^2$ value (which in this case can be clearly seen as $\chi^2 \approx 135$). The red stars represent median values. In the bottom-left panel, we compare the data and model over a wider time interval. The shaded region marks the time range used for fitting, and the red line shows the best-fitting model. The model predicts a self-lensing flare each orbital period, which for this observation is $P \approx 1.3$~d. In this case, the predicted second flare in the plotted window falls within a data gap, and the model remains reasonably consistent with the observations immediately outside the fitted window.

Fig.~\ref{fig:J183450_full} is the same but for J183450, which has the second lowest best-fitting compact object mass. In this case, there are no observational data outside of the fitted range. In Appendix~\ref{sec:others}, we present similar plots for the remaining 12 candidates. We see that the model does not reproduce the data outside of the immediate fitting window for all sources; for example J191320 exhibits a second flare roughly one day after the first flare that is not predicted by the model.

We conclude that J201150 and J183450 are the best two self-lensing candidates in the sample, given that the model reproduces the data, even outside of the immediate fitting window, and they have the lowest best-fitting compact object masses. However, the $\approx 44-47~M_\odot$ inferred compact object masses for these candidates are still at the upper end of the observed BH mass distribution \citep{Abbott2023,GaiaBH3}, and correspond to extreme binary mass ratios. Moreover, the orbital period is very short in both cases ($1.3$ and $0.6$ d). Lower-cadence ZTF observations span roughly $10^3$ days, yet only one flare has been observed from each source. Thus if these are the true orbital periods, it would require hundreds of self-lensing flares to have been missed during gaps in the survey cadence, which appears unlikely.

\section{Discussion}
\label{sec:discussion}

We have presented \texttt{COBALT-BLUE}, a fast, accurate and versatile public model for the photometric signatures resulting from a luminous star being in a binary system with a compact object. The download link for the model can be found in the Data Availability section. We include Doppler modulation and self-lensing for eccentric orbits, together with an approximate treatment of ellipsoidal variations.

\subsection{Model comparison}

Our model improves upon earlier treatments. For example, \citet{MasudaHotokezaka2019} considered Doppler modulation and ellipsoidal variations, but only for circular orbits. Moreover, their treatment of self-lensing was very simplistic, simply assuming a top hat profile with a calculated amplitude and duration. \citet{Sorabella2022} improved upon this by including elliptical orbits and calculating the self-lensing profile for a linear limb darkening law. \citet{Sajadian2024} also considered a linear limb darkening law, but no Doppler modulation or ellipsoidal variations.

We introduce several improvements, most notably in the self-lensing calculation for which we include non-linear limb darkening laws derived from stellar atmosphere models (e.g., \citealt{ParviainenLDTK2015,Claret2022}). Although the correction to the self-lensing flare profile is modest ($<1\%$) relative to the linear limb darkening law, our self-consistent limb darkening calculations enable the model to predict wavelength-dependent flare profiles for a small set of physical parameters (effective temperature and metallicity) as opposed to empirical limb darkening coefficients. This will hugely increase predictive power for multi-band, high signal to noise datasets (as already demonstrated for the case of exoplanet studies; e.g. \citealt{Knutson2007}). To balance speed and accuracy, we also offer the user the option to use simpler limb darkening laws. We recommend the use of the Eddington law for candidate screening, and detailed atmosphere-based laws for parameter inference.

Our Doppler boosting treatment also improves upon earlier approximations. Whereas \citet{Sorabella2022} do include elliptical orbits, they assume that the line of sight velocity is always ${\rm v}_{\rm los} = {\rm v} \sin i \sin(\phi-\phi_c)$ (in our notation), which is not formally true for $\varepsilon > 0$. We instead use the exact expression (Equation \ref{eqn:vlos}). \citet{Sorabella2022} also assume a unity Lorentz factor and employ a binomial expansion to evaluate the Doppler factor, whereas here we evaluate it exactly in special relativity (Equation \ref{eqn:doppler}). The difference is likely small in most cases, but the extra computational cost of the exact calculations is negligible.

Ellipsoidal variations remain treated approximately using formulae valid for circular orbits and linear limb darkening. This simplification is justified for most systems, as binaries close enough for strong tidal distortion are expected to circularize rapidly.

\subsection{Tests of ZTF candidates}

As a proof of principle, we fit our model to 14 candidate self-lensing sources identified in r-band ZTF data by \citet{Crossland2023}. Since each source only features one flare, several alternatives to a self-lensing origin remain plausible, most notably stellar flares. Our model fits well to most of the flare morphologies, but most of the inferred compact object masses are implausibly large, with the largest extending up to $M_{\rm co} \approx 800~M_\odot$. 
The lowest two inferred masses of $\approx 40-50~M_\odot$ are the closest to being consistent with the observational and theoretical upper limits of the black hole population \citep{Abbott2023}. Even these objects would be astrophysically very unusual, pairing very massive compact objects with low-mass main sequence companions. These two objects with the lowest best-fitting masses, J201150 and J183450, are thus the most compelling self-lensing candidates of the sample, particularly J201150 for which the best-fitting model reproduces high-cadence data outside of the immediate fitting window (no such data exist for J183450). Nevertheless, the self-lensing interpretation remains problematic. The inferred compact-object masses are still high ($M_{\rm co} = 44~M_\odot$ and $47~M_\odot$) and the inferred orbital periods are short ($P= 1.3$ d and $0.7$ d), implying that repeated flares should almost certainly have been observed in the lower-cadence ZTF baseline if the source were genuinely self-lensing.



We thus conclude, in agreement with \citet{Crossland2023}, that likely none of the explored flares are due to gravitational self-lensing. This is also in line with theoretical expectations. \citet{Crossland2023} estimated that around 1.3 self-lensing events should be theoretically present during the observations considered. To calculate this number, they scaled the population-synthesis estimate of \citet{Wiktorowicz2021}, who predicted that approximately 1800 sources would undergo at least one lensing flare while ZTF is observing over the full 5-year survey, by the reduced volume and time span of the analysed dataset. Even this estimate is likely optimistic, since most of the flares predicted by \citet{Wiktorowicz2021} are very low in amplitude, whereas the flares explored here are comparatively large. \citet{Wiktorowicz2025} estimate that the number of self-lensing flares that are \emph{observable}, once the flare amplitude is taken into account, is a few orders of magnitude lower than the approximately 1800 flares that are merely \emph{visible} to ZTF. We therefore theoretically expect far fewer than one self-lensing flare in this dataset with an amplitude at least as large as those observed here.

The candidate flares we analyse are more plausibly explained by stellar activity or unrelated transient phenomena than by self-lensing. Late-type stars, particularly active M dwarfs, exhibit flares with amplitudes comparable to or exceeding those observed in our candidate sample \citep[e.g.][]{Davenport2016}. These flares can mimic the symmetric morphology of lensing events when sampled sparsely, especially in surveys with limited cadence such as ZTF. Other possibilities include cataclysmic variables or interacting binaries observed during brief outbursts, which can produce large photometric excursions without leaving persistent signatures in sparse light curves. Micro-lensing by unrelated foreground objects cannot be ruled out for single-epoch events, although the probability of such alignments is low compared to intrinsic stellar variability.

\subsection{Future directions}

In future, it may be possible to discover self-lensing systems with larger surveys than what we consider here, in particular the full ZTF survey, TESS and LSST. Fitting a fast model such as \texttt{COBALT-BLUE} to many light curves will provide a key method to identify candidates, and ultimately to characterise confirmed self-lensing systems. Fitting the full model is a more powerful method than searching purely for self-lensing pulses, since it exploits the full suite of model-predicted signatures -- including Doppler boosting and ellipsoidal variations and the timing of the self-lensing flare relative to the phase of the Doppler modulation. As we have demonstrated here, fitting the model also gives an important plausibility check from the best fitting parameters even when the model statistically describes the data. Simultaneously fitting for multiple wavebands will provide an even more predictive test, given the strong wavelength dependence of the predicted lensing flare profile (Fig \ref{fig:LDmag}) and Doppler modulation (Fig \ref{fig:alleffects}), such that it may be possible to constrain multiple free parameters from photometry only, even in cases where the properties of the luminous star are not a priori known.

For the data we considered here, each candidate source only features one flare, which makes it very difficult to rule out alternative interpretations to self-lensing. Simultaneously fitting for multiple photometry bands will improve diagnostic power for cases with only a single flare, but it is clear that it is optimal to detect multiple flares, ideally enough to confirm a periodicity. For this reason, TESS data will realistically only be sensitive to systems with orbital period significantly shorter than the 27-day observing window. ZTF and LSST, on the other hand, have long enough baselines to be sensitive to sources with a wide range of orbital periods.

We note that it may be possible to use the model to discover binaries that do not exhibit self-lensing flares due to not being viewed sufficiently edge-on. For these systems, the Doppler modulation and ellipsoidal variations could provide the diagnostic. Although the self-lensing flare is the most distinctive feature, the vast majority of binaries will not display it, and so any sensitivity to sources without relying on self-lensing could enormously increase the number of binary systems that can potentially be discovered in photometric surveys. Fitting the model will provide greater sensitivity to eccentric binaries than searching for periodicities with e.g. Lomb-Scargle periodograms, since for eccentric orbits the variability power is shared over multiple harmonics. Ultimately, the most fruitful procedure may be to train machine learning algorithms on the model, and use the algorithms to search for binaries and return preliminary estimates of binary parameters. We will explore such methods in future work.

Such initial screening may yield a large list of candidate binaries that can then be narrowed down with detailed fitting considering realistic limb darkening laws. The best candidates from that procedure can then be followed up for confirmation with radial velocity measurements. Finding hidden compact objects in this way will provide diagnostics on the stellar and binary evolution processes that govern the birth and subsequent evolution of compact objects. The diagnostic provided by such photometric binaries is complementary with other methods of discovering binary systems such as X-ray detections of X-ray binaries \citep{blackcat}, astrometry \citep[e.g.]{GaiaBH3} and, in future, low frequency gravitational waves \citep[e.g.][]{McMillan2026}.

\section{Conclusions}
\label{sec:conclusions}

We have introduced \texttt{COBALT-BLUE}, a fast and physically accurate model for a luminous star in a binary system with a compact object that incorporates Doppler boosting, ellipsoidal variations, and gravitational self-lensing with (optionally) advanced limb-darkening prescriptions. Fitting the full model to stellar light curves, particularly in multiple wavebands, breaks the degeneracy associated with modelling lensing flares alone due to the orthogonal parameter dependencies of the Doppler modulation, ellipsoidal variations and self-lensing flares. In future, it may therefore be possible to use the model to measure parameters such as compact object mass from photometry alone. As a proof of principle, we applied our model to 14 self-lensing candidates from ZTF. While the model is able to reproduce the flare morphology, the inferred parameters were often unphysical (e.g. $M_{\rm co} > 100\,M_\odot$). We conclude that the flares in this sample are more likely stellar flares than self-lensing flares. In future, we plan to use the model to systematically search larger survey datasets (i.e. the full ZTF survey, TESS and LSST) for signatures of binary motion.

\section*{Acknowledgements}

This work was supported by the Science \& Technology Facilities Council [Grant Reference No ST/W006790/1]. AI acknowledges support from the Royal Society and by the European Union (ERC, X-MAPS, 101169908). Views and opinions expressed are however those of the author(s) only and do not necessarily reflect those of the European Union or the European Research Council. Neither the European Union nor the granting authority can be held responsible for them.

\section*{Data Availability}

The ZTF data analysed are publicly available \href{https://www.ztf.caltech.edu/ztf-public-releases.html}{https://www.ztf.caltech.edu/ztf-public-releases.html}. The \texttt{COBALT-BLUE} model can be downloaded from GitHub at \href{https://github.com/COBALT-lensing/BLUE}{https://github.com/COBALT-lensing/BLUE}.



\bibliographystyle{mnras}
\bibliography{example} 

@ARTICLE{Abbott2023,
       author = {{Abbott}, R. and {Abbott}, T.~D. and {Acernese}, F. and {Ackley}, K. and {Adams}, C. and {Adhikari}, N. and {Adhikari}, R.~X. and {Adya}, V.~B. and {Affeldt}, C. and {Agarwal}, D. and {Agathos}, M. and {Agatsuma}, K. and {Aggarwal}, N. and {Aguiar}, O.~D. and {Aiello}, L. and {Ain}, A. and {Ajith}, P. and {Akutsu}, T. and {de Alarc{\'o}n}, P.~F. and {Akcay}, S. and {Albanesi}, S. and {Allocca}, A. and {Altin}, P.~A. and {Amato}, A. and {Anand}, C. and {Anand}, S. and {Ananyeva}, A. and {Anderson}, S.~B. and {Anderson}, W.~G. and {Ando}, M. and {Andrade}, T. and {Andres}, N. and {Andri{\'c}}, T. and {Angelova}, S.~V. and {Ansoldi}, S. and {Antelis}, J.~M. and {Antier}, S. and {Antonini}, F. and {Appert}, S. and {Arai}, Koji and {Arai}, Koya and {Arai}, Y. and {Araki}, S. and {Araya}, A. and {Araya}, M.~C. and {Areeda}, J.~S. and {Ar{\`e}ne}, M. and {Aritomi}, N. and {Arnaud}, N. and {Arogeti}, M. and {Aronson}, S.~M. and {Arun}, K.~G. and {Asada}, H. and {Asali}, Y. and {Ashton}, G. and {Aso}, Y. and {Assiduo}, M. and {Aston}, S.~M. and {Astone}, P. and {Aubin}, F. and {Austin}, C. and {Babak}, S. and {Badaracco}, F. and {Bader}, M.~K.~M. and {Badger}, C. and {Bae}, S. and {Bae}, Y. and {Baer}, A.~M. and {Bagnasco}, S. and {Bai}, Y. and {Baiotti}, L. and {Baird}, J. and {Bajpai}, R. and {Ball}, M. and {Ballardin}, G. and {Ballmer}, S.~W. and {Balsamo}, A. and {Baltus}, G. and {Banagiri}, S. and {Bankar}, D. and {Barayoga}, J.~C. and {Barbieri}, C. and {Barish}, B.~C. and {Barker}, D. and {Barneo}, P. and {Barone}, F. and {Barr}, B. and {Barsotti}, L. and {Barsuglia}, M. and {Barta}, D. and {Bartlett}, J. and {Barton}, M.~A. and {Bartos}, I. and {Bassiri}, R. and {Basti}, A. and {Bawaj}, M. and {Bayley}, J.~C. and {Baylor}, A.~C. and {Bazzan}, M. and {B{\'e}csy}, B. and {Bedakihale}, V.~M. and {Bejger}, M. and {Belahcene}, I. and {Benedetto}, V. and {Beniwal}, D. and {Bennett}, T.~F. and {Bentley}, J.~D. and {Benyaala}, M. and {Bergamin}, F. and {Berger}, B.~K. and {Bernuzzi}, S. and {Berry}, C.~P.~L. and {Bersanetti}, D. and {Bertolini}, A. and {Betzwieser}, J. and {Beveridge}, D. and {Bhandare}, R. and {Bhardwaj}, U. and {Bhattacharjee}, D. and {Bhaumik}, S. and {Bilenko}, I.~A. and {Billingsley}, G. and {Bini}, S. and {Birney}, R. and {Birnholtz}, O. and {Biscans}, S. and {Bischi}, M. and {Biscoveanu}, S. and {Bisht}, A. and {Biswas}, B. and {Bitossi}, M. and {Bizouard}, M.-A. and {Blackburn}, J.~K. and {Blair}, C.~D. and {Blair}, D.~G. and {Blair}, R.~M. and {Bobba}, F. and {Bode}, N. and {Boer}, M. and {Bogaert}, G. and {Boldrini}, M. and {Bonavena}, L.~D. and {Bondu}, F. and {Bonilla}, E. and {Bonnand}, R. and {Booker}, P. and {Boom}, B.~A. and {Bork}, R. and {Boschi}, V. and {Bose}, N. and {Bose}, S. and {Bossilkov}, V. and {Boudart}, V. and {Bouffanais}, Y. and {Bozzi}, A. and {Bradaschia}, C. and {Brady}, P.~R. and {Bramley}, A. and {Branch}, A. and {Branchesi}, M. and {Brandt}, J. and {Brau}, J.~E. and {Breschi}, M. and {Briant}, T. and {Briggs}, J.~H. and {Brillet}, A. and {Brinkmann}, M. and {Brockill}, P. and {Brooks}, A.~F. and {Brooks}, J. and {Brown}, D.~D. and {Brunett}, S. and {Bruno}, G. and {Bruntz}, R. and {Bryant}, J. and {Bulik}, T. and {Bulten}, H.~J. and {Buonanno}, A. and {Buscicchio}, R. and {Buskulic}, D. and {Buy}, C. and {Byer}, R.~L. and {Cadonati}, L. and {Cagnoli}, G. and {Cahillane}, C. and {Bustillo}, J. Calder{\'o}n and {Callaghan}, J.~D. and {Callister}, T.~A. and {Calloni}, E. and {Cameron}, J. and {Camp}, J.~B. and {Canepa}, M. and {Canevarolo}, S. and {Cannavacciuolo}, M. and {Cannon}, K.~C. and {Cao}, H. and {Cao}, Z. and {Capocasa}, E. and {Capote}, E. and {Carapella}, G.},
        title = "{Population of Merging Compact Binaries Inferred Using Gravitational Waves through GWTC-3}",
      journal = {Physical Review X},
         year = 2023,
        month = jan,
       volume = {13},
       number = {1},
          eid = {011048},
        pages = {011048},
          doi = {10.1103/PhysRevX.13.011048},
archivePrefix = {arXiv},
       eprint = {2111.03634},
 primaryClass = {astro-ph.HE},
       adsurl = {https://ui.adsabs.harvard.edu/abs/2023PhRvX..13a1048A}
}

@ARTICLE{An2025,
       author = {{An}, Qian-Yu and {Huang}, Yang and {Gu}, Wei-Min and {Shao}, Yong and {Zhang}, Zhi-Xiang and {Yi}, Tuan and {Lailey}, B.~D. and {Sigut}, T.~A.~A. and {Akira Rocha}, Kyle and {Sun}, Meng and {Gossage}, Seth and {Gao}, Shi-Jie and {Weng}, Shan-Shan and {Wang}, Song and {Zhang}, Bowen and {Zhao}, Xinlin and {Qi}, Senyu and {Liao}, Shilong and {Ji}, Jianghui and {Wang}, Junfeng and {Wu}, Jianfeng and {Sun}, Mouyuan and {Li}, Xiang-Dong and {Liu}, Jifeng},
        title = "{A Be star-black hole binary with a wide orbit from LAMOST time-domain survey}",
      journal = {arXiv e-prints},
         year = 2025,
        month = may,
          eid = {arXiv:2505.23151},
        pages = {arXiv:2505.23151},
          doi = {10.48550/arXiv.2505.23151},
archivePrefix = {arXiv},
       eprint = {2505.23151},
 primaryClass = {astro-ph.SR},
       adsurl = {https://ui.adsabs.harvard.edu/abs/2025arXiv250523151A}
}

@ARTICLE{Sajadian2025,
       author = {{Sajadian}, Sedighe and {Kalantari}, Atousa and {Fatheddin}, Hossein and {Khakpash}, Somayeh},
        title = "{Simulating Gravitational Microlensing Events by TESS: Predictions on Statistics and Properties}",
      journal = {\aj},
         year = 2025,
        month = jan,
       volume = {169},
       number = {1},
          eid = {34},
        pages = {34},
          doi = {10.3847/1538-3881/ad88fb},
archivePrefix = {arXiv},
       eprint = {2408.14231},
 primaryClass = {astro-ph.EP},
       adsurl = {https://ui.adsabs.harvard.edu/abs/2025AJ....169...34S}
}

@ARTICLE{Wiktorowicz2025,
       author = {{Wiktorowicz}, Grzegorz and {Middleton}, Matthew and {Olejak}, Aleksandra and {Dashwood-Brown}, Cordelia and {Ward}, Madeleine-Mai and {Ingram}, Adam},
        title = "{Self-lensing binaries as probes of Supernova physics}",
      journal = {arXiv e-prints},
         year = 2025,
        month = sep,
          eid = {arXiv:2509.11726},
        pages = {arXiv:2509.11726},
          doi = {10.48550/arXiv.2509.11726},
archivePrefix = {arXiv},
       eprint = {2509.11726},
 primaryClass = {astro-ph.HE},
       adsurl = {https://ui.adsabs.harvard.edu/abs/2025arXiv250911726W}
}

@article{Davenport2016,
  author       = {James R. A. Davenport},
  title        = {The Kepler Catalog of Stellar Flares},
  journal      = {The Astrophysical Journal},
  volume       = {829},
  number       = {1},
  pages        = {23},
  year         = {2016},
  doi          = {10.3847/0004-637X/829/1/23},
  url          = {https://doi.org/10.3847/0004-637X/829/1/23},
  archivePrefix= {arXiv},
  eprint       = {1607.03494},
  primaryClass = {astro-ph.SR}
}

@ARTICLE{Han2016,
       author = {{Han}, Cheongho},
        title = "{Degeneracy between Lensing and Occultation in the Analysis of Self-lensing Phenomena}",
      journal = {\apj},
         year = 2016,
        month = mar,
       volume = {820},
       number = {1},
          eid = {53},
        pages = {53},
          doi = {10.3847/0004-637X/820/1/53},
archivePrefix = {arXiv},
       eprint = {1603.03500},
 primaryClass = {astro-ph.SR},
       adsurl = {https://ui.adsabs.harvard.edu/abs/2016ApJ...820...53H}
}

@ARTICLE{Kruse2014,
       author = {{Kruse}, Ethan and {Agol}, Eric},
        title = "{KOI-3278: A Self-Lensing Binary Star System}",
      journal = {Science},
         year = 2014,
        month = apr,
       volume = {344},
       number = {6181},
        pages = {275-277},
          doi = {10.1126/science.1251999},
archivePrefix = {arXiv},
       eprint = {1404.4379},
 primaryClass = {astro-ph.SR},
       adsurl = {https://ui.adsabs.harvard.edu/abs/2014Sci...344..275K}
}

@ARTICLE{Kupfer2021,
       author = {{Kupfer}, Thomas and {Prince}, Thomas A. and {van Roestel}, Jan and {Bellm}, Eric C. and {Bildsten}, Lars and {Coughlin}, Michael W. and {Drake}, Andrew J. and {Graham}, Matthew J. and {Klein}, Courtney and {Kulkarni}, Shrinivas R. and {Masci}, Frank J. and {Walters}, Richard and {Andreoni}, Igor and {Biswas}, Rahul and {Bradshaw}, Corey and {Duev}, Dmitry A. and {Dekany}, Richard and {Guidry}, Joseph A. and {Hermes}, J.~J. and {Laher}, Russ R. and {Riddle}, Reed},
        title = "{Year 1 of the ZTF high-cadence Galactic plane survey: strategy, goals, and early results on new single-mode hot subdwarf B-star pulsators}",
      journal = {\mnras},
         year = 2021,
        month = jul,
       volume = {505},
       number = {1},
        pages = {1254-1267},
          doi = {10.1093/mnras/stab1344},
archivePrefix = {arXiv},
       eprint = {2105.02758},
 primaryClass = {astro-ph.SR},
       adsurl = {https://ui.adsabs.harvard.edu/abs/2021MNRAS.505.1254K}
}

@ARTICLE{Gaggero2017,
       author = {{Gaggero}, Daniele and {Bertone}, Gianfranco and {Calore}, Francesca and {Connors}, Riley M.~T. and {Lovell}, Mark and {Markoff}, Sera and {Storm}, Emma},
        title = "{Searching for Primordial Black Holes in the Radio and X-Ray Sky}",
      journal = {\prl},
         year = 2017,
        month = jun,
       volume = {118},
       number = {24},
          eid = {241101},
        pages = {241101},
          doi = {10.1103/PhysRevLett.118.241101},
archivePrefix = {arXiv},
       eprint = {1612.00457},
 primaryClass = {astro-ph.HE},
       adsurl = {https://ui.adsabs.harvard.edu/abs/2017PhRvL.118x1101G}
}

@ARTICLE{GaiaBH1,
       author = {{El-Badry}, Kareem and {Rix}, Hans-Walter and {Quataert}, Eliot and {Howard}, Andrew W. and {Isaacson}, Howard and {Fuller}, Jim and {Hawkins}, Keith and {Breivik}, Katelyn and {Wong}, Kaze W.~K. and {Rodriguez}, Antonio C. and {Conroy}, Charlie and {Shahaf}, Sahar and {Mazeh}, Tsevi and {Arenou}, Fr{\'e}d{\'e}ric and {Burdge}, Kevin B. and {Bashi}, Dolev and {Faigler}, Simchon and {Weisz}, Daniel R. and {Seeburger}, Rhys and {Almada Monter}, Silvia and {Wojno}, Jennifer},
        title = "{A Sun-like star orbiting a black hole}",
      journal = {\mnras},
         year = 2023,
        month = jan,
       volume = {518},
       number = {1},
        pages = {1057-1085},
          doi = {10.1093/mnras/stac3140},
archivePrefix = {arXiv},
       eprint = {2209.06833},
 primaryClass = {astro-ph.SR},
       adsurl = {https://ui.adsabs.harvard.edu/abs/2023MNRAS.518.1057E}
}

@ARTICLE{GaiaBH2,
       author = {{El-Badry}, Kareem and {Rix}, Hans-Walter and {Cendes}, Yvette and {Rodriguez}, Antonio C. and {Conroy}, Charlie and {Quataert}, Eliot and {Hawkins}, Keith and {Zari}, Eleonora and {Hobson}, Melissa and {Breivik}, Katelyn and {Rau}, Arne and {Berger}, Edo and {Shahaf}, Sahar and {Seeburger}, Rhys and {Burdge}, Kevin B. and {Latham}, David W. and {Buchhave}, Lars A. and {Bieryla}, Allyson and {Bashi}, Dolev and {Mazeh}, Tsevi and {Faigler}, Simchon},
        title = "{A red giant orbiting a black hole}",
      journal = {\mnras},
         year = 2023,
        month = may,
       volume = {521},
       number = {3},
        pages = {4323-4348},
          doi = {10.1093/mnras/stad799},
archivePrefix = {arXiv},
       eprint = {2302.07880},
 primaryClass = {astro-ph.SR},
       adsurl = {https://ui.adsabs.harvard.edu/abs/2023MNRAS.521.4323E}
}

@ARTICLE{GaiaBH3,
       author = {{Gaia Collaboration} and {Panuzzo}, P. and {Mazeh}, T. and {Arenou}, F. and {Holl}, B. and {Caffau}, E. and {Jorissen}, A. and {Babusiaux}, C. and {Gavras}, P. and {Sahlmann}, J. and {Bastian}, U. and {Wyrzykowski}, {\L}. and {Eyer}, L. and {Leclerc}, N. and {Bauchet}, N. and {Bombrun}, A. and {Mowlavi}, N. and {Seabroke}, G.~M. and {Teyssier}, D. and {Balbinot}, E. and {Helmi}, A. and {Brown}, A.~G.~A. and {Vallenari}, A. and {Prusti}, T. and {de Bruijne}, J.~H.~J. and {Barbier}, A. and {Biermann}, M. and {Creevey}, O.~L. and {Ducourant}, C. and {Evans}, D.~W. and {Guerra}, R. and {Hutton}, A. and {Jordi}, C. and {Klioner}, S.~A. and {Lammers}, U. and {Lindegren}, L. and {Luri}, X. and {Mignard}, F. and {Nicolas}, C. and {Randich}, S. and {Sartoretti}, P. and {Smiljanic}, R. and {Tanga}, P. and {Walton}, N.~A. and {Aerts}, C. and {Bailer-Jones}, C.~A.~L. and {Cropper}, M. and {Drimmel}, R. and {Jansen}, F. and {Katz}, D. and {Lattanzi}, M.~G. and {Soubiran}, C. and {Th{\'e}venin}, F. and {van Leeuwen}, F. and {Andrae}, R. and {Audard}, M. and {Bakker}, J. and {Blomme}, R. and {Casta{\~n}eda}, J. and {De Angeli}, F. and {Fabricius}, C. and {Fouesneau}, M. and {Fr{\'e}mat}, Y. and {Galluccio}, L. and {Guerrier}, A. and {Heiter}, U. and {Masana}, E. and {Messineo}, R. and {Nienartowicz}, K. and {Pailler}, F. and {Riclet}, F. and {Roux}, W. and {Sordo}, R. and {Gracia-Abril}, G. and {Portell}, J. and {Altmann}, M. and {Benson}, K. and {Berthier}, J. and {Burgess}, P.~W. and {Busonero}, D. and {Busso}, G. and {Cacciari}, C. and {C{\'a}novas}, H. and {Carrasco}, J.~M. and {Carry}, B. and {Cellino}, A. and {Cheek}, N. and {Clementini}, G. and {Damerdji}, Y. and {Davidson}, M. and {de Teodoro}, P. and {Delchambre}, L. and {Dell'Oro}, A. and {Fraile Garcia}, E. and {Garabato}, D. and {Garc{\'\i}a-Lario}, P. and {Haigron}, R. and {Hambly}, N.~C. and {Harrison}, D.~L. and {Hatzidimitriou}, D. and {Hern{\'a}ndez}, J. and {Hestroffer}, D. and {Hodgkin}, S.~T. and {Jamal}, S. and {Jevardat de Fombelle}, G. and {Jordan}, S. and {Krone-Martins}, A. and {Lanzafame}, A.~C. and {L{\"o}ffler}, W. and {Lorca}, A. and {Marchal}, O. and {Marrese}, P.~M. and {Moitinho}, A. and {Muinonen}, K. and {Nu{\~n}ez Campos}, M. and {Oreshina-Slezak}, I. and {Osborne}, P. and {Pancino}, E. and {Pauwels}, T. and {Recio-Blanco}, A. and {Riello}, M. and {Rimoldini}, L. and {Robin}, A.~C. and {Roegiers}, T. and {Sarro}, L.~M. and {Schultheis}, M. and {Smith}, M. and {Sozzetti}, A. and {Utrilla}, E. and {van Leeuwen}, M. and {Weingrill}, K. and {Abbas}, U. and {{\'A}brah{\'a}m}, P. and {Abreu Aramburu}, A. and {Ahmed}, S. and {Altavilla}, G. and {{\'A}lvarez}, M.~A. and {Anders}, F. and {Anderson}, R.~I. and {Anglada Varela}, E. and {Antoja}, T. and {Baig}, S. and {Baines}, D. and {Baker}, S.~G. and {Balaguer-N{\'u}{\~n}ez}, L. and {Balog}, Z. and {Barache}, C. and {Barros}, M. and {Barstow}, M.~A. and {Bartolom{\'e}}, S. and {Bashi}, D. and {Bassilana}, J. -L. and {Baudeau}, N. and {Becciani}, U. and {Bedin}, L.~R. and {Bellas-Velidis}, I. and {Bellazzini}, M. and {Beordo}, W. and {Bernet}, M. and {Bertolotto}, C. and {Bertone}, S. and {Bianchi}, L. and {Binnenfeld}, A. and {Blanco-Cuaresma}, S. and {Bland-Hawthorn}, J. and {Blazere}, A. and {Boch}, T. and {Bossini}, D. and {Bouquillon}, S. and {Bragaglia}, A. and {Braine}, J. and {Bratsolis}, E. and {Breedt}, E. and {Bressan}, A. and {Brouillet}, N. and {Brugaletta}, E. and {Bucciarelli}, B. and {Butkevich}, A.~G. and {Buzzi}, R. and {Camut}, A. and {Cancelliere}, R. and {Cantat-Gaudin}, T. and {Capilla Guilarte}, D. and {Carballo}, R. and {Carlucci}, T. and {Carnerero}, M.~I. and {Carretero}, J. and {Carton}, S. and {Casamiquela}, L. and {Casey}, A. and {Castellani}, M. and {Castro-Ginard}, A. and {Ceraj}, L. and {Cesare}, V. and {Charlot}, P. and {Chaudet}, C. and {Chemin}, L. and {Chiavassa}, A. and {Chornay}, N. and {Chosson}, D.},
        title = "{Discovery of a dormant 33 solar-mass black hole in pre-release Gaia astrometry}",
      journal = {\aap},
         year = 2024,
        month = jun,
       volume = {686},
          eid = {L2},
        pages = {L2},
          doi = {10.1051/0004-6361/202449763},
archivePrefix = {arXiv},
       eprint = {2404.10486},
 primaryClass = {astro-ph.GA},
       adsurl = {https://ui.adsabs.harvard.edu/abs/2024A&A...686L...2G}
}

@ARTICLE{Yamaguchi2024,
       author = {{Yamaguchi}, Natsuko and {El-Badry}, Kareem and {Sorabella}, Nicholas M.},
        title = "{A Search for Self-lensing Binaries with TESS and Constraints on their Occurrence Rate}",
      journal = {\pasp},
         year = 2024,
        month = dec,
       volume = {136},
       number = {12},
          eid = {124202},
        pages = {124202},
          doi = {10.1088/1538-3873/ad9955},
archivePrefix = {arXiv},
       eprint = {2410.13939},
 primaryClass = {astro-ph.SR},
       adsurl = {https://ui.adsabs.harvard.edu/abs/2024PASP..136l4202Y}
}

@ARTICLE{Eggleton1983,
       author = {{Eggleton}, P.~P.},
        title = "{Aproximations to the radii of Roche lobes.}",
      journal = {\apj},
         year = 1983,
        month = may,
       volume = {268},
        pages = {368-369},
          doi = {10.1086/160960},
       adsurl = {https://ui.adsabs.harvard.edu/abs/1983ApJ...268..368E}
}

@ARTICLE{Sajadian2024,
       author = {{Sajadian}, Sedighe and {Afshordi}, Niayesh},
        title = "{Simulating Self-lensing and Eclipsing Signals due to Detached Compact Objects in the TESS Light Curves}",
      journal = {\aj},
         year = 2024,
        month = dec,
       volume = {168},
       number = {6},
          eid = {298},
        pages = {298},
          doi = {10.3847/1538-3881/ad7fdd},
archivePrefix = {arXiv},
       eprint = {2409.12441},
 primaryClass = {astro-ph.SR},
       adsurl = {https://ui.adsabs.harvard.edu/abs/2024AJ....168..298S}
}

@ARTICLE{Sahu2025,
       author = {{Sahu}, Kailash C. and {Anderson}, Jay and {Casertano}, Stefano and {Bond}, Howard E. and {Dominik}, Martin and {Calamida}, Annalisa and {Bellini}, Andrea and {Brown}, Thomas M. and {Ferguson}, Henry C. and {Rejkuba}, Marina},
        title = "{OGLE-2011-BLG-0462: An Isolated Stellar-mass Black Hole Confirmed Using New HST Astrometry and Updated Photometry}",
      journal = {\apj},
         year = 2025,
        month = apr,
       volume = {983},
       number = {2},
          eid = {104},
        pages = {104},
          doi = {10.3847/1538-4357/adbe6e},
archivePrefix = {arXiv},
       eprint = {2503.07820},
 primaryClass = {astro-ph.SR},
       adsurl = {https://ui.adsabs.harvard.edu/abs/2025ApJ...983..104S}
}

@ARTICLE{Olejak2020,
       author = {{Olejak}, A. and {Belczynski}, K. and {Bulik}, T. and {Sobolewska}, M.},
        title = "{Synthetic catalog of black holes in the Milky Way}",
      journal = {\aap},
         year = 2020,
        month = jun,
       volume = {638},
          eid = {A94},
        pages = {A94},
          doi = {10.1051/0004-6361/201936557},
archivePrefix = {arXiv},
       eprint = {1908.08775},
 primaryClass = {astro-ph.SR},
       adsurl = {https://ui.adsabs.harvard.edu/abs/2020A&A...638A..94O}
}

@ARTICLE{Hestroffer1997,
       author = {{Hestroffer}, D.},
        title = "{Centre to limb darkening of stars. New model and application to stellar interferometry.}",
      journal = {\aap},
         year = 1997,
        month = nov,
       volume = {327},
        pages = {199-206},
       adsurl = {https://ui.adsabs.harvard.edu/abs/1997A&A...327..199H}
}

@BOOK{eddington1926,
       author = {{Eddington}, A.~S.},
        title = "{The Internal Constitution of the Stars}",
    publisher = "{Cambridge University Press}",
         year = 1926,
       adsurl = {https://ui.adsabs.harvard.edu/abs/1926ics..book.....E},
    
}

@article{Espinoza_Jordán_2015, title={Limb darkening and exoplanets: testing stellar model atmospheres and identifying biases in transit parameters}, volume={450}, ISSN={1365-2966, 0035-8711}, url={http://academic.oup.com/mnras/article/450/2/1879/985166/Limb-darkening-and-exoplanets-testing-stellar}, DOI={10.1093/mnras/stv744}, number={2}, journal={Monthly Notices of the Royal Astronomical Society}, author={Espinoza, Néstor and Jordán, Andrés}, year={2015}, month=june, pages={1879–1899}, language={en} }

@BOOK{Mihalas1978,
       author = {{Mihalas}, Dimitri},
        title = "{Stellar atmospheres}",
    publisher = "{W. H. Freeman}",
         year = 1978,
       adsurl = {https://ui.adsabs.harvard.edu/abs/1978stat.book.....M}
}

@ARTICLE{MandelAgol2002,
       author = {{Mandel}, Kaisey and {Agol}, Eric},
        title = "{Analytic Light Curves for Planetary Transit Searches}",
      journal = {\apjl},
         year = 2002,
        month = dec,
       volume = {580},
       number = {2},
        pages = {L171-L175},
          doi = {10.1086/345520},
archivePrefix = {arXiv},
       eprint = {astro-ph/0210099},
 primaryClass = {astro-ph},
       adsurl = {https://ui.adsabs.harvard.edu/abs/2002ApJ...580L.171M}
}

@ARTICLE{Morris1993,
       author = {{Morris}, Steven L. and {Naftilan}, Stephen A.},
        title = "{The Equations of Ellipsoidal Star Variability Applied to HR 8427}",
      journal = {\apj},
         year = 1993,
        month = dec,
       volume = {419},
        pages = {344},
          doi = {10.1086/173488},
       adsurl = {https://ui.adsabs.harvard.edu/abs/1993ApJ...419..344M}
}

@ARTICLE{Husser2013,
       author = {{Husser}, T. -O. and {Wende-von Berg}, S. and {Dreizler}, S. and {Homeier}, D. and {Reiners}, A. and {Barman}, T. and {Hauschildt}, P.~H.},
        title = "{A new extensive library of PHOENIX stellar atmospheres and synthetic spectra}",
      journal = {\aap},
         year = 2013,
        month = may,
       volume = {553},
          eid = {A6},
        pages = {A6},
          doi = {10.1051/0004-6361/201219058},
archivePrefix = {arXiv},
       eprint = {1303.5632},
 primaryClass = {astro-ph.SR},
       adsurl = {https://ui.adsabs.harvard.edu/abs/2013A&A...553A...6H}
}

@ARTICLE{Sorabella2022,
       author = {{Sorabella}, Nicholas M. and {Bhattacharya}, Sayantan and {Laycock}, Silas G.~T. and {Christodoulou}, Dimitris M. and {Massarotti}, Alessandro},
        title = "{Modeling Long-term Variability in Stellar-compact Object Binary Systems for Mass Determinations}",
      journal = {\apj},
         year = 2022,
        month = sep,
       volume = {936},
       number = {1},
          eid = {63},
        pages = {63},
          doi = {10.3847/1538-4357/ac82b7},
       adsurl = {https://ui.adsabs.harvard.edu/abs/2022ApJ...936...63S}
}

@ARTICLE{Sorabella2024,
       author = {{Sorabella}, Nicholas M. and {Laycock}, Silas G.~T. and {Christodoulou}, Dimitris M. and {Bhattacharya}, Sayantan},
        title = "{The First TESS Self-lensing Pulses: Revisiting KIC 12254688}",
      journal = {\apjl},
         year = 2024,
        month = feb,
       volume = {961},
       number = {2},
          eid = {L45},
        pages = {L45},
          doi = {10.3847/2041-8213/ad19dc},
archivePrefix = {arXiv},
       eprint = {2401.01477},
 primaryClass = {astro-ph.SR},
       adsurl = {https://ui.adsabs.harvard.edu/abs/2024ApJ...961L..45S}
}

@ARTICLE{Fender2013,
       author = {{Fender}, R.~P. and {Maccarone}, T.~J. and {Heywood}, I.},
        title = "{The closest black holes}",
      journal = {\mnras},
         year = 2013,
        month = apr,
       volume = {430},
       number = {3},
        pages = {1538-1547},
          doi = {10.1093/mnras/sts688},
archivePrefix = {arXiv},
       eprint = {1301.1341},
 primaryClass = {astro-ph.HE},
       adsurl = {https://ui.adsabs.harvard.edu/abs/2013MNRAS.430.1538F}
}

@ARTICLE{OGLE,
       author = {{Udalski}, A. and {Szyma{\'n}ski}, M.~K. and {Szyma{\'n}ski}, G.},
        title = "{OGLE-IV: Fourth Phase of the Optical Gravitational Lensing Experiment}",
      journal = {\actaa},
         year = 2015,
        month = mar,
       volume = {65},
       number = {1},
        pages = {1-38},
archivePrefix = {arXiv},
       eprint = {1504.05966},
 primaryClass = {astro-ph.SR},
       adsurl = {https://ui.adsabs.harvard.edu/abs/2015AcA....65....1U}
}

@INPROCEEDINGS{Bond2001,
       author = {{Bond}, I.},
        title = "{The MOA Strategy for Microlensing Planet Searches}",
    booktitle = {Cosmological Physics with Gravitational Lensing},
         year = 2001,
       editor = {{Tran Thanh Van}, J. and {Mellier}, Yannick and {Moniez}, Marc},
        month = jan,
        pages = {11},
       adsurl = {https://ui.adsabs.harvard.edu/abs/2001cpgl.conf...11B}
}

@ARTICLE{KMTNet,
       author = {{Kim}, Hyoun-Woo and {Hwang}, Kyu-Ha and {Shvartzvald}, Yossi and {Yee}, Jennifer C. and {Albrow}, Michael D. and {Cha}, Sang-Mok and {Chung}, Sun-Ju and {Gould}, Andrew and {Han}, Cheongho and {Jung}, Youn Kil and {Kim}, Dong-Jin and {Kim}, Seung-Lee and {Lee}, Chung-Uk and {Lee}, Dong-Joo and {Lee}, Yongseok and {Park}, Byeong-Gon and {Pogge}, Richard W. and {Ryu}, Yoon-Hyun and {Shin}, In-Gu and {Zang}, Weicheng},
        title = "{The Korea Microlensing Telescope Network (KMTNet) Alert Algorithm and Alert System}",
      journal = {arXiv e-prints},
         year = 2018,
        month = jun,
          eid = {arXiv:1806.07545},
        pages = {arXiv:1806.07545},
archivePrefix = {arXiv},
       eprint = {1806.07545},
 primaryClass = {astro-ph.IM},
       adsurl = {https://ui.adsabs.harvard.edu/abs/2018arXiv180607545K}
}

@ARTICLE{Mao2002,
       author = {{Mao}, Shude and {Smith}, Martin C. and {Wo{\'z}niak}, P. and {Udalski}, A. and {Szyma{\'n}ski}, M. and {Kubiak}, M. and {Pietrzy{\'n}ski}, G. and {Soszy{\'n}ski}, I. and {{\.Z}ebru{\'n}}, K.},
        title = "{Optical Gravitational Lensing Experiment OGLE-1999-BUL-32: the longest ever microlensing event - evidence for a stellar mass black hole?}",
      journal = {\mnras},
         year = 2002,
        month = jan,
       volume = {329},
       number = {2},
        pages = {349-354},
          doi = {10.1046/j.1365-8711.2002.04986.x},
archivePrefix = {arXiv},
       eprint = {astro-ph/0108312},
 primaryClass = {astro-ph},
       adsurl = {https://ui.adsabs.harvard.edu/abs/2002MNRAS.329..349M}
}

@ARTICLE{Bennett2002,
       author = {{Bennett}, D.~P. and {Becker}, A.~C. and {Quinn}, J.~L. and {Tomaney}, A.~B. and {Alcock}, C. and {Allsman}, R.~A. and {Alves}, D.~R. and {Axelrod}, T.~S. and {Calitz}, J.~J. and {Cook}, K.~H. and {Drake}, A.~J. and {Fragile}, P.~C. and {Freeman}, K.~C. and {Geha}, M. and {Griest}, K. and {Johnson}, B.~R. and {Keller}, S.~C. and {Laws}, C. and {Lehner}, M.~J. and {Marshall}, S.~L. and {Minniti}, D. and {Nelson}, C.~A. and {Peterson}, B.~A. and {Popowski}, P. and {Pratt}, M.~R. and {Quinn}, P.~J. and {Rhie}, S.~H. and {Stubbs}, C.~W. and {Sutherland}, W. and {Vandehei}, T. and {Welch}, D. and {MACHO Collaboration} and {MPS Collaboration}},
        title = "{Gravitational Microlensing Events Due to Stellar-Mass Black Holes}",
      journal = {\apj},
         year = 2002,
        month = nov,
       volume = {579},
       number = {2},
        pages = {639-659},
          doi = {10.1086/342225},
archivePrefix = {arXiv},
       eprint = {astro-ph/0109467},
 primaryClass = {astro-ph},
       adsurl = {https://ui.adsabs.harvard.edu/abs/2002ApJ...579..639B}
}

@ARTICLE{Giesers2019,
       author = {{Giesers}, Benjamin and {Kamann}, Sebastian and {Dreizler}, Stefan and {Husser}, Tim-Oliver and {Askar}, Abbas and {G{\"o}ttgens}, Fabian and {Brinchmann}, Jarle and {Latour}, Marilyn and {Weilbacher}, Peter M. and {Wendt}, Martin and {Roth}, Martin M.},
        title = "{A stellar census in globular clusters with MUSE: Binaries in NGC 3201}",
      journal = {\aap},
         year = 2019,
        month = dec,
       volume = {632},
          eid = {A3},
        pages = {A3},
          doi = {10.1051/0004-6361/201936203},
archivePrefix = {arXiv},
       eprint = {1909.04050},
 primaryClass = {astro-ph.SR},
       adsurl = {https://ui.adsabs.harvard.edu/abs/2019A&A...632A...3G}
}

@ARTICLE{Thompson2019,
       author = {{Thompson}, Todd A. and {Kochanek}, Christopher S. and {Stanek}, Krzysztof Z. and {Badenes}, Carles and {Post}, Richard S. and {Jayasinghe}, Tharindu and {Latham}, David W. and {Bieryla}, Allyson and {Esquerdo}, Gilbert A. and {Berlind}, Perry and {Calkins}, Michael L. and {Tayar}, Jamie and {Lindegren}, Lennart and {Johnson}, Jennifer A. and {Holoien}, Thomas W. -S. and {Auchettl}, Katie and {Covey}, Kevin},
        title = "{A noninteracting low-mass black hole-giant star binary system}",
      journal = {Science},
         year = 2019,
        month = nov,
       volume = {366},
       number = {6465},
        pages = {637-640},
          doi = {10.1126/science.aau4005},
archivePrefix = {arXiv},
       eprint = {1806.02751},
 primaryClass = {astro-ph.HE},
       adsurl = {https://ui.adsabs.harvard.edu/abs/2019Sci...366..637T}
}

@ARTICLE{ZTF,
       author = {{Masci}, Frank J. and {Laher}, Russ R. and {Rusholme}, Ben and {Shupe}, David L. and {Groom}, Steven and {Surace}, Jason and {Jackson}, Edward and {Monkewitz}, Serge and {Beck}, Ron and {Flynn}, David and {Terek}, Scott and {Landry}, Walter and {Hacopians}, Eugean and {Desai}, Vandana and {Howell}, Justin and {Brooke}, Tim and {Imel}, David and {Wachter}, Stefanie and {Ye}, Quan-Zhi and {Lin}, Hsing-Wen and {Cenko}, S. Bradley and {Cunningham}, Virginia and {Rebbapragada}, Umaa and {Bue}, Brian and {Miller}, Adam A. and {Mahabal}, Ashish and {Bellm}, Eric C. and {Patterson}, Maria T. and {Juri{\'c}}, Mario and {Golkhou}, V. Zach and {Ofek}, Eran O. and {Walters}, Richard and {Graham}, Matthew and {Kasliwal}, Mansi M. and {Dekany}, Richard G. and {Kupfer}, Thomas and {Burdge}, Kevin and {Cannella}, Christopher B. and {Barlow}, Tom and {Van Sistine}, Angela and {Giomi}, Matteo and {Fremling}, Christoffer and {Blagorodnova}, Nadejda and {Levitan}, David and {Riddle}, Reed and {Smith}, Roger M. and {Helou}, George and {Prince}, Thomas A. and {Kulkarni}, Shrinivas R.},
        title = "{The Zwicky Transient Facility: Data Processing, Products, and Archive}",
      journal = {\pasp},
         year = 2019,
        month = jan,
       volume = {131},
       number = {995},
        pages = {018003},
          doi = {10.1088/1538-3873/aae8ac},
archivePrefix = {arXiv},
       eprint = {1902.01872},
 primaryClass = {astro-ph.IM},
       adsurl = {https://ui.adsabs.harvard.edu/abs/2019PASP..131a8003M}
}

@ARTICLE{LSST,
       author = {{Ivezi{\'c}}, {\v{Z}}eljko and {Kahn}, Steven M. and {Tyson}, J. Anthony and {Abel}, Bob and {Acosta}, Emily and {Allsman}, Robyn and {Alonso}, David and {AlSayyad}, Yusra and {Anderson}, Scott F. and {Andrew}, John and {Angel}, James Roger P. and {Angeli}, George Z. and {Ansari}, Reza and {Antilogus}, Pierre and {Araujo}, Constanza and {Armstrong}, Robert and {Arndt}, Kirk T. and {Astier}, Pierre and {Aubourg}, {\'E}ric and {Auza}, Nicole and {Axelrod}, Tim S. and {Bard}, Deborah J. and {Barr}, Jeff D. and {Barrau}, Aurelian and {Bartlett}, James G. and {Bauer}, Amanda E. and {Bauman}, Brian J. and {Baumont}, Sylvain and {Bechtol}, Ellen and {Bechtol}, Keith and {Becker}, Andrew C. and {Becla}, Jacek and {Beldica}, Cristina and {Bellavia}, Steve and {Bianco}, Federica B. and {Biswas}, Rahul and {Blanc}, Guillaume and {Blazek}, Jonathan and {Blandford}, Roger D. and {Bloom}, Josh S. and {Bogart}, Joanne and {Bond}, Tim W. and {Booth}, Michael T. and {Borgland}, Anders W. and {Borne}, Kirk and {Bosch}, James F. and {Boutigny}, Dominique and {Brackett}, Craig A. and {Bradshaw}, Andrew and {Brandt}, William Nielsen and {Brown}, Michael E. and {Bullock}, James S. and {Burchat}, Patricia and {Burke}, David L. and {Cagnoli}, Gianpietro and {Calabrese}, Daniel and {Callahan}, Shawn and {Callen}, Alice L. and {Carlin}, Jeffrey L. and {Carlson}, Erin L. and {Chandrasekharan}, Srinivasan and {Charles-Emerson}, Glenaver and {Chesley}, Steve and {Cheu}, Elliott C. and {Chiang}, Hsin-Fang and {Chiang}, James and {Chirino}, Carol and {Chow}, Derek and {Ciardi}, David R. and {Claver}, Charles F. and {Cohen-Tanugi}, Johann and {Cockrum}, Joseph J. and {Coles}, Rebecca and {Connolly}, Andrew J. and {Cook}, Kem H. and {Cooray}, Asantha and {Covey}, Kevin R. and {Cribbs}, Chris and {Cui}, Wei and {Cutri}, Roc and {Daly}, Philip N. and {Daniel}, Scott F. and {Daruich}, Felipe and {Daubard}, Guillaume and {Daues}, Greg and {Dawson}, William and {Delgado}, Francisco and {Dellapenna}, Alfred and {de Peyster}, Robert and {de Val-Borro}, Miguel and {Digel}, Seth W. and {Doherty}, Peter and {Dubois}, Richard and {Dubois-Felsmann}, Gregory P. and {Durech}, Josef and {Economou}, Frossie and {Eifler}, Tim and {Eracleous}, Michael and {Emmons}, Benjamin L. and {Fausti Neto}, Angelo and {Ferguson}, Henry and {Figueroa}, Enrique and {Fisher-Levine}, Merlin and {Focke}, Warren and {Foss}, Michael D. and {Frank}, James and {Freemon}, Michael D. and {Gangler}, Emmanuel and {Gawiser}, Eric and {Geary}, John C. and {Gee}, Perry and {Geha}, Marla and {Gessner}, Charles J.~B. and {Gibson}, Robert R. and {Gilmore}, D. Kirk and {Glanzman}, Thomas and {Glick}, William and {Goldina}, Tatiana and {Goldstein}, Daniel A. and {Goodenow}, Iain and {Graham}, Melissa L. and {Gressler}, William J. and {Gris}, Philippe and {Guy}, Leanne P. and {Guyonnet}, Augustin and {Haller}, Gunther and {Harris}, Ron and {Hascall}, Patrick A. and {Haupt}, Justine and {Hernandez}, Fabio and {Herrmann}, Sven and {Hileman}, Edward and {Hoblitt}, Joshua and {Hodgson}, John A. and {Hogan}, Craig and {Howard}, James D. and {Huang}, Dajun and {Huffer}, Michael E. and {Ingraham}, Patrick and {Innes}, Walter R. and {Jacoby}, Suzanne H. and {Jain}, Bhuvnesh and {Jammes}, Fabrice and {Jee}, M. James and {Jenness}, Tim and {Jernigan}, Garrett and {Jevremovi{\'c}}, Darko and {Johns}, Kenneth and {Johnson}, Anthony S. and {Johnson}, Margaret W.~G. and {Jones}, R. Lynne and {Juramy-Gilles}, Claire and {Juri{\'c}}, Mario and {Kalirai}, Jason S. and {Kallivayalil}, Nitya J. and {Kalmbach}, Bryce and {Kantor}, Jeffrey P. and {Karst}, Pierre and {Kasliwal}, Mansi M. and {Kelly}, Heather and {Kessler}, Richard and {Kinnison}, Veronica and {Kirkby}, David and {Knox}, Lloyd and {Kotov}, Ivan V. and {Krabbendam}, Victor L. and {Krughoff}, K. Simon and {Kub{\'a}nek}, Petr and {Kuczewski}, John and {Kulkarni}, Shri and {Ku}, John and {Kurita}, Nadine R. and {Lage}, Craig S. and {Lambert}, Ron and {Lange}, Travis and {Langton}, J. Brian and {Le Guillou}, Laurent and {Levine}, Deborah and {Liang}, Ming and {Lim}, Kian-Tat and {Lintott}, Chris J. and {Long}, Kevin E. and {Lopez}, Margaux and {Lotz}, Paul J. and {Lupton}, Robert H. and {Lust}, Nate B. and {MacArthur}, Lauren A. and {Mahabal}, Ashish and {Mandelbaum}, Rachel and {Markiewicz}, Thomas W. and {Marsh}, Darren S. and {Marshall}, Philip J. and {Marshall}, Stuart and {May}, Morgan and {McKercher}, Robert and {McQueen}, Michelle and {Meyers}, Joshua and {Migliore}, Myriam and {Miller}, Michelle and {Mills}, David J. and {Miraval}, Connor and {Moeyens}, Joachim and {Moolekamp}, Fred E. and {Monet}, David G. and {Moniez}, Marc and {Monkewitz}, Serge and {Montgomery}, Christopher and {Morrison}, Christopher B. and {Mueller}, Fritz and {Muller}, Gary P. and {Mu{\~n}oz Arancibia}, Freddy and {Neill}, Douglas R. and {Newbry}, Scott P. and {Nief}, Jean-Yves and {Nomerotski}, Andrei and {Nordby}, Martin and {O'Connor}, Paul and {Oliver}, John and {Olivier}, Scot S. and {Olsen}, Knut and {O'Mullane}, William and {Ortiz}, Sandra and {Osier}, Shawn and {Owen}, Russell E. and {Pain}, Reynald and {Palecek}, Paul E. and {Parejko}, John K. and {Parsons}, James B. and {Pease}, Nathan M. and {Peterson}, J. Matt and {Peterson}, John R. and {Petravick}, Donald L. and {Libby Petrick}, M.~E. and {Petry}, Cathy E. and {Pierfederici}, Francesco and {Pietrowicz}, Stephen and {Pike}, Rob and {Pinto}, Philip A. and {Plante}, Raymond and {Plate}, Stephen and {Plutchak}, Joel P. and {Price}, Paul A. and {Prouza}, Michael and {Radeka}, Veljko and {Rajagopal}, Jayadev and {Rasmussen}, Andrew P. and {Regnault}, Nicolas and {Reil}, Kevin A. and {Reiss}, David J. and {Reuter}, Michael A. and {Ridgway}, Stephen T. and {Riot}, Vincent J. and {Ritz}, Steve and {Robinson}, Sean and {Roby}, William and {Roodman}, Aaron and {Rosing}, Wayne and {Roucelle}, Cecille and {Rumore}, Matthew R. and {Russo}, Stefano and {Saha}, Abhijit and {Sassolas}, Benoit and {Schalk}, Terry L. and {Schellart}, Pim and {Schindler}, Rafe H. and {Schmidt}, Samuel and {Schneider}, Donald P. and {Schneider}, Michael D. and {Schoening}, William and {Schumacher}, German and {Schwamb}, Megan E. and {Sebag}, Jacques and {Selvy}, Brian and {Sembroski}, Glenn H. and {Seppala}, Lynn G. and {Serio}, Andrew and {Serrano}, Eduardo and {Shaw}, Richard A. and {Shipsey}, Ian and {Sick}, Jonathan and {Silvestri}, Nicole and {Slater}, Colin T. and {Smith}, J. Allyn and {Smith}, R. Chris and {Sobhani}, Shahram and {Soldahl}, Christine and {Storrie-Lombardi}, Lisa and {Stover}, Edward and {Strauss}, Michael A. and {Street}, Rachel A. and {Stubbs}, Christopher W. and {Sullivan}, Ian S. and {Sweeney}, Donald and {Swinbank}, John D. and {Szalay}, Alexander and {Takacs}, Peter and {Tether}, Stephen A. and {Thaler}, Jon J. and {Thayer}, John Gregg and {Thomas}, Sandrine and {Thornton}, Adam J. and {Thukral}, Vaikunth and {Tice}, Jeffrey and {Trilling}, David E. and {Turri}, Max and {Van Berg}, Richard and {Vanden Berk}, Daniel and {Vetter}, Kurt and {Virieux}, Francoise and {Vucina}, Tomislav and {Wahl}, William and {Walkowicz}, Lucianne and {Walsh}, Brian and {Walter}, Christopher W. and {Wang}, Daniel L. and {Wang}, Shin-Yawn and {Warner}, Michael and {Wiecha}, Oliver and {Willman}, Beth and {Winters}, Scott E. and {Wittman}, David and {Wolff}, Sidney C. and {Wood-Vasey}, W. Michael and {Wu}, Xiuqin and {Xin}, Bo and {Yoachim}, Peter and {Zhan}, Hu},
        title = "{LSST: From Science Drivers to Reference Design and Anticipated Data Products}",
      journal = {\apj},
         year = 2019,
        month = mar,
       volume = {873},
       number = {2},
          eid = {111},
        pages = {111},
          doi = {10.3847/1538-4357/ab042c},
archivePrefix = {arXiv},
       eprint = {0805.2366},
 primaryClass = {astro-ph},
       adsurl = {https://ui.adsabs.harvard.edu/abs/2019ApJ...873..111I}
}

@ARTICLE{Kawahara2018,
       author = {{Kawahara}, Hajime and {Masuda}, Kento and {MacLeod}, Morgan and {Latham}, David W. and {Bieryla}, Allyson and {Benomar}, Othman},
        title = "{Discovery of Three Self-lensing Binaries from Kepler}",
      journal = {\aj},
         year = 2018,
        month = mar,
       volume = {155},
       number = {3},
          eid = {144},
        pages = {144},
          doi = {10.3847/1538-3881/aaaaaf},
archivePrefix = {arXiv},
       eprint = {1801.07874},
 primaryClass = {astro-ph.SR},
       adsurl = {https://ui.adsabs.harvard.edu/abs/2018AJ....155..144K}
}

@ARTICLE{TESS,
       author = {{Ricker}, George R. and {Winn}, Joshua N. and {Vanderspek}, Roland and {Latham}, David W. and {Bakos}, G{\'a}sp{\'a}r {\'A}. and {Bean}, Jacob L. and {Berta-Thompson}, Zachory K. and {Brown}, Timothy M. and {Buchhave}, Lars and {Butler}, Nathaniel R. and {Butler}, R. Paul and {Chaplin}, William J. and {Charbonneau}, David and {Christensen-Dalsgaard}, J{\o}rgen and {Clampin}, Mark and {Deming}, Drake and {Doty}, John and {De Lee}, Nathan and {Dressing}, Courtney and {Dunham}, Edward W. and {Endl}, Michael and {Fressin}, Francois and {Ge}, Jian and {Henning}, Thomas and {Holman}, Matthew J. and {Howard}, Andrew W. and {Ida}, Shigeru and {Jenkins}, Jon M. and {Jernigan}, Garrett and {Johnson}, John Asher and {Kaltenegger}, Lisa and {Kawai}, Nobuyuki and {Kjeldsen}, Hans and {Laughlin}, Gregory and {Levine}, Alan M. and {Lin}, Douglas and {Lissauer}, Jack J. and {MacQueen}, Phillip and {Marcy}, Geoffrey and {McCullough}, Peter R. and {Morton}, Timothy D. and {Narita}, Norio and {Paegert}, Martin and {Palle}, Enric and {Pepe}, Francesco and {Pepper}, Joshua and {Quirrenbach}, Andreas and {Rinehart}, Stephen A. and {Sasselov}, Dimitar and {Sato}, Bun'ei and {Seager}, Sara and {Sozzetti}, Alessandro and {Stassun}, Keivan G. and {Sullivan}, Peter and {Szentgyorgyi}, Andrew and {Torres}, Guillermo and {Udry}, Stephane and {Villasenor}, Joel},
        title = "{Transiting Exoplanet Survey Satellite (TESS)}",
      journal = {Journal of Astronomical Telescopes, Instruments, and Systems},
         year = 2015,
        month = jan,
       volume = {1},
          eid = {014003},
        pages = {014003},
          doi = {10.1117/1.JATIS.1.1.014003},
       adsurl = {https://ui.adsabs.harvard.edu/abs/2015JATIS...1a4003R}
}

@ARTICLE{Jayasinghe2021,
       author = {{Jayasinghe}, T. and {Stanek}, K.~Z. and {Thompson}, Todd A. and {Kochanek}, C.~S. and {Rowan}, D.~M. and {Vallely}, P.~J. and {Strassmeier}, K.~G. and {Weber}, M. and {Hinkle}, J.~T. and {Hambsch}, F. -J. and {Martin}, D.~V. and {Prieto}, J.~L. and {Pessi}, T. and {Huber}, D. and {Auchettl}, K. and {Lopez}, L.~A. and {Ilyin}, I. and {Badenes}, C. and {Howard}, A.~W. and {Isaacson}, H. and {Murphy}, S.~J.},
        title = "{A unicorn in monoceros: the 3 M$_{{\ensuremath{\odot}}}$ dark companion to the bright, nearby red giant V723 Mon is a non-interacting, mass-gap black hole candidate}",
      journal = {\mnras},
         year = 2021,
        month = jun,
       volume = {504},
       number = {2},
        pages = {2577-2602},
          doi = {10.1093/mnras/stab907},
archivePrefix = {arXiv},
       eprint = {2101.02212},
 primaryClass = {astro-ph.SR},
       adsurl = {https://ui.adsabs.harvard.edu/abs/2021MNRAS.504.2577J}
}

@ARTICLE{Wiktorowicz2019,
       author = {{Wiktorowicz}, Grzegorz and {Wyrzykowski}, {\L}ukasz and {Chruslinska}, Martyna and {Klencki}, Jakub and {Rybicki}, Krzysztof A. and {Belczynski}, Krzysztof},
        title = "{Populations of Stellar-mass Black Holes from Binary Systems}",
      journal = {\apj},
         year = 2019,
        month = nov,
       volume = {885},
       number = {1},
          eid = {1},
        pages = {1},
          doi = {10.3847/1538-4357/ab45e6},
archivePrefix = {arXiv},
       eprint = {1907.11431},
 primaryClass = {astro-ph.HE},
       adsurl = {https://ui.adsabs.harvard.edu/abs/2019ApJ...885....1W}
}

@ARTICLE{Agol2003,
       author = {{Agol}, Eric},
        title = "{Microlensing of Large Sources}",
      journal = {\apj},
         year = 2003,
        month = sep,
       volume = {594},
       number = {1},
        pages = {449-455},
          doi = {10.1086/376833},
archivePrefix = {arXiv},
       eprint = {astro-ph/0303457},
 primaryClass = {astro-ph},
       adsurl = {https://ui.adsabs.harvard.edu/abs/2003ApJ...594..449A}
}

@ARTICLE{Agol2002,
       author = {{Agol}, Eric},
        title = "{Occultation and Microlensing}",
      journal = {\apj},
         year = "2002",
        month = "Nov",
       volume = {579},
       number = {1},
        pages = {430-436},
          doi = {10.1086/342880},
archivePrefix = {arXiv},
       eprint = {astro-ph/0207228},
 primaryClass = {astro-ph},
       adsurl = {https://ui.adsabs.harvard.edu/abs/2002ApJ...579..430A}
}

@ARTICLE{Beskin2002,
       author = {{Beskin}, G.~M. and {Tuntsov}, A.~V.},
        title = "{Detection of compact objects by means of gravitational lensing in binary systems}",
      journal = {\aap},
         year = 2002,
        month = nov,
       volume = {394},
        pages = {489-503},
          doi = {10.1051/0004-6361:20021150},
archivePrefix = {arXiv},
       eprint = {astro-ph/0208095},
 primaryClass = {astro-ph},
       adsurl = {https://ui.adsabs.harvard.edu/abs/2002A&A...394..489B}
}

@ARTICLE{blackcat,
       author = {{Corral-Santana}, J.~M. and {Casares}, J. and {Mu{\~n}oz-Darias}, T. and {Bauer}, F.~E. and {Mart{\'\i}nez-Pais}, I.~G. and {Russell}, D.~M.},
        title = "{BlackCAT: A catalogue of stellar-mass black holes in X-ray transients}",
      journal = {\aap},
         year = 2016,
        month = mar,
       volume = {587},
          eid = {A61},
        pages = {A61},
          doi = {10.1051/0004-6361/201527130},
archivePrefix = {arXiv},
       eprint = {1510.08869},
 primaryClass = {astro-ph.HE},
       adsurl = {https://ui.adsabs.harvard.edu/abs/2016A&A...587A..61C}
}

@ARTICLE{Crossland2023,
       author = {{Crossland}, Allison and {Bellm}, Eric C. and {Klein}, Courtney and {Davenport}, James R.~A. and {Kupfer}, Thomas and {Groom}, Steven L. and {Laher}, Russ R. and {Riddle}, Reed},
        title = "{A Pilot Search for Gravitational Self-Lensing Binaries with the Zwicky Transient Facility}",
      journal = {The Open Journal of Astrophysics},
         year = 2024,
        month = aug,
       volume = {7},
          eid = {67},
        pages = {67},
          doi = {10.33232/001c.122349},
archivePrefix = {arXiv},
       eprint = {2311.17862},
 primaryClass = {astro-ph.HE},
       adsurl = {https://ui.adsabs.harvard.edu/abs/2024OJAp....7E..67C}
}

@ARTICLE{DemircanKahraman1991,
       author = {{Demircan}, Osman and {Kahraman}, Goksel},
        title = "{Stellar Mass / Luminosity and Mass / Radius Relations}",
      journal = {\apss},
         year = 1991,
        month = jul,
       volume = {181},
       number = {2},
        pages = {313-322},
          doi = {10.1007/BF00639097},
       adsurl = {https://ui.adsabs.harvard.edu/abs/1991Ap&SS.181..313D}
}

@ARTICLE{Gould1995,
       author = {{Gould}, Andrew},
        title = "{Self-lensing by Binaries}",
      journal = {\apj},
         year = 1995,
        month = jun,
       volume = {446},
        pages = {541},
          doi = {10.1086/175812},
archivePrefix = {arXiv},
       eprint = {astro-ph/9409057},
 primaryClass = {astro-ph},
       adsurl = {https://ui.adsabs.harvard.edu/abs/1995ApJ...446..541G}
}

@ARTICLE{Knutson2007,
       author = {{Knutson}, Heather A. and {Charbonneau}, David and {Noyes}, Robert W. and {Brown}, Timothy M. and {Gilliland}, Ronald L.},
        title = "{Using Stellar Limb-Darkening to Refine the Properties of HD 209458b}",
      journal = {\apj},
         year = 2007,
        month = jan,
       volume = {655},
       number = {1},
        pages = {564-575},
          doi = {10.1086/510111},
archivePrefix = {arXiv},
       eprint = {astro-ph/0603542},
 primaryClass = {astro-ph},
       adsurl = {https://ui.adsabs.harvard.edu/abs/2007ApJ...655..564K}
}

@ARTICLE{Leahy1983,
  author = {{Leahy}, D.~A. and {Elsner}, R.~F. and {Weisskopf}, M.~C.},
  title = {{On searches for periodic pulsed emission - The Rayleigh test compared
	to epoch folding}},
  journal = {\apj},
  year = {1983},
  volume = {272},
  pages = {256-258},
  month = sep,
  adsurl = {http://adsabs.harvard.edu/abs/1983ApJ...272..256L},
  doi = {10.1086/161288}
}

@ARTICLE{Maeder1973,
       author = {{Maeder}, A.},
        title = "{Light Curves of the Gravitational Lens-like Action for Binaries with Degenerate Stars}",
      journal = {\aap},
         year = 1973,
        month = jul,
       volume = {26},
        pages = {215},
       adsurl = {https://ui.adsabs.harvard.edu/abs/1973A&A....26..215M}
}

@ARTICLE{Hawking1975,
       author = {{Hawking}, S.~W.},
        title = "{Particle creation by black holes}",
      journal = {Communications in Mathematical Physics},
         year = 1975,
        month = aug,
       volume = {43},
       number = {3},
        pages = {199-220},
          doi = {10.1007/BF02345020},
       adsurl = {https://ui.adsabs.harvard.edu/abs/1975CMaPh..43..199H}
}

@ARTICLE{MasudaHotokezaka2019,
       author = {{Masuda}, Kento and {Hotokezaka}, Kenta},
        title = "{Prospects of Finding Detached Black Hole-Star Binaries with TESS}",
      journal = {\apj},
         year = 2019,
        month = oct,
       volume = {883},
       number = {2},
          eid = {169},
        pages = {169},
          doi = {10.3847/1538-4357/ab3a4f},
archivePrefix = {arXiv},
       eprint = {1808.10856},
 primaryClass = {astro-ph.HE},
       adsurl = {https://ui.adsabs.harvard.edu/abs/2019ApJ...883..169M}
}

@ARTICLE{Masuda_etal2019,
       author = {{Masuda}, Kento and {Kawahara}, Hajime and {Latham}, David W. and {Bieryla}, Allyson and {Kunitomo}, Masanobu and {MacLeod}, Morgan and {Aoki}, Wako},
        title = "{Self-lensing Discovery of a 0.2 M $_{{\ensuremath{\odot}}}$ White Dwarf in an Unusually Wide Orbit around a Sun-like Star}",
      journal = {\apjl},
         year = 2019,
        month = aug,
       volume = {881},
       number = {1},
          eid = {L3},
        pages = {L3},
          doi = {10.3847/2041-8213/ab321b},
archivePrefix = {arXiv},
       eprint = {1907.07656},
 primaryClass = {astro-ph.SR},
       adsurl = {https://ui.adsabs.harvard.edu/abs/2019ApJ...881L...3M}
}

@ARTICLE{Marsh2001,
       author = {{Marsh}, T.~R.},
        title = "{Gravitational lensing in eclipsing binary stars}",
      journal = {\mnras},
         year = 2001,
        month = jun,
       volume = {324},
       number = {3},
        pages = {547-552},
          doi = {10.1046/j.1365-8711.2001.04293.x},
archivePrefix = {arXiv},
       eprint = {astro-ph/0012390},
 primaryClass = {astro-ph},
       adsurl = {https://ui.adsabs.harvard.edu/abs/2001MNRAS.324..547M}
}

@ARTICLE{McMillan2026,
       author = {{McMillan}, Jake and {Ingram}, Adam and {Brown}, Cordelia Dashwood and {Igoshev}, Andrei and {Middleton}, Matthew and {Wiktorowicz}, Grzegorz and {Scaringi}, Simone},
        title = "{Population synthesis predictions of the Galactic compact binary gravitational wave foreground detectable by LISA}",
      journal = {\mnras},
         year = 2026,
        month = mar,
       volume = {546},
       number = {4},
          eid = {stag117},
        pages = {stag117},
          doi = {10.1093/mnras/stag117},
archivePrefix = {arXiv},
       eprint = {2602.11765},
 primaryClass = {astro-ph.HE},
       adsurl = {https://ui.adsabs.harvard.edu/abs/2026MNRAS.546ag117M}
}

@BOOK{Misner1973,
   author = {{Misner}, C.~W. and {Thorne}, K.~S. and {Wheeler}, J.~A.},
    title = "{Gravitation}",
publisher = {San Francisco: W.H.~Freeman and Co., 1973},
     year = 1973,
   adsurl = {http://adsabs.harvard.edu/abs/1973grav.book.....M}
}

@ARTICLE{Qin1997,
       author = {{Qin}, Bo and {Wu}, Xiang-ping and {Zou}, Zhen-long},
        title = "{Self-Microlensing in Compact Binary Systems}",
      journal = {Chinese Physics Letters},
         year = 1997,
        month = feb,
       volume = {14},
       number = {2},
        pages = {155-157},
          doi = {10.1088/0256-307X/14/2/022},
archivePrefix = {arXiv},
       eprint = {astro-ph/9611118},
 primaryClass = {astro-ph},
       adsurl = {https://ui.adsabs.harvard.edu/abs/1997ChPhL..14..155Q}
}

@ARTICLE{Sahu2003,
       author = {{Sahu}, Kailash C. and {Gilliland}, Ronald L.},
        title = "{Near-Field Microlensing and Its Effects on Stellar Transit Observations by Kepler}",
      journal = {\apj},
         year = 2003,
        month = feb,
       volume = {584},
       number = {2},
        pages = {1042-1052},
          doi = {10.1086/345776},
archivePrefix = {arXiv},
       eprint = {astro-ph/0210554},
 primaryClass = {astro-ph},
       adsurl = {https://ui.adsabs.harvard.edu/abs/2003ApJ...584.1042S}
}

@ARTICLE{Wang2024,
       author = {{Wang}, Song and {Zhao}, Xinlin and {Feng}, Fabo and {Ge}, Hongwei and {Shao}, Yong and {Cui}, Yingzhen and {Gao}, Shijie and {Zhang}, Lifu and {Wang}, Pei and {Li}, Xue and {Bai}, Zhongrui and {Yuan}, Hailong and {Huang}, Yang and {Yuan}, Haibo and {Zhang}, Zhixiang and {Yi}, Tuan and {Xiang}, Maosheng and {Li}, Zhenwei and {Li}, Tanda and {Zhang}, Junbo and {Zhang}, Meng and {Han}, Henggeng and {Fan}, Dongwei and {Li}, Xiangdong and {Chen}, Xuefei and {Liu}, Zhengwei and {Meng}, Xiangcun and {Liu}, Qingzhong and {Zhang}, Haotong and {Gu}, Wei-Min and {Liu}, Jifeng},
        title = "{A potential mass-gap black hole in a wide binary with a circular orbit}",
      journal = {Nature Astronomy},
         year = 2024,
        month = dec,
       volume = {8},
        pages = {1583-1591},
          doi = {10.1038/s41550-024-02359-9},
archivePrefix = {arXiv},
       eprint = {2409.06352},
 primaryClass = {astro-ph.SR},
       adsurl = {https://ui.adsabs.harvard.edu/abs/2024NatAs...8.1583W}
}

@ARTICLE{Wiktorowicz2021,
       author = {{Wiktorowicz}, Grzegorz and {Middleton}, Matthew and {Khan}, Norman and {Ingram}, Adam and {Gandhi}, Poshak and {Dickinson}, Hugh},
        title = "{Predicting the self-lensing population in optical surveys}",
      journal = {\mnras},
         year = 2021,
        month = oct,
       volume = {507},
       number = {1},
        pages = {374-384},
          doi = {10.1093/mnras/stab2135},
archivePrefix = {arXiv},
       eprint = {2104.12666},
 primaryClass = {astro-ph.HE},
       adsurl = {https://ui.adsabs.harvard.edu/abs/2021MNRAS.507..374W}
}

@ARTICLE{Witt1994,
       author = {{Witt}, Hans J. and {Mao}, Shude},
        title = "{Can Lensed Stars Be Regarded as Pointlike for Microlensing by MACHOs?}",
      journal = {\apj},
         year = 1994,
        month = aug,
       volume = {430},
        pages = {505},
          doi = {10.1086/174426},
       adsurl = {https://ui.adsabs.harvard.edu/abs/1994ApJ...430..505W}
}

@ARTICLE{Yamaguchi2018,
       author = {{Yamaguchi}, Masaki S. and {Kawanaka}, Norita and {Bulik}, Tomasz and {Piran}, Tsvi},
        title = "{Detecting Black Hole Binaries by Gaia}",
      journal = {\apj},
         year = 2018,
        month = jul,
       volume = {861},
       number = {1},
          eid = {21},
        pages = {21},
          doi = {10.3847/1538-4357/aac5ec},
archivePrefix = {arXiv},
       eprint = {1710.09839},
 primaryClass = {astro-ph.SR},
       adsurl = {https://ui.adsabs.harvard.edu/abs/2018ApJ...861...21Y}
}

@ARTICLE{GaiaDR3,
       author = {{Gaia Collaboration} and {Vallenari}, A. and {Brown}, A.~G.~A. and {Prusti}, T. and {de Bruijne}, J.~H.~J. and {Arenou}, F. and {Babusiaux}, C. and {Biermann}, M. and {Creevey}, O.~L. and {Ducourant}, C. and {Evans}, D.~W. and {Eyer}, L. and {Guerra}, R. and {Hutton}, A. and {Jordi}, C. and {Klioner}, S.~A. and {Lammers}, U.~L. and {Lindegren}, L. and {Luri}, X. and {Mignard}, F. and {Panem}, C. and {Pourbaix}, D. and {Randich}, S. and {Sartoretti}, P. and {Soubiran}, C. and {Tanga}, P. and {Walton}, N.~A. and {Bailer-Jones}, C.~A.~L. and {Bastian}, U. and {Drimmel}, R. and {Jansen}, F. and {Katz}, D. and {Lattanzi}, M.~G. and {van Leeuwen}, F. and {Bakker}, J. and {Cacciari}, C. and {Casta{\~n}eda}, J. and {De Angeli}, F. and {Fabricius}, C. and {Fouesneau}, M. and {Fr{\'e}mat}, Y. and {Galluccio}, L. and {Guerrier}, A. and {Heiter}, U. and {Masana}, E. and {Messineo}, R. and {Mowlavi}, N. and {Nicolas}, C. and {Nienartowicz}, K. and {Pailler}, F. and {Panuzzo}, P. and {Riclet}, F. and {Roux}, W. and {Seabroke}, G.~M. and {Sordo}, R. and {Th{\'e}venin}, F. and {Gracia-Abril}, G. and {Portell}, J. and {Teyssier}, D. and {Altmann}, M. and {Andrae}, R. and {Audard}, M. and {Bellas-Velidis}, I. and {Benson}, K. and {Berthier}, J. and {Blomme}, R. and {Burgess}, P.~W. and {Busonero}, D. and {Busso}, G. and {C{\'a}novas}, H. and {Carry}, B. and {Cellino}, A. and {Cheek}, N. and {Clementini}, G. and {Damerdji}, Y. and {Davidson}, M. and {de Teodoro}, P. and {Nu{\~n}ez Campos}, M. and {Delchambre}, L. and {Dell'Oro}, A. and {Esquej}, P. and {Fern{\'a}ndez-Hern{\'a}ndez}, J. and {Fraile}, E. and {Garabato}, D. and {Garc{\'\i}a-Lario}, P. and {Gosset}, E. and {Haigron}, R. and {Halbwachs}, J.-L. and {Hambly}, N.~C. and {Harrison}, D.~L. and {Hern{\'a}ndez}, J. and {Hestroffer}, D. and {Hodgkin}, S.~T. and {Holl}, B. and {Jan{\ss}en}, K. and {Jevardat de Fombelle}, G. and {Jordan}, S. and {Krone-Martins}, A. and {Lanzafame}, A.~C. and {L{\"o}ffler}, W. and {Marchal}, O. and {Marrese}, P.~M. and {Moitinho}, A. and {Muinonen}, K. and {Osborne}, P. and {Pancino}, E. and {Pauwels}, T. and {Recio-Blanco}, A. and {Reyl{\'e}}, C. and {Riello}, M. and {Rimoldini}, L. and {Roegiers}, T. and {Rybizki}, J. and {Sarro}, L.~M. and {Siopis}, C. and {Smith}, M. and {Sozzetti}, A. and {Utrilla}, E. and {van Leeuwen}, M. and {Abbas}, U. and {{\'A}brah{\'a}m}, P. and {Abreu Aramburu}, A. and {Aerts}, C. and {Aguado}, J.~J. and {Ajaj}, M. and {Aldea-Montero}, F. and {Altavilla}, G. and {{\'A}lvarez}, M.~A. and {Alves}, J. and {Anders}, F. and {Anderson}, R.~I. and {Anglada Varela}, E. and {Antoja}, T. and {Baines}, D. and {Baker}, S.~G. and {Balaguer-N{\'u}{\~n}ez}, L. and {Balbinot}, E. and {Balog}, Z. and {Barache}, C. and {Barbato}, D. and {Barros}, M. and {Barstow}, M.~A. and {Bartolom{\'e}}, S. and {Bassilana}, J.-L. and {Bauchet}, N. and {Becciani}, U. and {Bellazzini}, M. and {Berihuete}, A. and {Bernet}, M. and {Bertone}, S. and {Bianchi}, L. and {Binnenfeld}, A. and {Blanco-Cuaresma}, S. and {Blazere}, A. and {Boch}, T. and {Bombrun}, A. and {Bossini}, D. and {Bouquillon}, S. and {Bragaglia}, A. and {Bramante}, L. and {Breedt}, E. and {Bressan}, A. and {Brouillet}, N. and {Brugaletta}, E. and {Bucciarelli}, B. and {Burlacu}, A. and {Butkevich}, A.~G. and {Buzzi}, R. and {Caffau}, E. and {Cancelliere}, R. and {Cantat-Gaudin}, T. and {Carballo}, R. and {Carlucci}, T. and {Carnerero}, M.~I. and {Carrasco}, J.~M. and {Casamiquela}, L. and {Castellani}, M. and {Castro-Ginard}, A. and {Chaoul}, L. and {Charlot}, P. and {Chemin}, L. and {Chiaramida}, V. and {Chiavassa}, A. and {Chornay}, N. and {Comoretto}, G. and {Contursi}, G. and {Cooper}, W.~J. and {Cornez}, T. and {Cowell}, S. and {Crifo}, F. and {Cropper}, M. and {Crosta}, M. and {Crowley}, C. and {Dafonte}, C. and {Dapergolas}, A. and {David}, M. and {David}, P. and {de Laverny}, P. and {De Luise}, F. and {De March}, R.},
        title = "{Gaia Data Release 3. Summary of the content and survey properties}",
      journal = {\aap},
         year = 2023,
        month = jun,
       volume = {674},
          eid = {A1},
        pages = {A1},
          doi = {10.1051/0004-6361/202243940},
archivePrefix = {arXiv},
       eprint = {2208.00211},
 primaryClass = {astro-ph.GA},
       adsurl = {https://ui.adsabs.harvard.edu/abs/2023A&A...674A...1G}
}

@ARTICLE{Stassun2019,
       author = {{Stassun}, Keivan G. and {Oelkers}, Ryan J. and {Paegert}, Martin and {Torres}, Guillermo and {Pepper}, Joshua and {De Lee}, Nathan and {Collins}, Kevin and {Latham}, David W. and {Muirhead}, Philip S. and {Chittidi}, Jay and {Rojas-Ayala}, B{\'a}rbara and {Fleming}, Scott W. and {Rose}, Mark E. and {Tenenbaum}, Peter and {Ting}, Eric B. and {Kane}, Stephen R. and {Barclay}, Thomas and {Bean}, Jacob L. and {Brassuer}, C.~E. and {Charbonneau}, David and {Ge}, Jian and {Lissauer}, Jack J. and {Mann}, Andrew W. and {McLean}, Brian and {Mullally}, Susan and {Narita}, Norio and {Plavchan}, Peter and {Ricker}, George R. and {Sasselov}, Dimitar and {Seager}, S. and {Sharma}, Sanjib and {Shiao}, Bernie and {Sozzetti}, Alessandro and {Stello}, Dennis and {Vanderspek}, Roland and {Wallace}, Geoff and {Winn}, Joshua N.},
        title = "{The Revised TESS Input Catalog and Candidate Target List}",
      journal = {\aj},
         year = 2019,
        month = oct,
       volume = {158},
       number = {4},
          eid = {138},
        pages = {138},
          doi = {10.3847/1538-3881/ab3467},
archivePrefix = {arXiv},
       eprint = {1905.10694},
 primaryClass = {astro-ph.SR},
       adsurl = {https://ui.adsabs.harvard.edu/abs/2019AJ....158..138S}
}

@ARTICLE{Claret2025,
       author = {{Claret}, A. and {Hauschildt}, P.~H. and {Torres}, G.},
        title = "{Limb-darkening coefficients for four-term and power-2 laws for the JWST mission adopting spherical PHOENIX models at high resolution: NIRCam, NIRISS, and NIRSpec passbands}",
      journal = {\aap},
         year = 2025,
        month = jul,
       volume = {699},
          eid = {A97},
        pages = {A97},
          doi = {10.1051/0004-6361/202554770},
archivePrefix = {arXiv},
       eprint = {2506.07265},
 primaryClass = {astro-ph.SR},
       adsurl = {https://ui.adsabs.harvard.edu/abs/2025A&A...699A..97C}
}

@ARTICLE{Claret2023,
       author = {{Claret}, A. and {Southworth}, J.},
        title = "{Power-2 limb-darkening coefficients for the uvby, UBVRIJHK, SDSS ugriz, Gaia, Kepler, TESS, and CHEOPS photometric systems. II. PHOENIX spherically symmetric stellar atmosphere models}",
      journal = {\aap},
         year = 2023,
        month = jun,
       volume = {674},
          eid = {A63},
        pages = {A63},
          doi = {10.1051/0004-6361/202346478},
archivePrefix = {arXiv},
       eprint = {2305.01704},
 primaryClass = {astro-ph.SR},
       adsurl = {https://ui.adsabs.harvard.edu/abs/2023A&A...674A..63C}
}

@ARTICLE{Claret2022,
       author = {{Claret}, A. and {Southworth}, J.},
        title = "{Power-2 limb-darkening coefficients for the uvby, UBVRIJHK, SDSS ugriz, Gaia, Kepler, and TESS photometric systems. I. ATLAS stellar atmosphere models}",
      journal = {\aap},
         year = 2022,
        month = aug,
       volume = {664},
          eid = {A128},
        pages = {A128},
          doi = {10.1051/0004-6361/202243827},
archivePrefix = {arXiv},
       eprint = {2206.11098},
 primaryClass = {astro-ph.SR},
       adsurl = {https://ui.adsabs.harvard.edu/abs/2022A&A...664A.128C}
}

@ARTICLE{Claret2000,
       author = {{Claret}, A.},
        title = "{A new non-linear limb-darkening law for LTE stellar atmosphere models. Calculations for -5.0 <= log[M/H] <= +1, 2000 K <= T$_{eff}$ <= 50000 K at several surface gravities}",
      journal = {\aap},
         year = 2000,
        month = nov,
       volume = {363},
        pages = {1081-1190},
       adsurl = {https://ui.adsabs.harvard.edu/abs/2000A&A...363.1081C}
}

@ARTICLE{ParviainenLDTK2015,
       author = {{Parviainen}, H. and {Aigrain}, S.},
        title = "{LDTK: Limb Darkening Toolkit}",
      journal = {\mnras},
         year = 2015,
        month = nov,
       volume = {453},
       number = {4},
        pages = {3821-3826},
          doi = {10.1093/mnras/stv1857},
archivePrefix = {arXiv},
       eprint = {1508.02634},
 primaryClass = {astro-ph.EP},
       adsurl = {https://ui.adsabs.harvard.edu/abs/2015MNRAS.453.3821P}
}




\appendix

\section{Plots for the remaining candidates}
\label{sec:others}

\begin{figure*}
\centering
\includegraphics[width=0.49\textwidth]{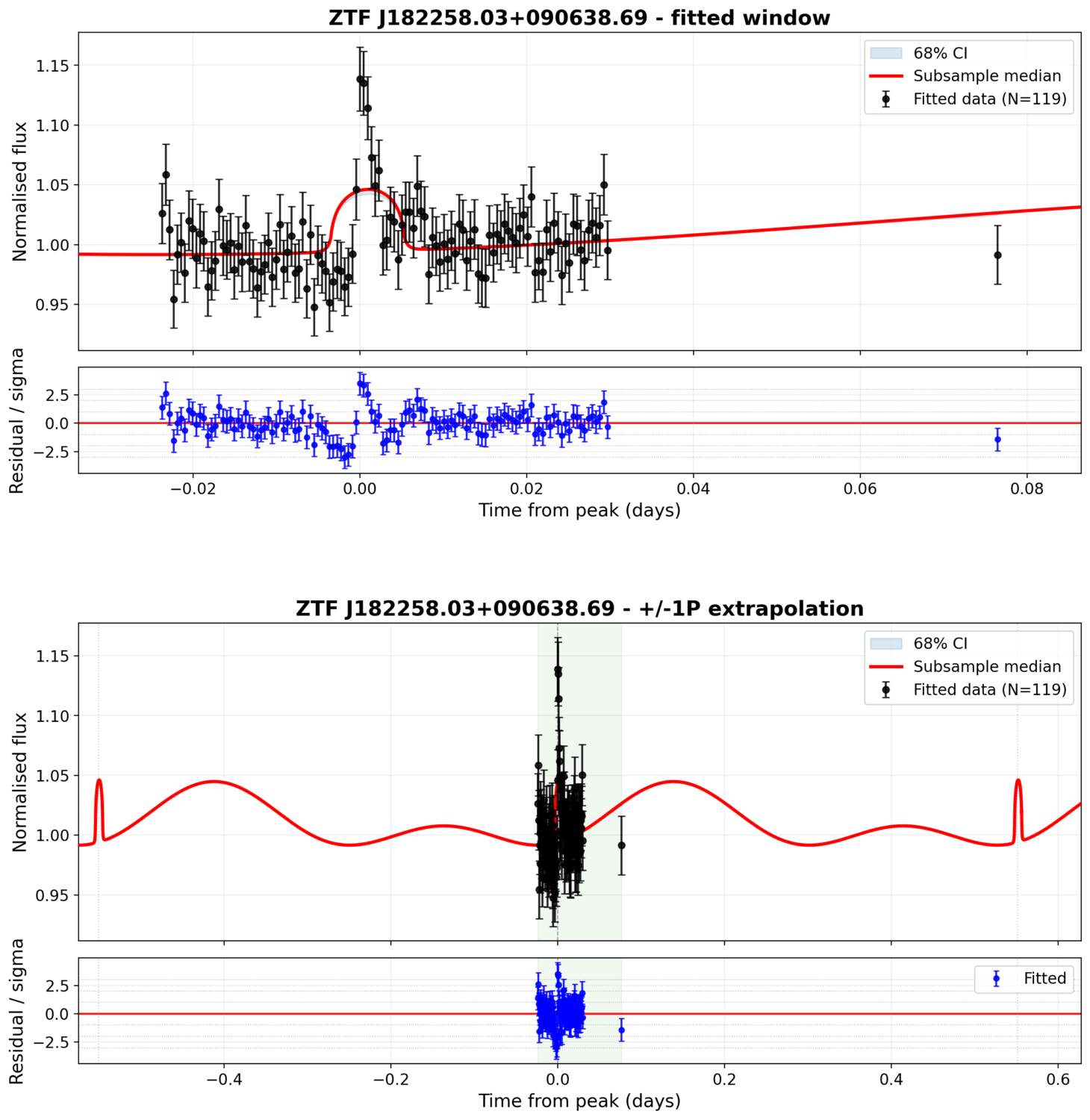}
\hfill
\includegraphics[width=0.47\textwidth]{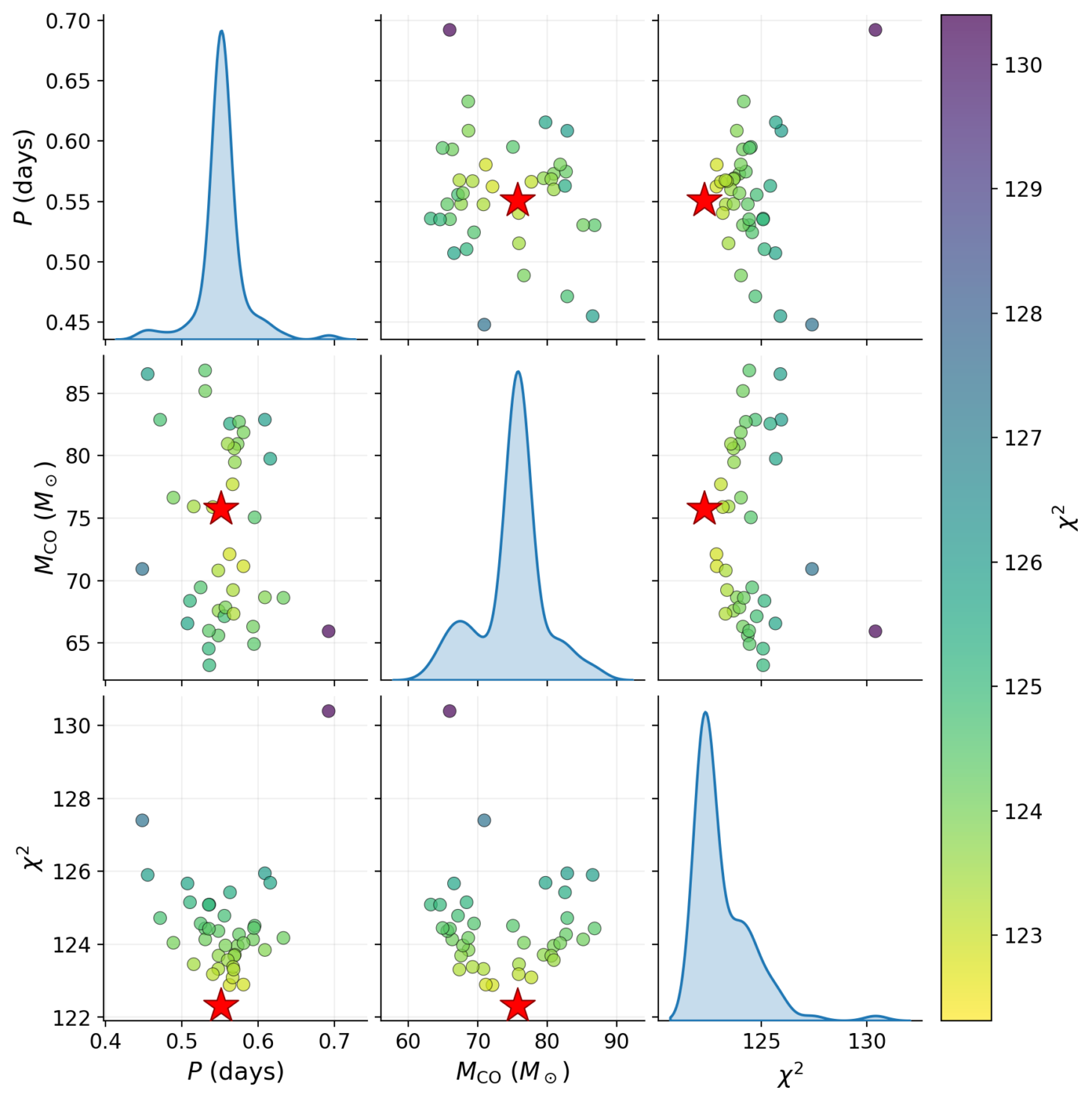}

\caption{
Results of fitting the model to ZTF J182258.03+090638.69. See the Caption of Fig \ref{fig:J201150_full} for details.
}
\label{fig:J182258_full}
\end{figure*}

\begin{figure*}
\centering
\includegraphics[width=0.49\textwidth]{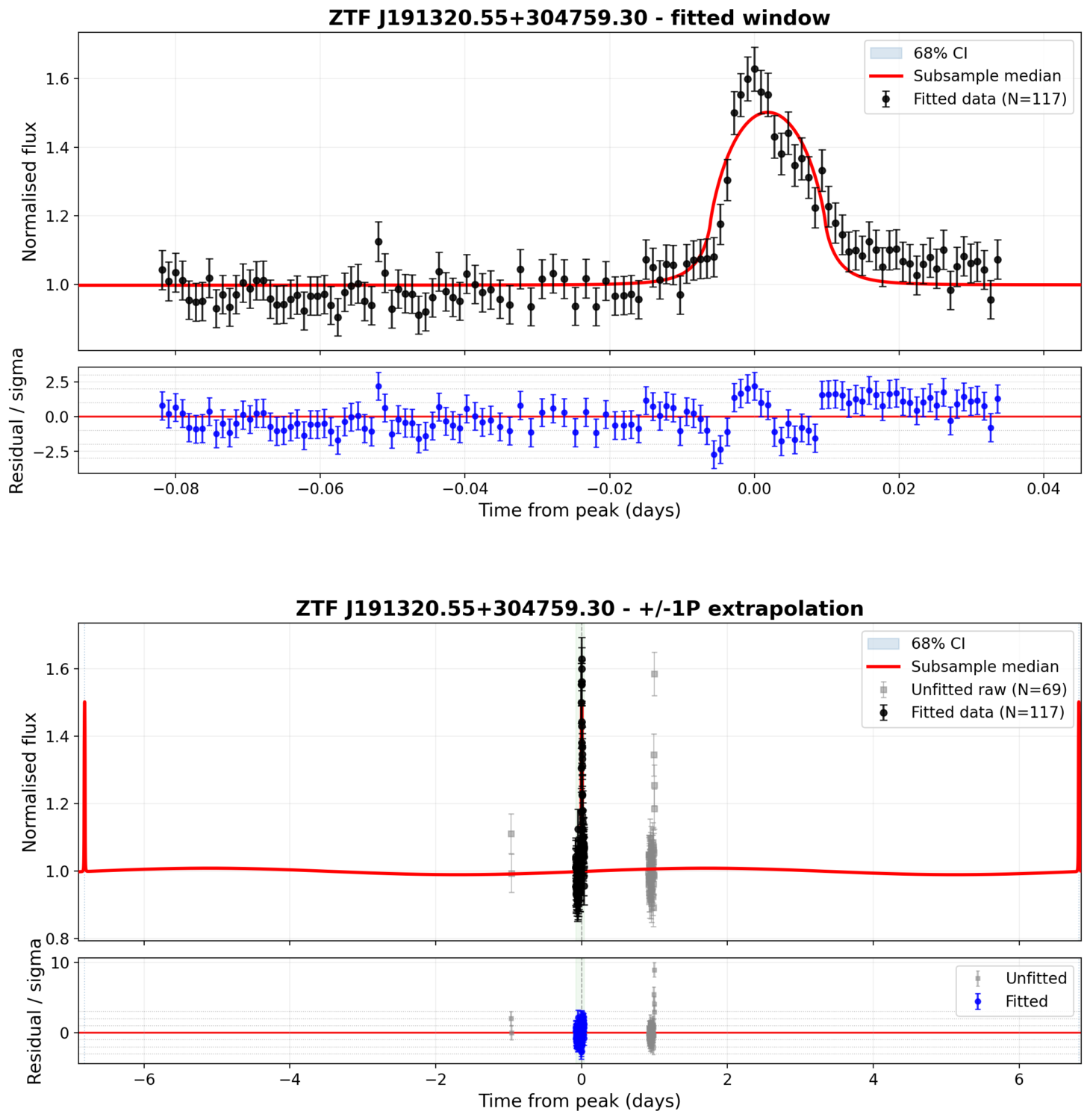}
\hfill
\includegraphics[width=0.47\textwidth]{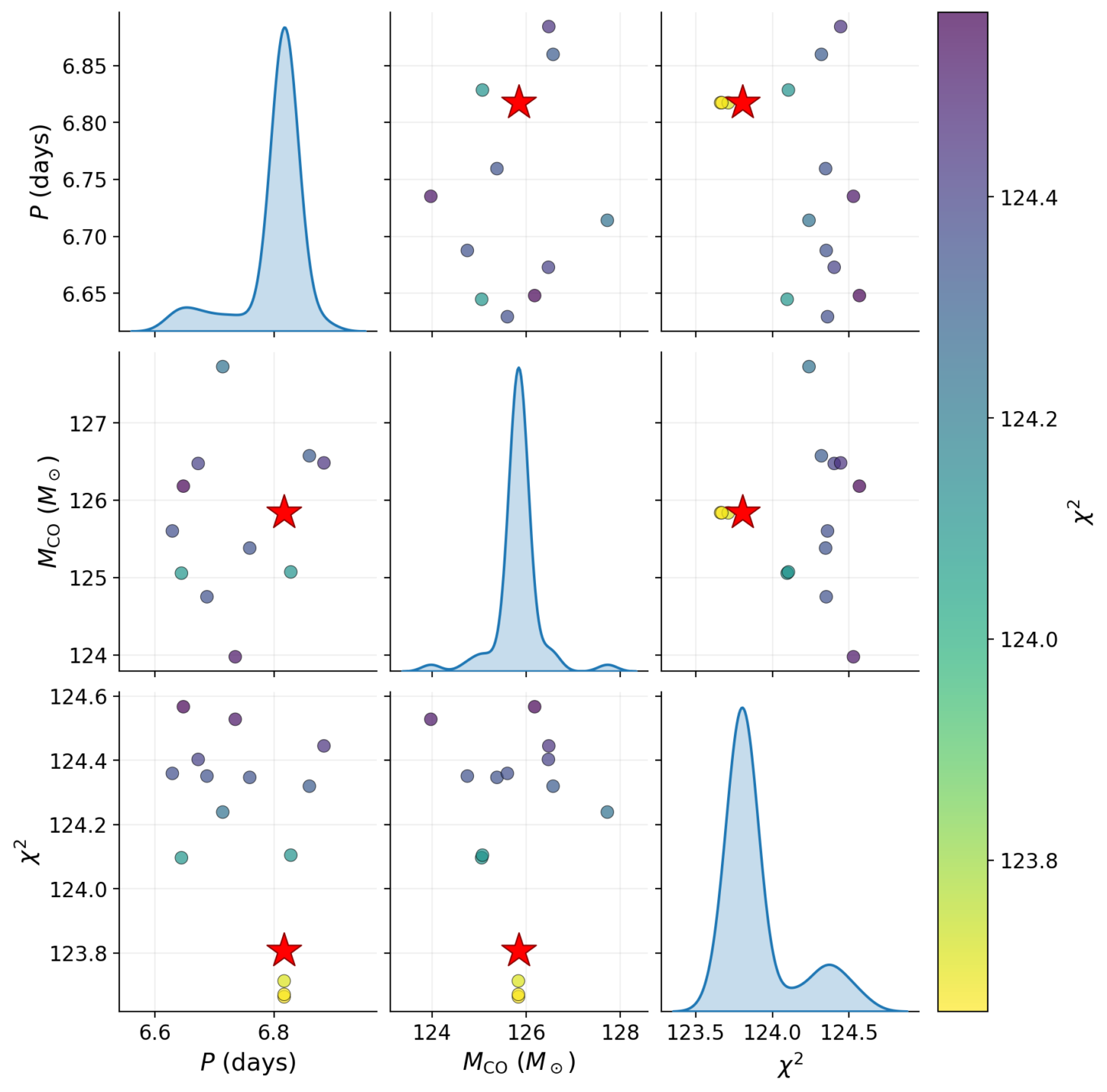}

\caption{
Results of fitting the model to ZTF J191320.55+304759.30.
}
\label{fig:J191320_full}
\end{figure*}

\begin{figure*}
\centering
\includegraphics[width=0.49\textwidth]{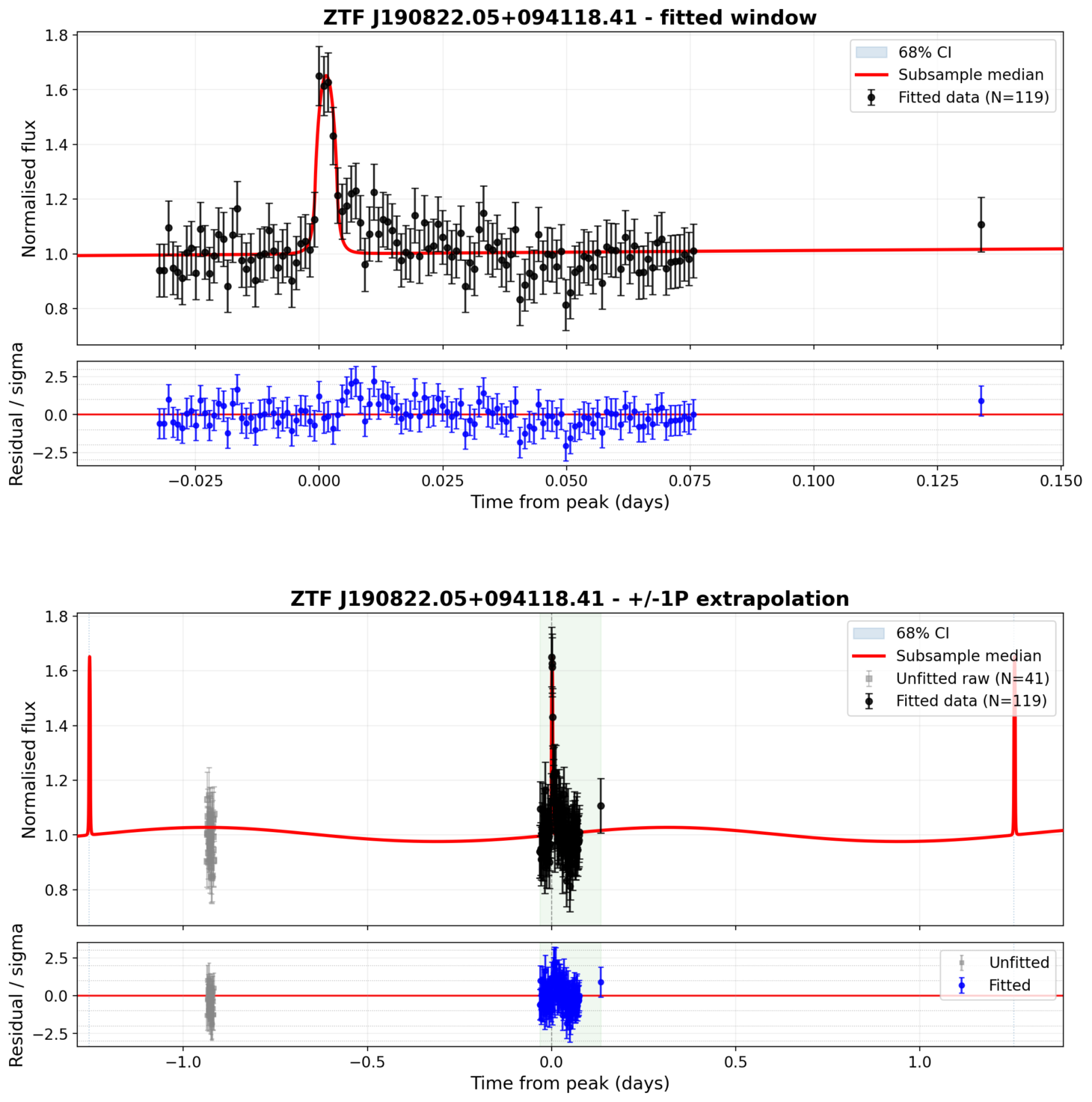}
\hfill
\includegraphics[width=0.47\textwidth]{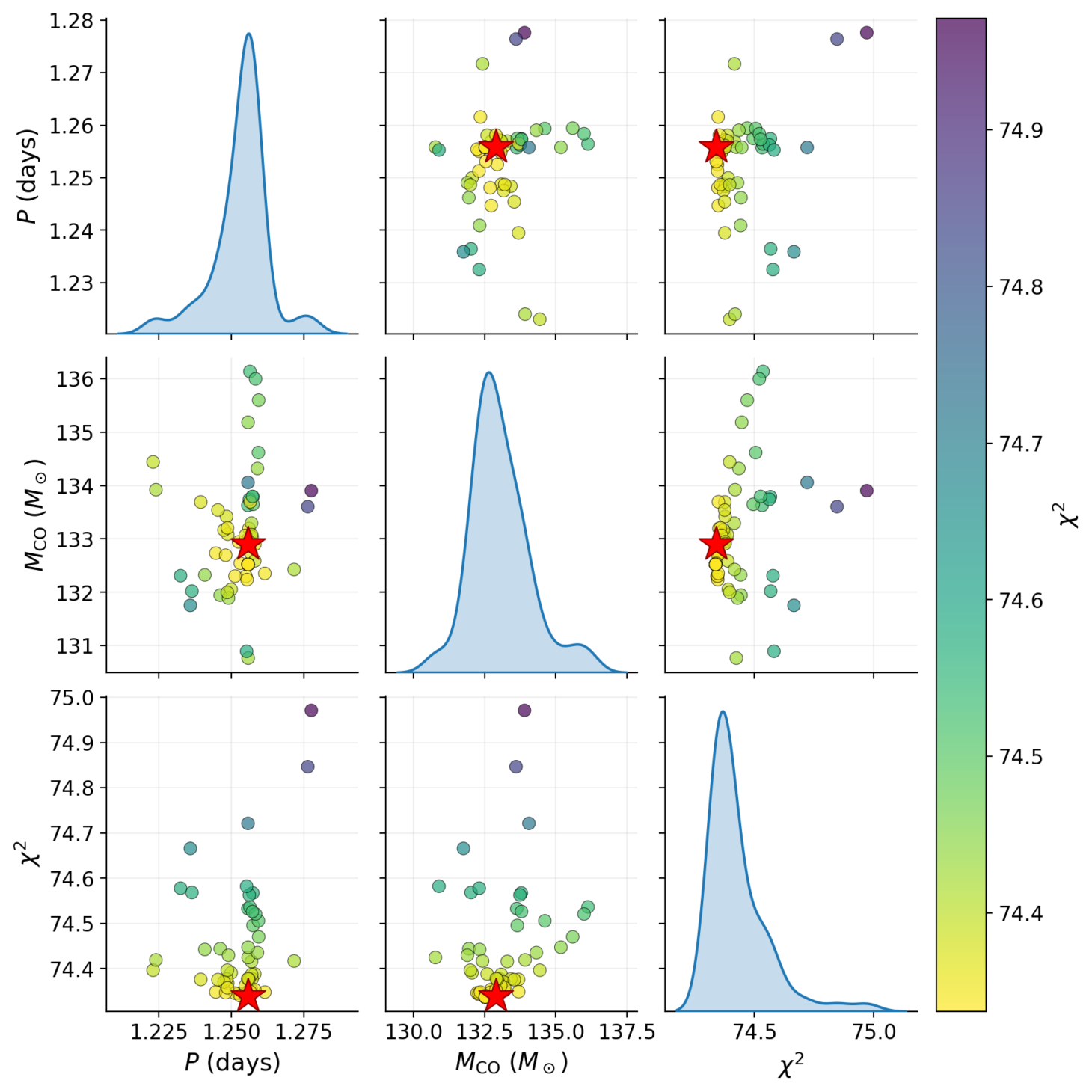}

\caption{Results of fitting the model to ZTF J190822.05+094118.41.}
\label{fig:J190822_full}
\end{figure*}

\begin{figure}
\centering
\includegraphics[width=0.47\textwidth]{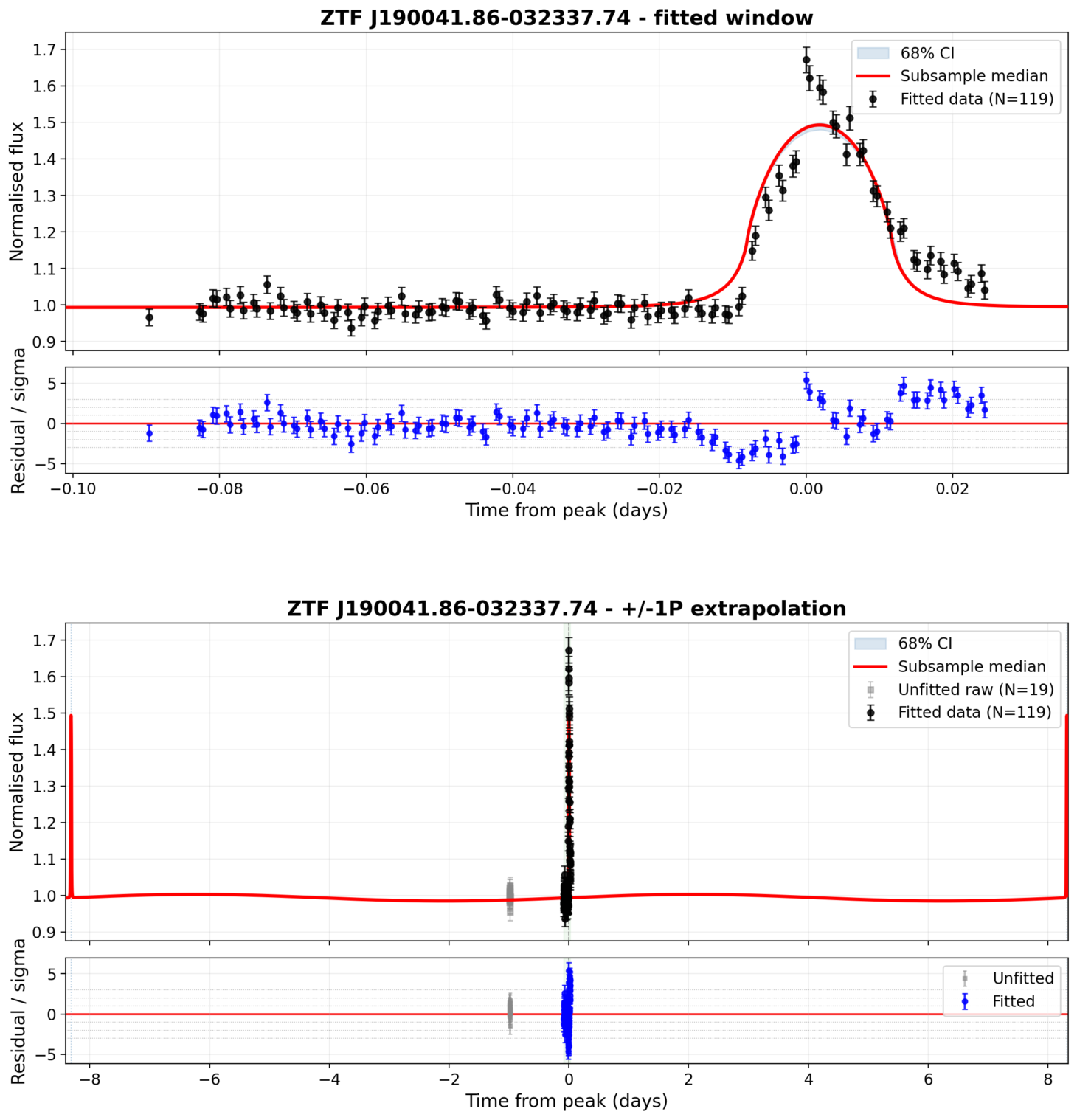}
\caption{Results of fitting the model to J190041}
\label{fig:J190041}
\end{figure}

\begin{figure}
\centering
\includegraphics[width=0.47\textwidth]{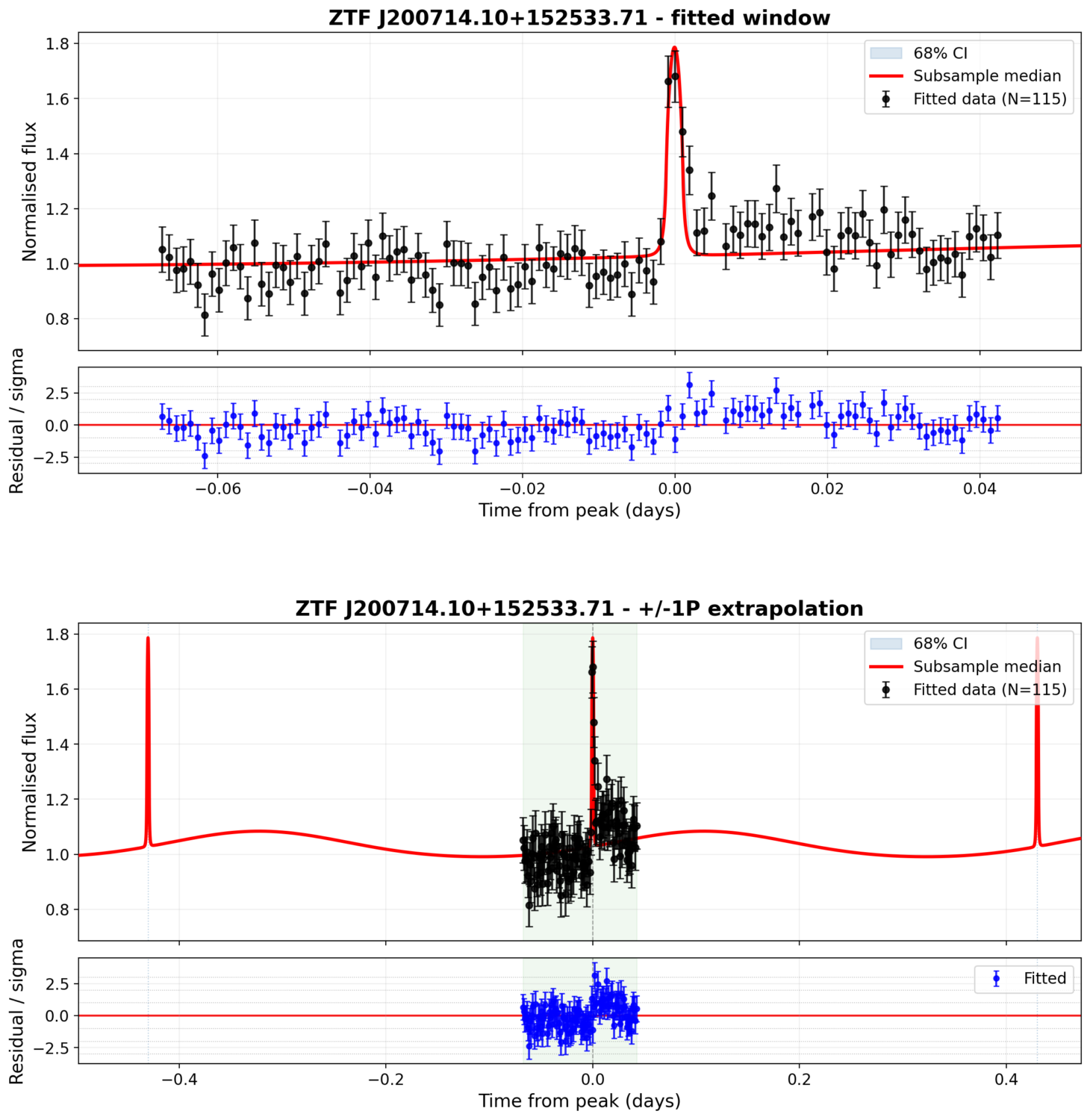}
\caption{Results of fitting the model to J200714}
\label{fig:J200714}
\end{figure}

\begin{figure}
\centering
\includegraphics[width=0.47\textwidth]{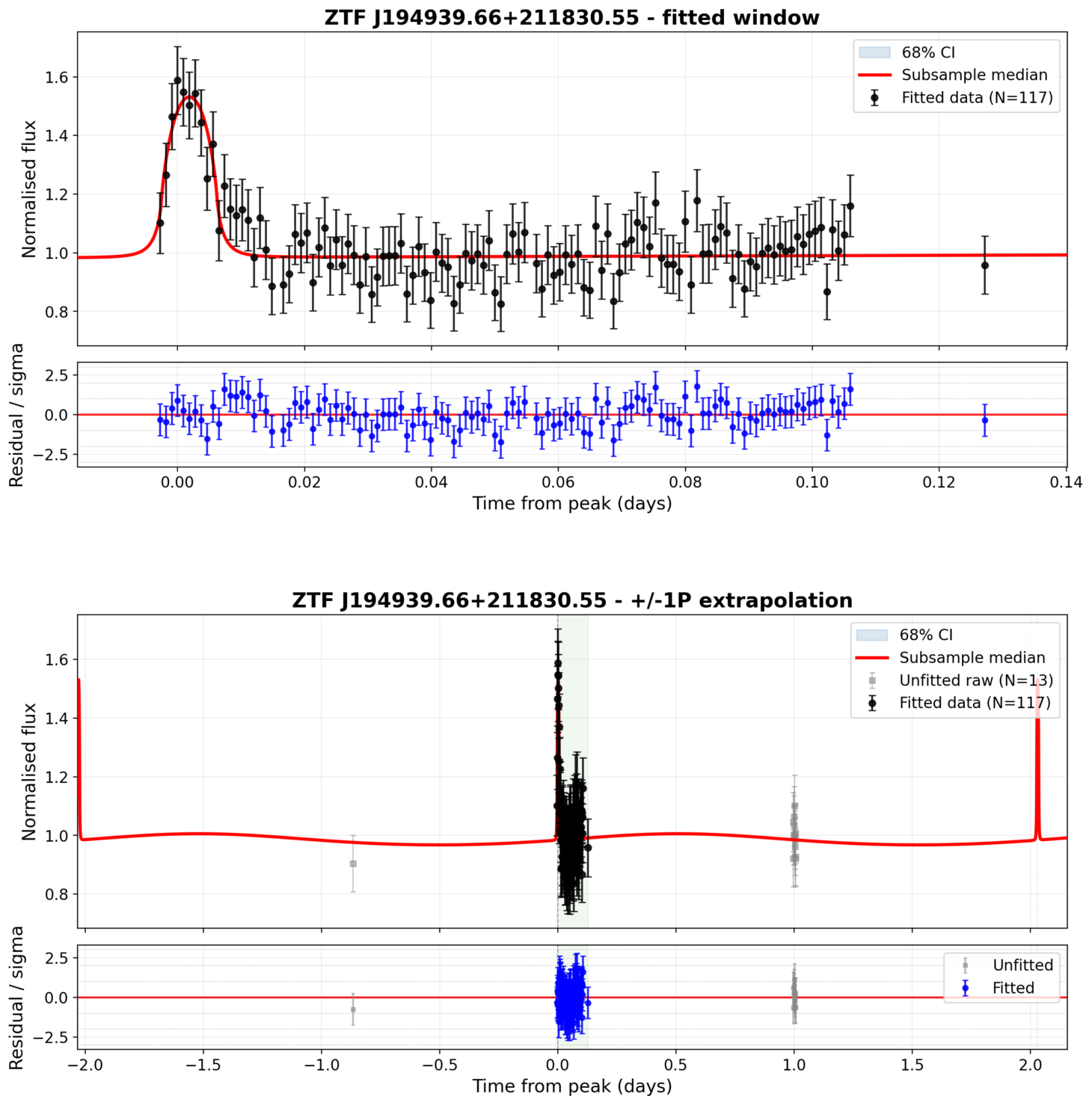}
\caption{Results of fitting the model to J194939}
\label{fig:J194939}
\end{figure}

\begin{figure}
\centering
\includegraphics[width=0.47\textwidth]{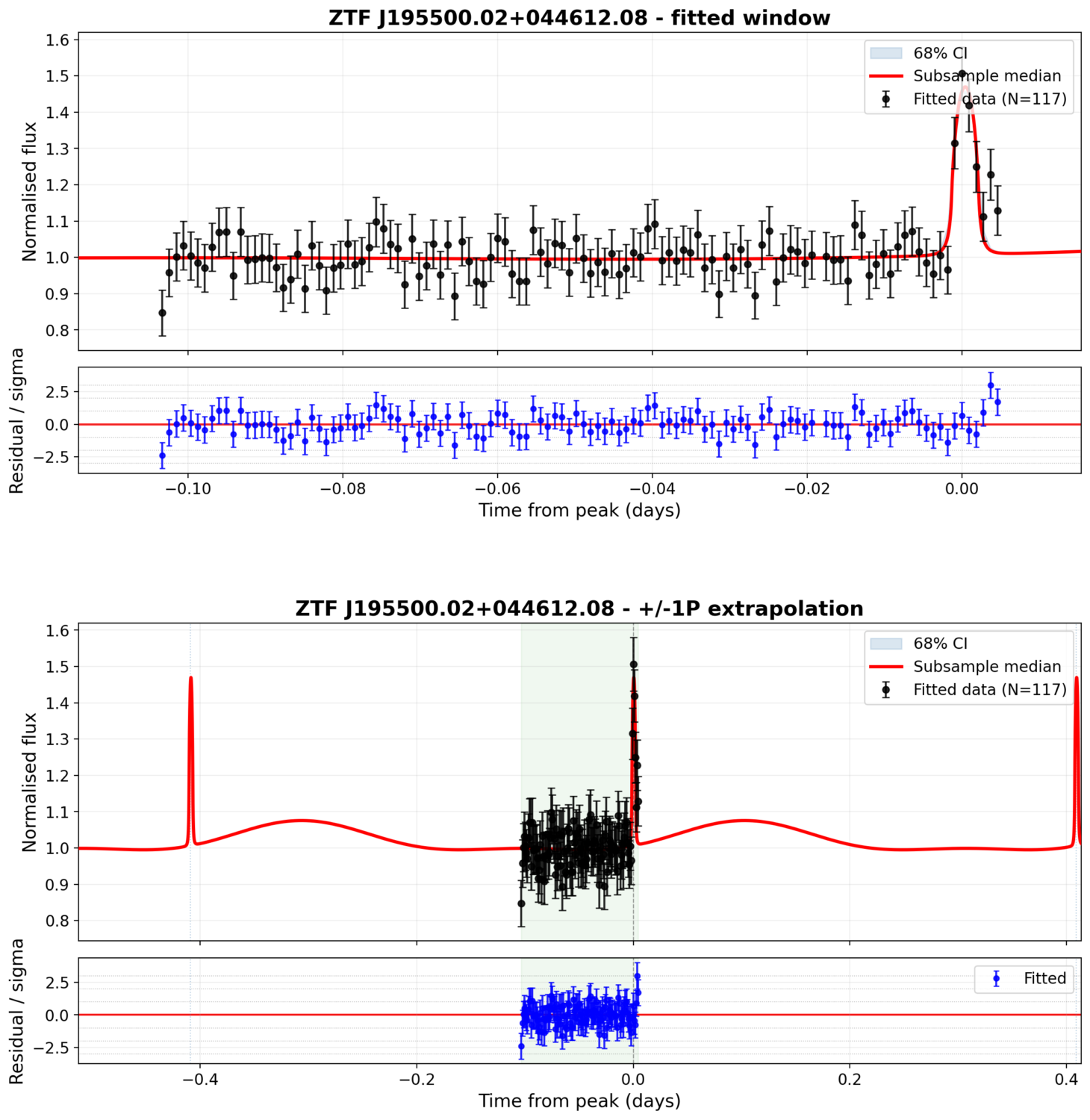}
\caption{Results of fitting the model to J195500}
\label{fig:J195500}
\end{figure}

\begin{figure}
\centering
\includegraphics[width=0.47\textwidth]{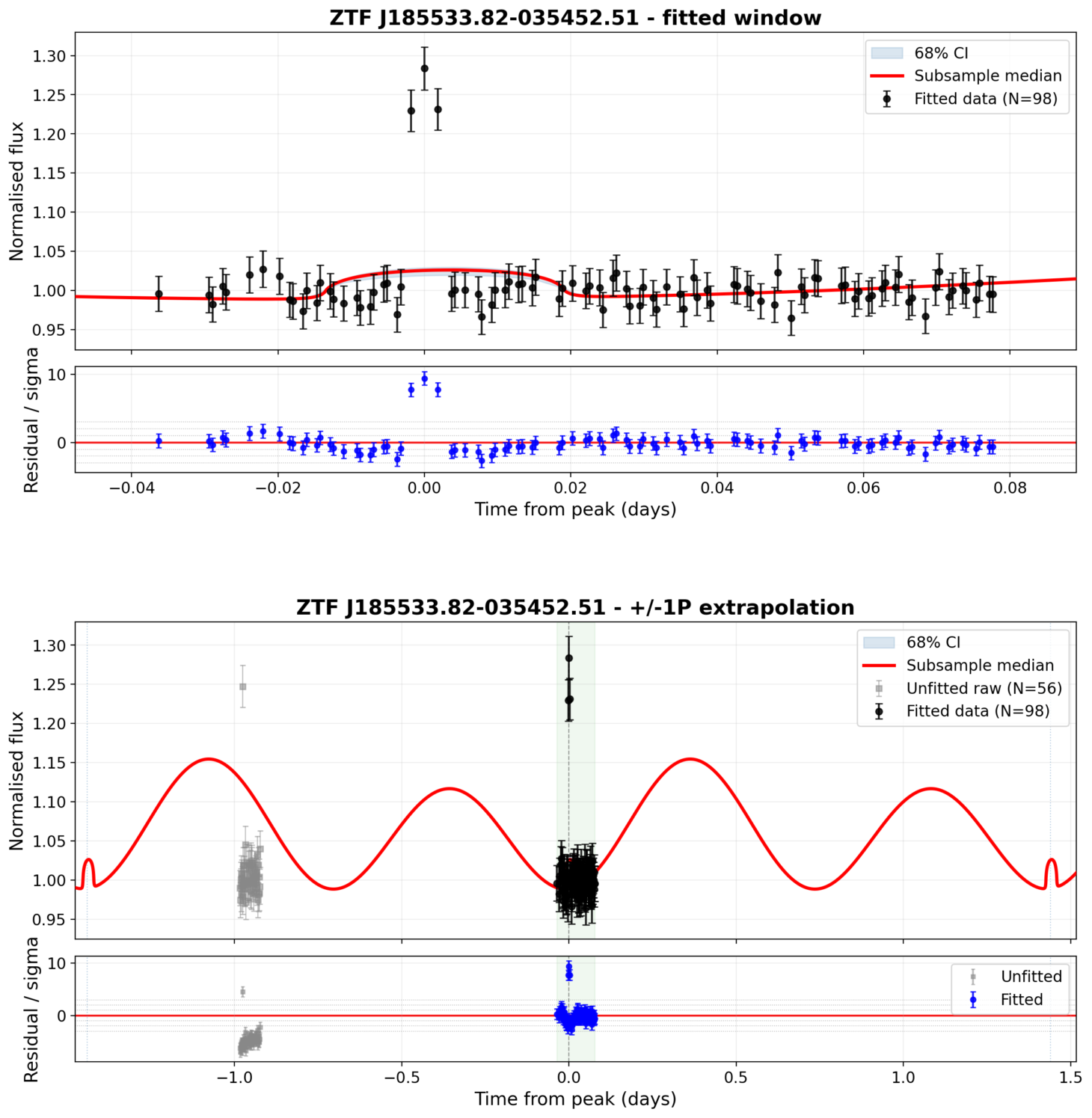}
\caption{Results of fitting the model to J185533}
\label{fig:J185533}
\end{figure}

\begin{figure}
\centering
\includegraphics[width=0.47\textwidth]{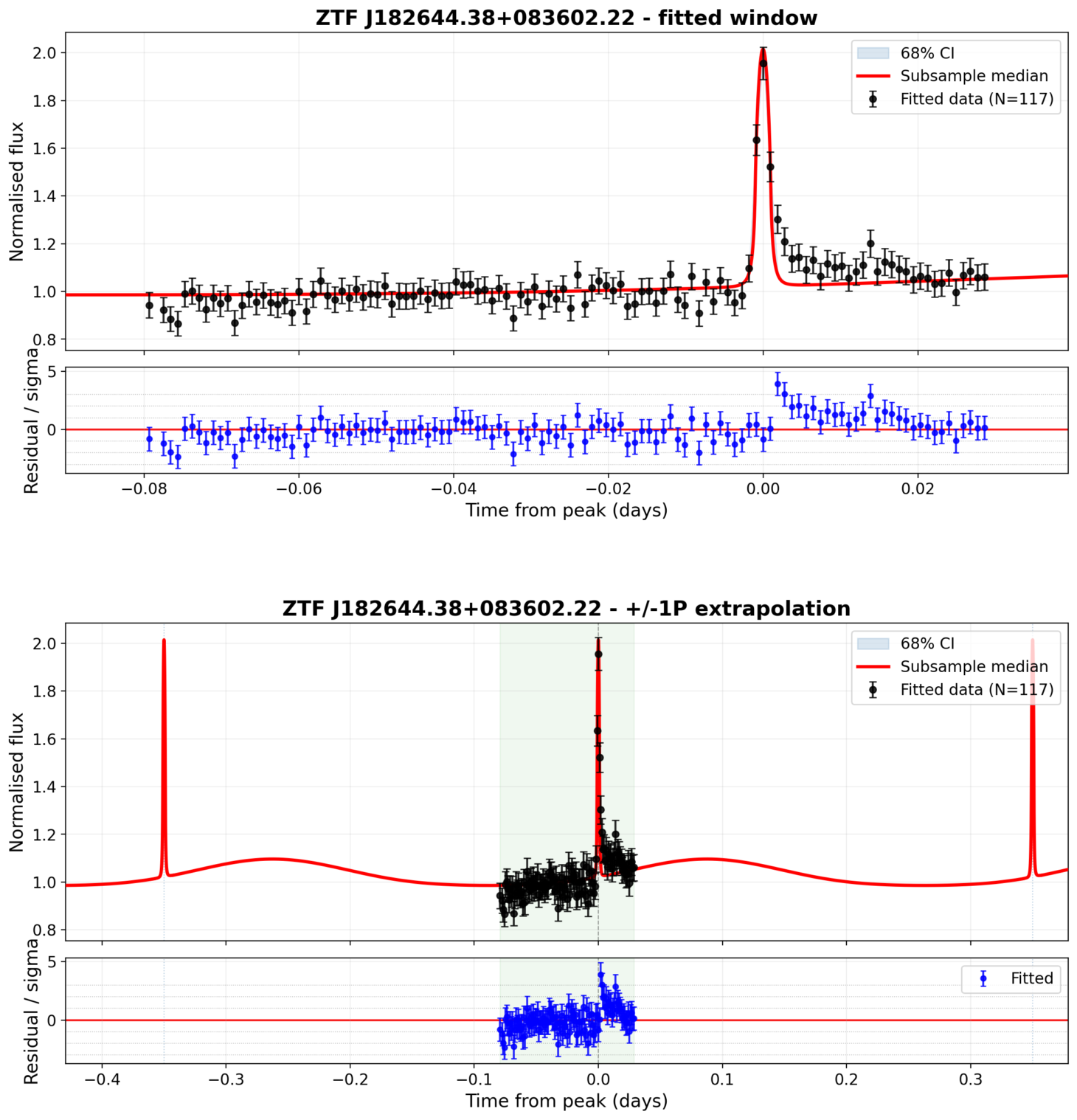}
\caption{Results of fitting the model to J182644}
\label{fig:J182644}
\end{figure}

\begin{figure}
\centering
\includegraphics[width=0.47\textwidth]{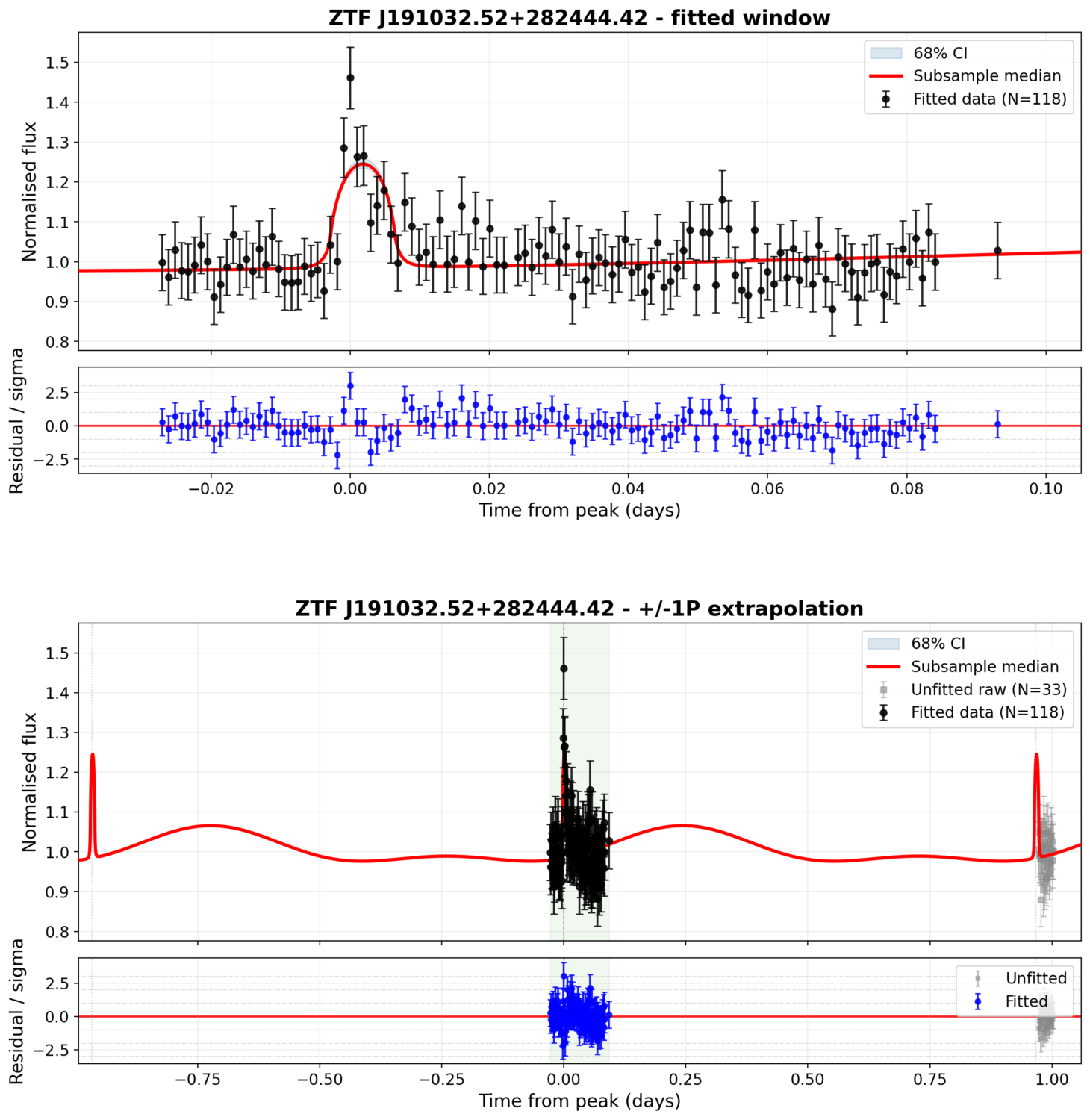}
\caption{Results of fitting the model to J191032}
\label{fig:J191032}
\end{figure}

\begin{figure}
\centering
\includegraphics[width=0.47\textwidth]{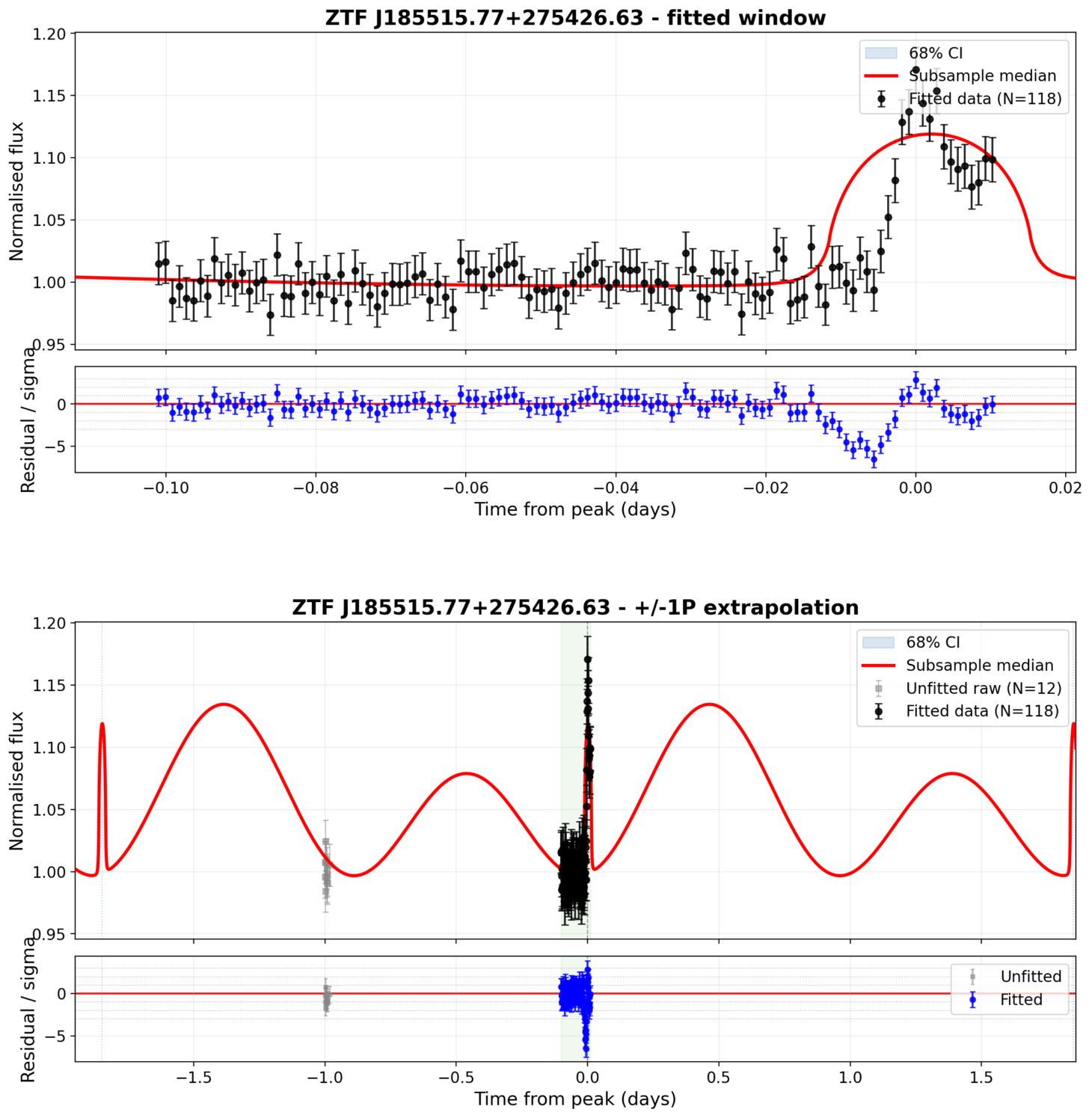}
\caption{Results of fitting the model to J185515}
\label{fig:J185515}
\end{figure}

\begin{figure}
\centering
\includegraphics[width=0.47\textwidth]{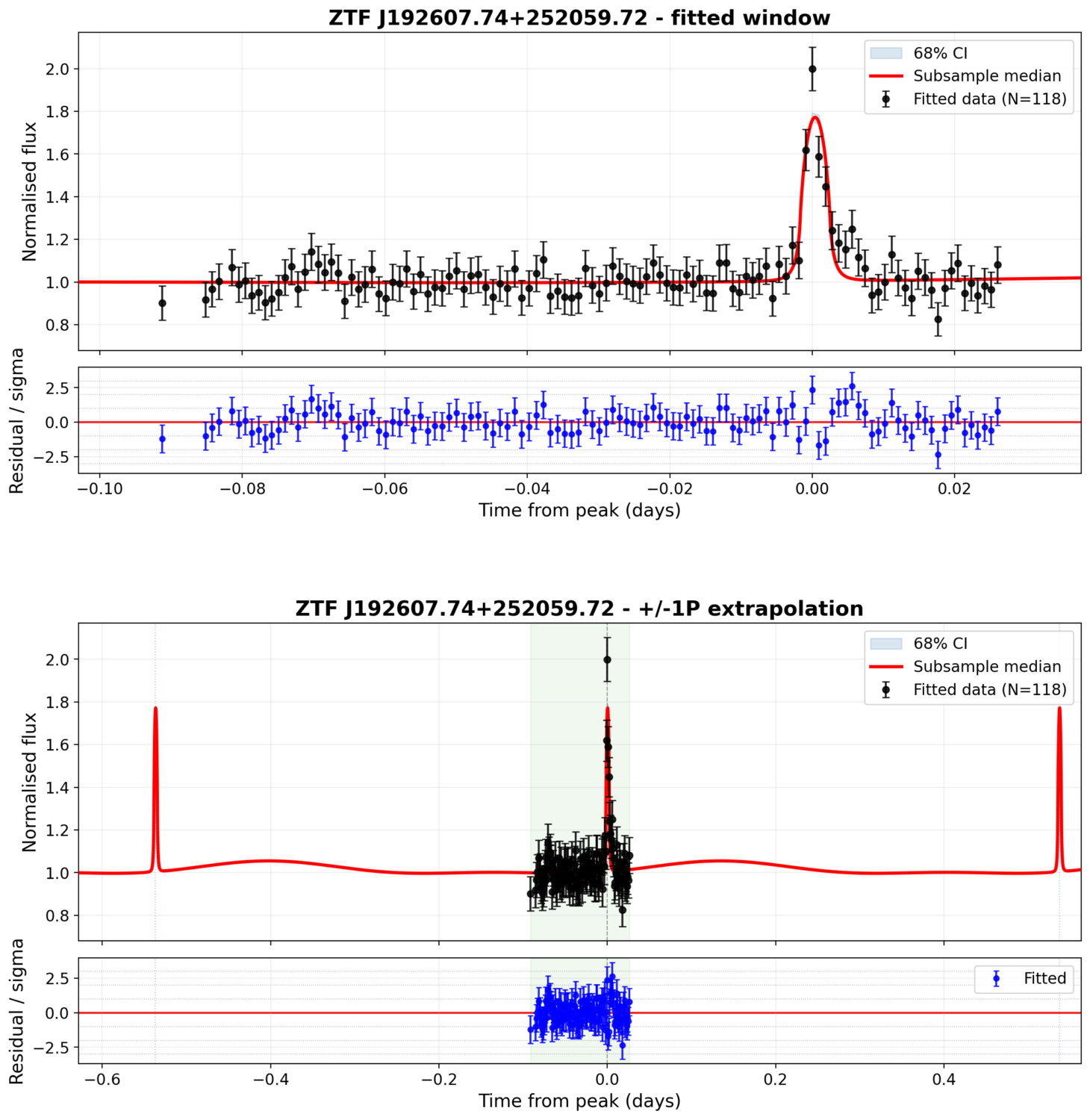}
\caption{Results of fitting the model to J192607}
\label{fig:J192607}
\end{figure}

Here we plot the data and best-fitting model for the remaining 12 self-lensing candidates not shown in the main text. Figures ~\ref{fig:J182258_full}--\ref{fig:J190822_full} have the same format as Figures \ref{fig:J201150_full} and \ref{fig:J183450_full}, in that we plot the data and model in the fitted range (top left) and an extended range (bottom left) and the parameter posterior distribution (right). For the remaining 9 candidates (Figures \ref{fig:J190041}--\ref{fig:J192607}), which have the highest best-fitting compact object masses, we omit the posterior distribution for brevity.





\bsp	
\label{lastpage}
\end{document}